\pdfoutput=1
\documentclass[prl,twocolumn,superscriptaddress,nofootinbib]{revtex4-2}
\usepackage{amsmath,amssymb,bm,graphicx,xcolor}
\usepackage{longtable,amsthm}
\newtheorem{lemma}{Lemma}
\newtheorem{theorem}{Theorem}
\usepackage[hidelinks]{hyperref}
\usepackage{dsfont}
\newcommand{\kc}{k_{\mathrm{c}}}
\newcommand{\ks}{k_{\mathrm{s}}}
\newcommand{\net}[1]{\textbf{#1}}

\newcommand{\one}{\mathds{1}}
\newcommand{\gainWinNbo}{255}
\newcommand{\gainPosNbo}{256}
\newcommand{\gainWinFlu}{252}
\newcommand{\gainPosFlu}{256}
\newcommand{\gainWinHcb}{244}
\newcommand{\gainPosHcb}{251}

\newcommand{\gainPosDia}{91}
\newcommand{\gainNq}{256}

\newcommand{\gainGammaFlu}{0.0018}
\newcommand{\gainGammaFluPct}{7}
\newcommand{\lvtSplit}{1.801}
\newcommand{\otherPairs}{7}
\newcommand{\otherNq}{256}
\newcommand{\otherFracMin}{94}
\newcommand{\otherFracMax}{100}
\newcommand{\otherMeanMin}{0.925}
\newcommand{\otherMeanMax}{0.999}
\newcommand{\otherLow}{\net{pcu}$\otimes$\net{urk}}
\newcommand{\zvBcu}{5.5}
\newcommand{\zvHex}{3.0}
\newcommand{\zvT}[2]{\ifcsname zv@#1@#2\endcsname\csname zv@#1@#2\endcsname\else--\fi}
\expandafter\def\csname zv@acs@gra\endcsname{37.4}
\expandafter\def\csname zv@bcu@nbo\endcsname{5.5}
\expandafter\def\csname zv@hex@bnn\endcsname{3.0}
\expandafter\def\csname zv@bsn@lvt\endcsname{59.8}
\expandafter\def\csname zv@hxg-d@hxg\endcsname{4.2}
\expandafter\def\csname zv@nia@nia-d\endcsname{31.7}
\expandafter\def\csname zv@pcu@urk\endcsname{34.0}
\expandafter\def\csname zv@qtz-x@uom\endcsname{75.7}
\expandafter\def\csname zv@fcu@flu\endcsname{1.5}
\expandafter\def\csname zv@she-d@she\endcsname{0.7}
\newcommand{\diaLo}{0.14}
\newcommand{\diaHi}{2.6}
\newcommand{\falseCallBound}{63}
\newcommand{\gtOne}{2.6}
\newcommand{\gtSix}{9.6}
\newcommand{\gtThree}{42.9}
\newcommand{\gtC}{3.9}
\newcommand{\gtSlope}{2.1}
\newcommand{\gtFitLo}{0.2}
\newcommand{\gtFitHi}{0.6}
\newcommand{\gtFitRes}{12}
\newcommand{\hcbMedian}{1.4989}
\newcommand{\hcbFigMedian}{1.4964}

\newcommand{\orbAnisoLo}{0.9}
\newcommand{\orbAnisoHi}{2.1}
\newcommand{\orbFluWidthBroken}{0.43}
\newcommand{\orbSlopeU}{2.00}
\newcommand{\orbSlopeL}{2.00}
\newcommand{\orbLvalue}{0.8}
\newcommand{\orbSL}{\begin{pmatrix}1.5&0&0.4\\0&1.5&0.2\\0.4&0.2&0.8\end{pmatrix}}

\newcommand{\pctDual}{66}
\newcommand{\pctOther}{19}
\newcommand{\exactWidth}{4\times10^{-14}}
\newcommand{\exactNq}{1,032}
\newcommand{\exactKcs}{0.1, 1\text{ and }3}
\newcommand{\lvtCtwo}{2.80}
\newcommand{\uomCtwo}{4.38}
\newcommand{\protEligible}{88}
\newcommand{\protFlat}{14}
\newcommand{\protFlatCalled}{8}
\newcommand{\missGroup}{1.930712}
\newcommand{\missWidth}{2.2\times10^{-14}}
\newcommand{\znContactMean}{0.44}

\newcommand{\znContactRange}{0.91}
\newcommand{\znIntrinsicMean}{31}
\newcommand{\znRatio}{70}
\newcommand{\znOperatorMean}{0.78}
\newcommand{\znOperatorRel}{4.7}
\newcommand{\metaNboLock}{1.8}
\newcommand{\metaNboFree}{35}
\newcommand{\metaNboLockTen}{0.18}

\newcommand{\metaNboRel}{0.040}
\newcommand{\metaNboLockHundred}{0.023}
\newcommand{\metaNboCutZero}{9.6}
\newcommand{\metaNboCutTwo}{0.31}
\newcommand{\metaNiaLock}{0.19}
\newcommand{\metaNiaFree}{4.0}

\newcommand{\metaNiaLockRaw}{0.34}

\newcommand{\metaNboLockMid}{0.31}
\newcommand{\metaNboFreeMid}{17}
\newcommand{\metaNboGapOne}{5}
\newcommand{\metaNboGapFive}{16}
\newcommand{\metaNq}{48}
\newcommand{\metaGapMin}{0.05}
\newcommand{\metaOtherLockMin}{13}
\newcommand{\metaOtherLockMax}{105}
\newcommand{\metaFreeMin}{4.0}
\newcommand{\metaFreeMax}{135}
\newcommand{\metaIdentity}{1\times10^{-15}}
\newcommand{\metaDotMin}{0.33}
\newcommand{\metaDotMax}{0.88}
\newcommand{\metaHzPoly}{282}
\newcommand{\metaHzSteel}{1129}
\newcommand{\metaMainWidth}{0.48}

\newcommand{\metaMainRatio}{15.7}
\newcommand{\metaMainKb}{30}
\newcommand{\metaMainKa}{25.1}
\newcommand{\metaSteelFactor}{4.006}
\newcommand{\metaRelHz}{216.5}
\newcommand{\metaRelWidthPhys}{6.5\times10^{-4}}
\newcommand{\metaWeldHz}{263}
\newcommand{\metaWeldWidth}{17}
\newcommand{\metaWeldRot}{360}
\newcommand{\metaWeldRatio}{1.28}
\newcommand{\metaLockFourRatio}{4.8}
\newcommand{\metaLockFourHz}{702}
\newcommand{\metaLockFourWidth}{4.8}
\newcommand{\metaNiaHz}{254}
\newcommand{\metaNiaWidthPhys}{0.072}

\newcommand{\metaFluRelWidth}{1.3}

\newcommand{\metaConvBand}{8\times10^{-7}}
\newcommand{\metaConvStrut}{0.12}
\newcommand{\metaPRon}{1.6}
\newcommand{\metaPRoff}{33}
\newcommand{\metaPRfree}{199}
\newcommand{\metaPRbnn}{236}
\newcommand{\metaPRnia}{1.0}
\newcommand{\metaPRniaFree}{20}

\newcommand{\metaPRbnnRel}{6.4}
\newcommand{\metaLocCells}{1.02}

\newcommand{\metaRfree}{2.7}
\newcommand{\metaFracOne}{85}
\newcommand{\metaPRonQfifty}{1.6}
\newcommand{\metaPRfreeOne}{199.2}
\newcommand{\metaPRfreeQfifty}{207.4}
\newcommand{\metaAfter}{60}
\newcommand{\metaEoffExp}{-10}
\newcommand{\metaWoffRel}{126}
\newcommand{\metaPRnboRelOff}{1.0}
\newcommand{\metaBurstBW}{4}
\newcommand{\metaNdof}{5{,}184}
\newcommand{\metaWonNbo}{6.00}
\newcommand{\metaWoffNbo}{259}
\newcommand{\metaTolLockZero}{0.27}
\newcommand{\metaTolPRLockZero}{38}

\newcommand{\metaTolLockOne}{0.36}

\newcommand{\metaTolLockThree}{0.76}

\newcommand{\metaTolLockTen}{2.4}

\newcommand{\metaTolXiLockTen}{0.67}
\newcommand{\metaTolFreeZero}{13}

\newcommand{\metaTolPRFreeTen}{32}

\newcommand{\metaTolSeeds}{12}
\newcommand{\metaTolSlopeLock}{0.24}
\newcommand{\metaTolNodesLock}{1.8}
\newcommand{\metaTolNodesOneLock}{9.5}
\newcommand{\metaTolSlopeRel}{0.42}

\newcommand{\metaTolNodesOneRel}{1.4}
\newcommand{\metaTolSlopeNia}{0.26}

\newcommand{\metaTolNodesOneNia}{3.7}

\newcommand{\metaTolPRBnnTen}{19}

\newcommand{\metaTolPRFluTen}{57}

\newcommand{\metaTolNmc}{162}
\newcommand{\metaTolFluGap}{0.003}
\newcommand{\xcHighDualRigid}{21}
\newcommand{\xcHighDual}{22}
\newcommand{\xcHighOtherRigid}{18}
\newcommand{\xcHighOther}{21}
\newcommand{\xcLowDual}{10}
\newcommand{\xcLowOther}{75}
\newcommand{\xcFisherP}{0.34}
\newcommand{\xcAtSix}{11}
\newcommand{\xcRows}{128}
\newcommand{\urkNq}{40}
\newcommand{\urkNused}{20}
\newcommand{\urkNqLarge}{400}
\newcommand{\urkNusedLarge}{204}
\newcommand{\urkCoef}{0.67}
\newcommand{\urkCoefLarge}{0.87}
\newcommand{\urkSlope}{1.95}
\newcommand{\urkKcMin}{0.003}
\newcommand{\urkKcMax}{0.1}
\newcommand{\urkGap}{0.1}
\newcommand{\znNbranch}{8}
\newcommand{\znBranchWidthMin}{6.4}
\newcommand{\znBranchWidthMax}{25.3}
\newcommand{\znOnsiteWidth}{7\times10^{-14}}
\newcommand{\xeSheN}{12}
\newcommand{\xeSheE}{24}
\newcommand{\xeSheDim}{4}
\newcommand{\xfFlatMin}{7.8\times10^{-16}}
\newcommand{\xfFlatMax}{1.3\times10^{-12}}
\newcommand{\xfNearMin}{0.010}
\newcommand{\papertitle}{Lifting mechanical zero modes without dispersion}
\makeatletter
\def\arx@refi#1#2#3\@nil{\ifx-#2\expandafter\@firstoftwo\else\expandafter\@secondoftwo\fi
  {\arx@origref{#3}}{\arx@origref{#1#2#3}}}
\AtBeginDocument{\let\arx@origref\ref\protected\def\ref#1{\arx@refi#1\@nil}}
\makeatother

\begin{document}
\title{\papertitle}
\author{Joohan Nam}
\affiliation{Technische Universit\"at Dresden, Dresden, Germany}
\begin{abstract}Floppy frameworks carry flat bands of zero-frequency mechanisms with no finite-frequency resonance, and couplings that lift them usually spread them. We show that an on-site tensor lifts a mechanism band without first-order dispersion exactly when, projected onto the mechanisms, it acts as wavevector-independent scalars on orthogonal pieces. Three structures achieve this: local normal modes, invariant symmetry sectors, and a balance law. A mechanism moves both ends of every bond between two vertex classes equally along it, and where the inter-class bond tensors are class scalars this balance law fixes its norm shares and hence the flat value from geometry alone. For the fluorite net the band lies outside the tensor spectrum, at $\tfrac{14}{9}$ of the contact stiffness to first order, and stays flat to all orders in a fixed cage. A placement inside a tiling dual that keeps the common symmetry supplies such tensors, because each contact star inherits its site symmetry. Ten catalogued dual pairs lift flat; half are eigenvalue cases a resolved count matches, and the balance law explains flat bands in random cages too. Duality makes the rigid cage common. In a printable beam-frame lattice, contacts that lock joint rotation remove first-order bending dispersion whenever bonded mechanisms are orthogonal local normals.\end{abstract}
\maketitle
\begin{figure*}[!tbp]\centering
\includegraphics[width=\textwidth]{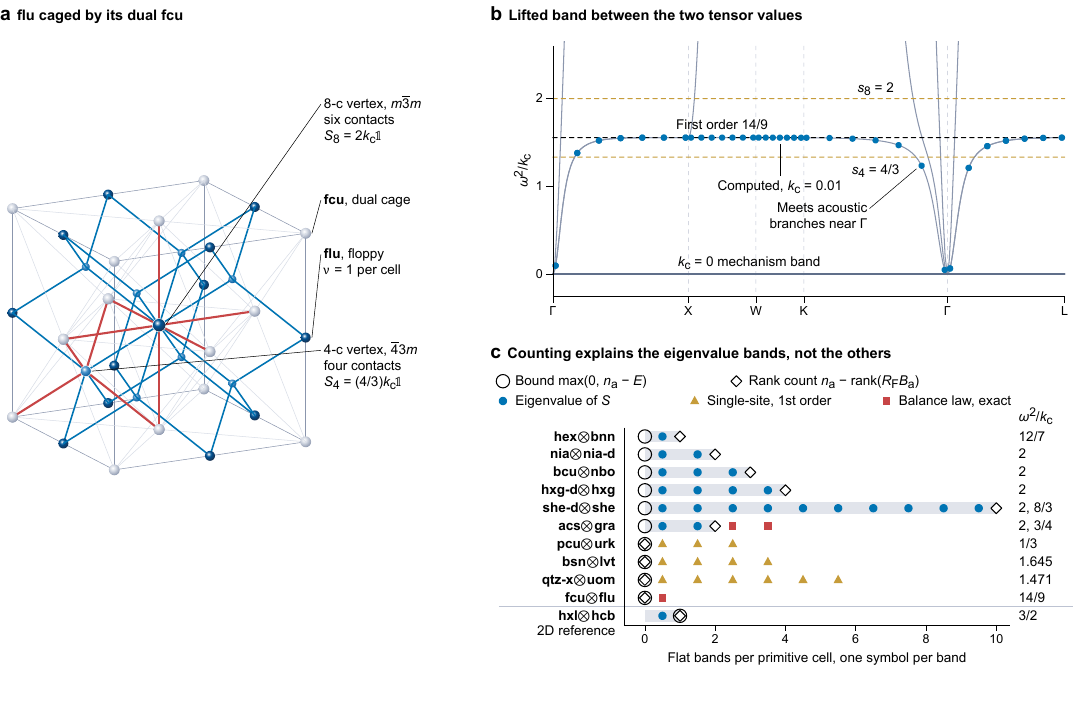}
\caption{\textbf{A flat band at a value the contact tensor does not have.} \textbf{a}, The fluorite net \net{flu} ($\nu=1$ per primitive cell) inside its dual cage, the face-centred cubic net \net{fcu}, in one conventional cubic cell. Both vertex classes of \net{flu} sit at cubic sites, so each contact tensor is isotropic. Four contacts give $S_4=\tfrac43\kc\one$ at a 4-coordinated vertex (site $\bar43m$), and six give $S_8=2\kc\one$ at an 8-coordinated vertex (site $m\bar3m$). Contacts are drawn for the two highlighted vertices only. \textbf{b}, Band structure of the full \net{fcu}$\otimes$\net{flu} composite along $\Gamma$--X--W--K--$\Gamma$--L at $\kc=0.01$, with $\ks=1$ and unit masses. Markers show modes with more than half their weight in the \net{flu} mechanism space. The computed band lies close to the first-order value $\tfrac{14}{9}\kc$ (black dashed), between the two tensor eigenvalues $\tfrac43\kc$ and $2\kc$ (gold dashed). Every bond of \net{flu} joins the two vertex classes, and a mechanism moves its two ends equally along it. This balance fixes the share of the norm of every mechanism on the 4-coordinated class at $f_4=\tfrac23$. Near $\Gamma$ the band meets the acoustic branches of the composite, as marked, an effect of finite $\kc$ outside first order. The line at zero is the mechanism band at $\kc=0$. \textbf{c}, In these pairs, counting matches the bands at eigenvalues of $S$ and none of the others. Rows are the ten three-dimensional flat dual pairs and the planar pair \net{hxl}$\otimes$\net{hcb}, with counts per common primitive cell. Open circles give the all-edge bound $\sum_a\max(0,n_a-E)$ \cite{calugaru_general_2022}, over the eigenvalues $a$ of $S$ with multiplicity $n_a$ and the number $E$ of floppy-net edges. Open diamonds and shaded bars give the rank-resolved count $\sum_a\dim(\mathcal E_a\cap\mathcal K)=\sum_a[n_a-\operatorname{rank}(R_FB_a)]$, with $B_a$ a basis of the eigenspace $\mathcal E_a$, evaluated at generic wavevectors. Filled symbols give the flat bands found, one symbol per band. Circles mark eigenvalues of $S$, triangles mark single-site projections $\mathbf n^{\mathsf T}S\mathbf n$ that are flat at first order only, and squares mark balance-law values that stay flat beyond first order in a fixed cage. The right column lists the flat values $\omega^2/\kc$. The bound is zero for every three-dimensional pair. In every row the rank-resolved count equals the number of eigenvalue bands, so it matches \net{hex}$\otimes$\net{bnn}, \net{nia}$\otimes$\net{nia-d}, \net{bcu}$\otimes$\net{nbo}, \net{hxg-d}$\otimes$\net{hxg} and \net{she-d}$\otimes$\net{she} completely and the branch at $2\kc$ of \net{acs}$\otimes$\net{gra}. It assigns no band to a value outside the spectrum of $S$.}
\label{fig:concept}
\end{figure*}

\begin{figure}[!tbp]\centering
\includegraphics[width=\columnwidth,height=0.5\textheight,keepaspectratio]{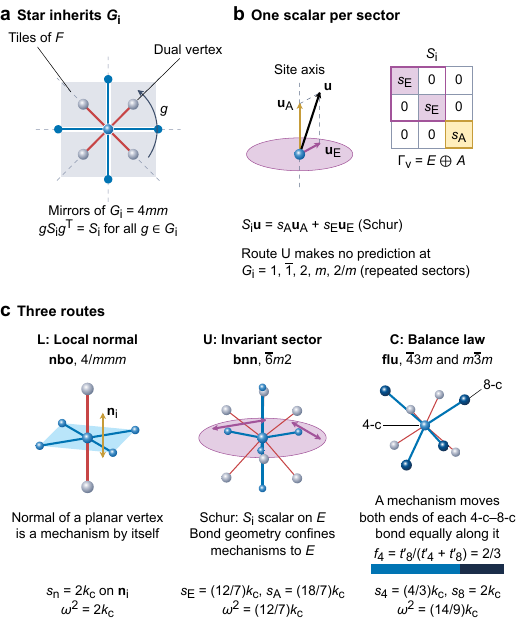}
\caption{\textbf{Three structures make the projected contact tensor scalar on orthogonal pieces of the mechanism space.} \textbf{a}, The dual places a vertex in every tile of the floppy net $F$. For a symmetry-preserving alignment, the contact star of vertex $i$ is invariant under its site group $G_i$, here the in-plane part $4mm$ of $4/mmm$ with its mirrors (dashed) and fourfold rotation $g$, and $gS_ig^{\mathsf T}=S_i$. \textbf{b}, Schur's lemma at a uniaxial site. The vector representation splits into an axial sector $A$ and an in-plane sector $E$, each occurring once, and $S_i$ acts on each as one scalar, $s_A$ or $s_E$. A displacement $\mathbf u$ splits into the same sectors. The sector rule of route U makes no prediction where a sector repeats ($G_i=1$, $\bar1$, $2$, $m$, $2/m$), where the alignment lowers the site group, or where the mechanisms are not confined to single sectors. Routes L and C do not rely on this rule. \textbf{c}, The three routes as local mechanism shapes. L, local normal: a square-planar \net{nbo} vertex moves along its normal, which carries two contacts, so $\omega^2=2\kc$. U, invariant sector: Schur's lemma makes $S_i$ a scalar on the in-plane sector $E$ of the hexagonal site of \net{bnn}. The bond geometry, with every edge along or perpendicular to the axis and rigid axial chains, then confines the mechanisms to $E$. The six contacts to \net{hex} give $s_E=\tfrac{12}{7}\kc$, so $\omega^2=\tfrac{12}{7}\kc$. C, balance law: every bond of \net{flu} joins a 4-coordinated and an 8-coordinated vertex, and a mechanism moves its two ends equally along it. The inter-class bond tensors $t'_4=\tfrac43k_F$ and $t'_8=\tfrac83k_F$ then fix the share $f_4=t'_8/(t'_4+t'_8)=\tfrac23$ of the norm of every mechanism on the 4-coordinated class. With the contact scalars $s_4=\tfrac43\kc$ and $s_8=2\kc$, the band then sits to first order at $\omega^2=\tfrac23\cdot\tfrac43\kc+\tfrac13\cdot2\kc=\tfrac{14}{9}\kc$. Contacts are drawn for the central vertex only, and the axial bonds in U are shortened.}
\label{fig:rule}
\end{figure}

\begin{figure*}[!tbp]\centering
\includegraphics[width=\textwidth]{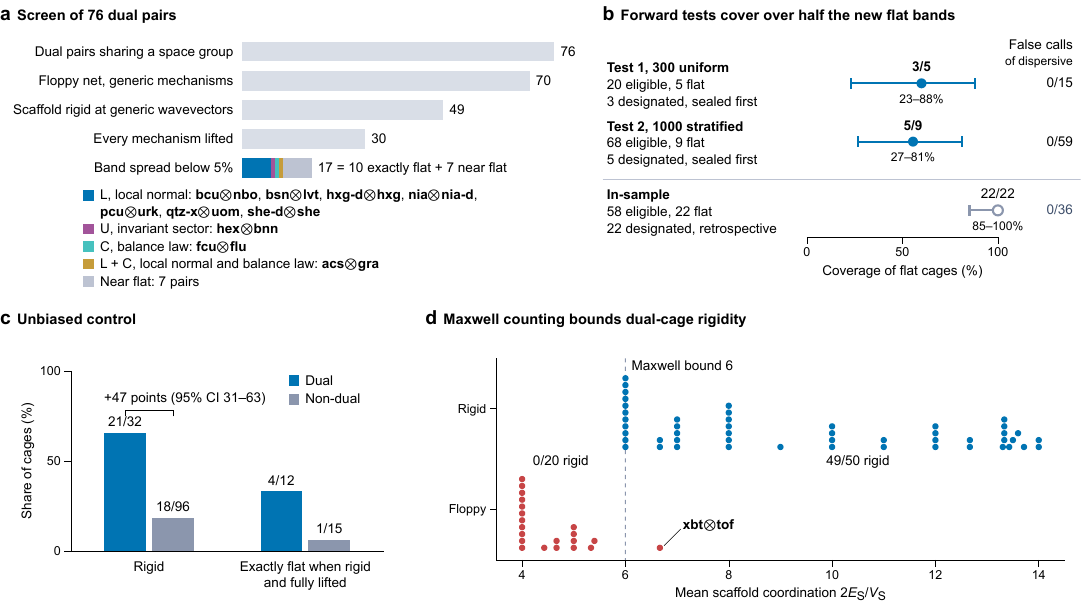}
\caption{\textbf{Ten of 76 dual pairs are exactly flat at first order, and the frozen routes find over half of the flat cages in two forward tests.} \textbf{a}, Screen of the 76 non-self-dual pairs in the Reticular Chemistry Structure Resource whose two nets share a space group. Each bar keeps the pairs that pass one more condition at generic wavevectors. Of the 17 pairs with a band spread below 5\%, 10 are exactly flat at first order and 7 are near flat. The 10 are coloured by the route that flattens them (L, local normal; U, invariant sector; C, balance law; L + C, local normals and balance law on different branches). \textbf{b}, Coverage of the frozen classifier in two forward tests, shown separately because their sampling differs. Test 1 drew 300 non-dual pairs uniformly: 20 had a rigid cage that lifts every mechanism, 5 of these were flat, and the classifier designated 3. Test 2 drew 1000 pairs stratified by crystal system: 68 were eligible, 9 were flat, and 5 were designated. In both tests the predictions were sealed before any spectrum was computed. Points give the coverage, the share of flat eligible cages that the classifier designated, with Wilson 95\% intervals. The right column gives the false calls, designated pairs among the dispersive eligible ones, which are zero in both tests. The in-sample row pools the dual, non-dual and unbiased sets on which the routes were built, counted once per ordered pair. \textbf{c}, Unbiased control. Floppy nets were drawn uniformly from the 78 nets with $\nu>0$ and a listed dual of the same space group. Each was caged by its dual and by three non-dual nets of that group, without selection on rigidity or flatness. Bars give the share of rigid scaffolds and, among cages that are rigid and lift every mechanism, the share with an exactly flat band. The bracket gives the difference in rigid share with its 95\% cluster bootstrap interval, resampled over the 32 floppy nets. \textbf{d}, Rigidity of each dual scaffold at generic wavevectors against its mean coordination $2E_S/V_S$. The dashed line marks the Maxwell bound 6. Every scaffold below the bound keeps mechanisms, and all but one at or above it are rigid.}
\label{fig:atlas}
\end{figure*}

\begin{figure}[!tbp]\centering
\includegraphics[width=\columnwidth,height=0.44\textheight,keepaspectratio]{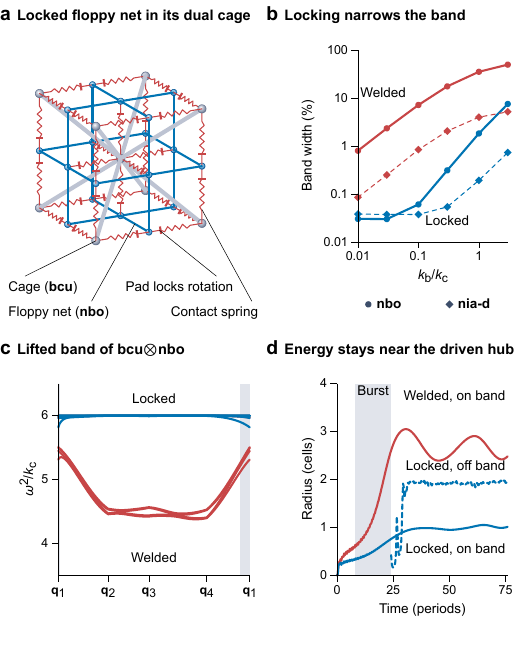}
\caption{\textbf{Architected realization.} Holding hub rotations at the contacts removes the first-order bending dispersion of the lifted band in a beam-frame metamaterial. \textbf{a}, One conventional cell of \net{bcu}$\otimes$\net{nbo} in the RCSR embedding: cage struts (grey), floppy struts and hubs (blue), and contact springs whose pads (red) hold the rotation of each hub to the cage. \textbf{b}, Largest relative branch width of the lifted band against $k_b/\kc$ for \net{bcu}$\otimes$\net{nbo} (circles, solid) and \net{nia}$\otimes$\net{nia-d} (diamonds, dashed), with locked (blue) and welded (red) joints. Units are the floppy axial stiffness $k_a$ and the node mass. The parameters are $\kc=10^{-3}k_a$ and cage struts with axial and bending stiffness $10k_a$ and $k_a$, sampled at 48 random wavevectors where the stiffness gap of the floppy net is at least $0.05k_a$. At $k_b=\kc$ locking cuts the width from 35\% to 1.8\% for \net{nbo} and from 4.0\% to 0.19\% for \net{nia-d}. With a fixed cage and rigid locking, the leading residual width is second order in $k_b$ and falls as $1/k_a$. Finite rotational springs and cage motion add corrections (Section~\ref{S-sec:meta}). \textbf{c}, Lifted band of \net{bcu}$\otimes$\net{nbo} at $k_b=\kc$ along a closed path through four generic wavevectors, locked (blue) and welded (red); grey bands mark path segments where the gap of the floppy net falls below $0.05k_a$. \textbf{d}, Energy-weighted RMS distance from the driven hub, in conventional cells, after a Gaussian burst (envelope width four periods, relative bandwidth 4\%) on one \net{nbo} hub of a $6\times6\times6$-cell specimen with massless struts and $k_b=\kc$. The cage is clamped outside the block (grey: burst envelope, $\pm2\sigma$). Over the last 20 periods, an on-band burst in the locked specimen has a mean RMS radius of \metaLocCells{} cells and a participation number of \metaPRon{} nodes, against \metaPRfree{} nodes in the welded specimen. An off-band burst at $\omega^2=259\kc$ stores only about $10^{-10}$ of the on-band energy and spreads it over 33 nodes. In a 20~mm polymer cell with locked hubs and struts of radius 0.2~mm, the band sits at 282~Hz, 15.7 times below the first strut resonance.}
\label{fig:meta}
\end{figure}
Flat bands concentrate a whole band at one frequency, which is why they are sought from interacting electrons \cite{lieb_two_1989,regnault_catalogue_2022} to wave control in metamaterials \cite{karki_non-singular_2023,mao_giant_2026}. Frameworks with fewer constraints than degrees of freedom carry such bands for free, as zero-frequency mechanisms \cite{maxwell_calculation_1864,calladine_buckminster_1978,lubensky_phonons_2015,kane_topological_2014}. In network solids their analogues, the rigid-unit modes, can fill the Brillouin zone \cite{giddy_determination_1993,tan_rigid_2024}. At zero frequency, however, a band offers no resonance to drive, and the couplings that lift it generically spread it \cite{ma_spin-orbit-induced_2020}. Two exceptions are known. Counting guarantees flat bands at an eigenvalue of an on-site term when its eigenspace holds more degrees of freedom than the constraints acting on it \cite{calugaru_general_2022}. A perturbation that acts as one common scalar on the compact localized states of an electronic flat band shifts it without spreading it \cite{bodyfelt_flatbands_2014,leykam_localization_2017}. Both constructions place the lifted band at an eigenvalue of the on-site term. A bipartite operator with scalar on-site blocks places bands elsewhere through its singular values \cite{calugaru_general_2022}, but such a band is flat only where its singular value does not depend on the wavevector. What remains open is which structures make an on-site tensor keep a collective mechanism band flat, and at what value. We show that the lifted band is flat at first order exactly when the tensor, projected onto the mechanisms, acts as a wavevector-independent scalar on each of a set of orthogonal pieces. Three structures achieve this: local normal modes, invariant symmetry sectors and a balance law between vertex classes. The balance law fixes the value of a collective flat band from the coordination geometry alone, before any diagonalization, and places it outside the spectrum of the tensor, as in the fluorite net.

Reticular chemistry supplies the tensors. Most nets have a tiling dual, the net whose vertices sit inside the tiles of the natural tiling \cite{delgado_friedrichs_three-periodic_2003,blatov_three-periodic_2007,okeeffe_reticular_2008}. When a net is placed inside its dual so that their common space group is kept, each contact star inherits the site group of its floppy vertex. Among the dual pairs of the Reticular Chemistry Structure Resource, ten three-dimensional pairs lift their mechanisms into exactly flat first-order bands. Half of them are eigenvalue cases that a rank-resolved count matches. The others are set by single-site projections or by the balance law. The balance law gives first-order values of $\tfrac{14}{9}$ and $\tfrac34$ of the contact stiffness, and three more such bands in randomly drawn cages. In a fixed cage all five stay flat beyond first order in the computed spectra, and an invariant two-class subspace explains four of them. The same duality makes the required rigid cage common, because the dual of a net with few, many-faced tiles is highly coordinated: in an unbiased draw, \pctDual\% of dual scaffolds are rigid, against \pctOther\% of other scaffolds of the same space group. In an architected lattice joint bending is the remaining obstacle, and contacts that lock the rotation of the joints remove its first-order dispersion whenever the mechanisms are local normals and bonded normals are orthogonal.

\emph{Setup.} We treat each net as an unstressed central-force bar framework with pin joints and unit masses. Let $F$ be the floppy subnet, with compatibility matrix $R_F(\mathbf q)$ at Bloch wavevector $\mathbf q$ and mechanisms $\mathcal K(\mathbf q)=\ker R_F(\mathbf q)$. The second subnet, the scaffold, has no mechanisms at generic wavevectors, that is, away from a set of $\mathbf q$ of measure zero. Contacts of stiffness $\kc$ join each vertex $i$ of $F$ to its nearest scaffold vertices along unit vectors $\hat{\mathbf c}$. For a scaffold much stiffer than the contacts, they act on $F$ as an on-site tensor
\begin{equation}
S_i=\kc\sum_{c\ni i}\hat{\mathbf c}\,\hat{\mathbf c}^{\mathsf T}.
\label{eq:cage}
\end{equation}
Let $S=\bigoplus_iS_i$ be the contact operator on the floppy net. Degenerate perturbation theory on $\mathcal K(\mathbf q)$ then gives the lifted squared frequencies
\begin{equation}
\begin{aligned}\omega^2_\alpha(\mathbf q)={}&\operatorname{eig}\big[\Pi(\mathbf q)\,S\,\Pi(\mathbf q)\big]_{\mathcal K(\mathbf q)}\\&+O\!\left(\frac{\kc^2}{g_F(\mathbf q)}\right)+O\!\left(\frac{\kc^2}{g_S(\mathbf q)}\right),\end{aligned}
\label{eq:lift}
\end{equation}
where $\Pi(\mathbf q)$ projects onto $\mathcal K(\mathbf q)$. The floppy net and the scaffold have bond stiffnesses $k_F$ and $\ks$, and $g_F(\mathbf q)$ and $g_S(\mathbf q)$ are their squared-frequency gaps, that of $F$ taken above its mechanisms. Both gaps are of the order of the bond stiffness away from the sets where they close, and the expansion needs $\kc\ll g_F,g_S$ at each wavevector (Section~S1). The band is flat at first order exactly when this projected spectrum, with its multiplicities, does not depend on $\mathbf q$ throughout a generic domain of constant mechanism dimension. Equivalently, $\Pi S\Pi=\sum_\sigma\tilde s_\sigma\Pi_\sigma$, where the $\Pi_\sigma(\mathbf q)$ project onto orthogonal pieces of $\mathcal K(\mathbf q)$ of fixed dimension and the scalars $\tilde s_\sigma$ do not depend on $\mathbf q$ (Section~S1.4). The three routes are three structural ways of building these pieces (Fig.~\ref{fig:rule}).

\emph{Local mechanisms.} The first route needs no symmetry. Let $\mathbf n_i$ be the unit normal at a vertex $i$ with planar coordination. Every edge at $i$ is orthogonal to $\mathbf n_i$, so the displacement $\mathbf n_i$ of vertex $i$ alone, with Bloch phase $e^{i\mathbf q\cdot\mathbf R}$ in cell $\mathbf R$, is a mechanism at every $\mathbf q$. If these normal displacements exhaust $\mathcal K(\mathbf q)$ at generic wavevectors, the pieces are single vertices, and $\tilde s_i=\mathbf n_i^{\mathsf T}S_i\mathbf n_i$ whatever the contact geometry (Section~S2.2). This is route L, a local resonator \cite{liu_locally_2000,huang_negative_2009,mao_giant_2026} whose direction the planar coordination fixes. In the language of flat-band theory, each piece is a compact localized state of the smallest class, confined to one vertex \cite{sutherland_localization_1986,bergman_band_2008,maimaiti_compact_2017}. Its value is an eigenvalue of $S_i$ when the normal is an eigenvector, as for square-planar \net{nbo} inside \net{bcu} at $2\kc$. At a site $2/m$ in \net{lvt} it is the projection $1.645\kc$ instead. That band is flat at first order and, in a fixed cage, disperses at order $\kc^2$ (Section~S2.6).

\emph{Collective mechanisms.} Most mechanisms are not local, and for them flatness needs a projected spectrum that does not depend on $\mathbf q$. Two structural facts provide it. In the first, route U, a dual that keeps the common space group makes the contact star of every floppy vertex invariant under its site group $G_i$. By Schur's lemma, $S_i$ is then a scalar on each irreducible sector that occurs once \cite{fowler_symmetry_2000,kangwai_symmetry-adapted_2000,schulze_block-diagonalized_2009,guest_symmetry-extended_2014}, whereas a repeated sector permits non-scalar blocks \cite{chen_scalar-wave_2026,fung_exact_2026}. Route U also needs the bond geometry of the floppy net to confine the mechanisms to one such sector. In \net{hex}$\otimes$\net{bnn}, all edges are along or perpendicular to the hexagonal axis, and the axial chains are rigid. The in-plane scalar therefore places the band at $\tfrac{12}{7}\kc$, an eigenvalue of $S$. The second, route C, needs no common scalar at all. Partition the floppy vertices into two classes, and write the compatibility rows of the bonds that join them as $R'_F=[R'_A\mid R'_B]$. A mechanism annihilates these rows as well, so $R'_A\mathbf u_A=-R'_B\mathbf u_B$, and the two ends of each such bond move equally along it. Squaring gives the balance law $\sum_{i\in A}\mathbf u_i^\dagger T'_i\mathbf u_i=\sum_{j\in B}\mathbf u_j^\dagger T'_j\mathbf u_j$, with $T'_i$ the tensor of inter-class bonds at vertex $i$, whatever the bonds within a class do (Section~S2.4). Where $k_FT'$ acts as positive scalars $t'$ and $S$ as scalars $s$ on the piece that carries these mechanisms, every mechanism puts the fixed share $f_A=t'_B/(t'_A+t'_B)$ of its norm on class $A$, and $f_B=1-f_A$ on class $B$. Hence $\Pi_\sigma S\Pi_\sigma=(f_As_A+f_Bs_B)\Pi_\sigma$ on the piece $\mathcal K_\sigma$ that these mechanisms span. Because these scalars act on the two class components of each mechanism themselves, the components span an invariant subspace of the dynamical matrix in a fixed cage. The band then stays flat at every order in $\kc$ wherever such mechanisms exist, and its value departs nonlinearly from the first-order one unless the two contact scalars are equal (Section~S2.6).

\emph{A flat band outside the tensor spectrum.} The fluorite net shows the balance law at work (Fig.~\ref{fig:concept}). Inside its dual, the face-centred cubic net, the 4-coordinated vertices of \net{flu} see $S_4=\tfrac43\kc\one$ and the 8-coordinated ones see $S_8=2\kc\one$. The isotropic bond stars then give $f_4=\tfrac23$. To first order the band therefore sits at $\omega^2=\tfrac{14}{9}\kc$, between the two eigenvalues of $S$ and equal to neither (Fig.~\ref{fig:concept}a,b). In a fixed cage it stays flat at finite $\kc$ (Section~S2.6). Edge counting adds a simple reading. Because $N_4z_4=N_8z_8$, the two classes carry equal mean squared amplitude per vertex, so $\tfrac{14}{9}$ is the vertex-weighted average of the contact scalars over the three vertices of the primitive cell. The algebra is the singular-value reduction of a bipartite operator with scalar blocks \cite{calugaru_general_2022}, in which a band is flat only where its singular value is wavevector independent. The balance law supplies this condition for every mechanism, and the value follows from the coordination geometry alone.

\emph{Counting, resolved.} Counting and the routes overlap on the eigenvalue bands, and resolving the count shows where they part. Three counts of increasing resolution bear on flat bands at an eigenvalue $a$ of $S$ with multiplicity $n_a$. With all $E$ constraints of the floppy net, the bound $\max(0,n_a-E)$ is zero for all ten pairs (Fig.~\ref{fig:concept}c). Keeping only the $E_{\rm eff}$ edges whose rows act on the eigenspace $\mathcal E_a$ sharpens it to $\max(0,n_a-E_{\rm eff})$ \cite{calugaru_general_2022}, which captures every eigenvalue band of these pairs except the level at $\tfrac83\kc$ of \net{she-d}$\otimes$\net{she} (Section~S2.7). Resolving the rank of those rows gives the exact dimension $\dim(\mathcal E_a\cap\mathcal K)=n_a-\operatorname{rank}(R_FB_a)$, with $B_a$ a basis of $\mathcal E_a$. For these pairs the rank-resolved count equals the number of eigenvalue bands. It thus matches five pairs completely, the route-L and route-U cases at eigenvalues of $S$, and the branch at $2\kc$ of \net{acs}$\otimes$\net{gra} (Section~S2.7 and Table~\ref{S-tab:counting}). No eigenspace count reaches a value outside the spectrum of $S$, and five pairs have such values. Three are single-site projections of route L, flat at first order only. The other two, \net{fcu}$\otimes$\net{flu} at $\tfrac{14}{9}\kc$ and \net{acs}$\otimes$\net{gra} at $\tfrac34\kc$, are balance-law bands that stay flat beyond first order in a fixed cage.

\emph{Atlas.} These ten pairs come from a complete screen. We took all non-self-dual pairs in the Reticular Chemistry Structure Resource whose two nets share a space group \cite{okeeffe_reticular_2008}. For each pair we optimized the relative origin of the two nets for clearance, applied a nearest-contact rule and evaluated Eq.~(\ref{eq:lift}). Of 76 pairs, 70 have a floppy subnet with mechanisms throughout the zone. Of these, 49 have a scaffold without mechanisms at generic wavevectors, and in 30 of the 49 the contacts lift every mechanism. Ten of the 30 are exactly flat at first order and seven more keep their spread below 5\% (Fig.~\ref{fig:atlas}a and Table~\ref{tab:atlas}). Flatness holds at generic wavevectors. On sets of zero measure where the scaffold or the floppy net gains mechanisms, eight of the ten split their band, and only \net{fcu}$\otimes$\net{flu} and \net{she-d}$\otimes$\net{she} stay flat at every sampled wavevector away from the zone centre (Table~\ref{S-tab:split}). At a finite stiffness ratio the band also departs from its first-order value near the zone centre and near the soft sets. At $\kc/\ks=10^{-3}$ it deviates by more than 1\% on \zvBcu\% of the zone for \net{bcu}$\otimes$\net{nbo} and \zvHex\% for \net{hex}$\otimes$\net{bnn}. It deviates on 32 to 76\% for pairs with soft modes across much of the zone, such as \net{qtz-x}$\otimes$\net{uom}, for which flatness is a limit of weak coupling (Section~S3.5).

\begin{table*}[!tbp]\centering
\caption{Dual pairs with exactly flat first-order lifted bands at generic wavevectors (scaffold $\otimes$ floppy net). First-order $\omega^2/\kc$: value of the first-order band. $G_i$: site groups of the floppy vertices in the composite. Route: L, local normal; U, invariant sector; C, inter-class balance; L + C, local normals and in-plane balance on different branches. Exact: band flat at every order in $\kc$ with the cage fixed (Section~S2.6); its value may then depend nonlinearly on $\kc$. Split: band splits on a measure-zero set of wavevectors where the scaffold or the floppy net gains mechanisms (Table~\ref{S-tab:split}). Deviating zone: share of the zone where the lifted band of the full composite deviates from its first-order value by more than 1\% in $\omega^2$ at $\kc/\ks=10^{-3}$, from 8,192 wavevectors (Table~\ref{S-tab:zerovol}). $^a$Value of the embedding, not fixed by symmetry. $^b$The screened alignment lowers the site group; the symmetric alignment overlaps the nets.}
\label{tab:atlas}
\begin{tabular}{llllccc}\hline
Pair & First-order $\omega^2/\kc$ & $G_i$ & Route & Exact & Split & Deviating zone (\%) \\ \hline
\net{bcu}$\otimes$\net{nbo} & 2 & $4/mmm$ & L & yes & yes & \zvT{bcu}{nbo} \\
\net{hxg-d}$\otimes$\net{hxg} & 2 & $\bar3m$ & L & yes & yes & \zvT{hxg-d}{hxg} \\
\net{nia}$\otimes$\net{nia-d} & 2 & $\bar6m2$ & L & yes & yes & \zvT{nia}{nia-d} \\
\net{she-d}$\otimes$\net{she} & 2, 8/3 & $\bar3m$, $\bar42m$ & L & yes & no & \zvT{she-d}{she} \\
\net{bsn}$\otimes$\net{lvt} & 1.645$^a$ & $2/m$ & L & no & yes & \zvT{bsn}{lvt} \\
\net{qtz-x}$\otimes$\net{uom} & 1.471$^a$ & $2$ & L & no & yes & \zvT{qtz-x}{uom} \\
\net{pcu}$\otimes$\net{urk} & 1/3$^b$ & $3m$, $m$ & L & no & yes & \zvT{pcu}{urk} \\
\net{hex}$\otimes$\net{bnn} & 12/7$^a$ & $\bar6m2$ & U & yes & yes & \zvT{hex}{bnn} \\
\net{fcu}$\otimes$\net{flu} & 14/9 & $\bar43m$, $m\bar3m$ & C & yes & no & \zvT{fcu}{flu} \\
\net{acs}$\otimes$\net{gra} & 3/4, 2 & $\bar6m2$ & L + C & yes & yes & \zvT{acs}{gra} \\ \hline
\end{tabular}
\end{table*}

\emph{Prospective tests.} We then tested whether the routes find flat bands in cages they were not built on. We froze a classifier built on the routes and recorded its calls for random three-dimensional same-group non-dual pairs. Only then did we compute their bands, first for 300 pairs drawn uniformly and then for 1000 pairs stratified by crystal system. Of the \protEligible{} pairs with a rigid cage that lifts every mechanism, \protFlat{} were flat. The classifier designated \protFlatCalled{} of them, all by route L, and designated no dispersive pair (Fig.~\ref{fig:atlas}b, Fig.~\ref{S-fig:s_classifier} and Section~S3.6). Three of the six misses are an eigenvalue case and two compact multi-vertex bands flat at first order only (Section~S3.7). The other three carry collective bands outside the spectrum of $S$, with first-order values $1.598\kc$ in \net{pdb}$\otimes$\net{xav}, $1.906\kc$ in \net{ith-e}$\otimes$\net{urg} and $\tfrac{28}{15}\kc$ in \net{ttr}$\otimes$\net{omy}. With the cage fixed, their computed squared-frequency widths are at most $\missWidth\,k_F$ on ten sampled wavevectors at $\kc/k_F=0.1$, 1 and 3. These bands are therefore numerically flat beyond first order. Their first-order values follow from the balance law, applied with inter-class bond tensors within in-plane sectors (\net{pdb}$\otimes$\net{xav}, \net{ith-e}$\otimes$\net{urg}) or with a rigidly translating tetrahedron counted as one vertex (\net{ttr}$\otimes$\net{omy}) (Section~S3.7). The invariant two-class subspace of the balance law proves the exactness for \net{ith-e}$\otimes$\net{urg} at every generic wavevector, and for \net{pdb}$\otimes$\net{xav} wherever its sector mechanism exists, which includes every sampled wavevector. The frozen route C, which required bipartite cubic sites, was thus narrower than the law it rests on. This diagnosis was made after scoring, so the prospective coverage remains \protFlatCalled{} of \protFlat.

\emph{Duality supplies the cage.} The construction needs a scaffold that is rigid at generic wavevectors, and duality makes such scaffolds common. We drew 32 nets uniformly from the 78 nets of positive Maxwell index $\nu=3V-E$ that appear in a same-group dual pair, with $V$ vertices and $E$ edges per cell. Each was caged by its dual and by three random non-dual nets of the same space group. The dual was rigid for 21 of 32 nets (66\%), and the other nets for 18 of 96 (19\%). The difference is 47 percentage points, with a 95\% cluster-bootstrap interval of 31 to 63 (Fig.~\ref{fig:atlas}c). Coordination carries most of this effect. A scaffold whose mean coordination is below six cannot be rigid, and in the unbiased draw none of the \xcLowDual{} dual and \xcLowOther{} other scaffolds below six is rigid. At or above six, \xcHighDualRigid{} of \xcHighDual{} dual and \xcHighOtherRigid{} of \xcHighOther{} other scaffolds are rigid, a difference within sampling error (Fisher $p=\xcFisherP$, Section~S3.4). Across the atlas, likewise, none of the 20 dual scaffolds below six is rigid, against 49 of the 50 at or above six (Fig.~\ref{fig:atlas}d). Duals are highly coordinated because of the tiling. For a three-periodic tiling with $C$ tiles per cell, Euler's relation gives $\nu(G^*)=\nu(G)+2(C-V)$, so a net with few, many-faced tiles has a highly coordinated dual. The relation applies to the 59 atlas pairs whose scaffold passes both checks for being that tiling dual (Section~S3.4). Duality is therefore a reliable source of rigid cages with inherited site symmetry rather than a cause of flatness. Indeed, the same floppy nets are flat in rigid non-dual cages too (Section~S3.3). This tiling duality, a physical cage built on the tiles of a net, differs from the dualities of mechanical metamaterials, which map one configuration or hinge structure onto another \cite{fruchart_dualities_2020,lei_duality_2022}. It also differs from the self-duality of a single isostatic lattice, which gives critical flat bands \cite{lei_self-assembly_2021}.

\emph{Homo-interpenetration.} Without a rigid scaffold the single-net reduction of Eq.~(\ref{eq:lift}) no longer applies, as the commonest interpenetration motif shows. Two interpenetrated \net{dia} nets, the structure of Zn(CN)$_2$ with its low-energy rigid-unit modes \cite{batten_interpenetrating_1998,goodwin_negative_2005,goodwin_rigid_2006,mittal_relationship_2011}, lift every mechanism in the bar model. But neither subnet is rigid, so the first-order operator acts on the joint mechanism space of both nets, and here it is dispersive. Across the tested $\kc=0.003$ to $0.03$, the pooled 5--95\% interval of the lifted modes stays near $\diaLo\,\kc$ to $\diaHi\,\kc$ (Table~\ref{S-tab:control}). The ordered crystal adds a second failure. With a machine-learned interatomic potential, the rigid-unit modes of one net carry a dispersive intrinsic spectrum of mean \znIntrinsicMean{} THz$^2$, with branch widths of \znBranchWidthMin{} to \znBranchWidthMax{} THz$^2$. The on-site contact tensors of the other net, projected onto these modes, are wavevector independent to $\znOnsiteWidth$ THz$^2$ but about \znRatio{} times weaker (Section~S5.3). The \net{dia}$\otimes$\net{dia} pair therefore supplies neither a rigid cage nor a flat lifted band.

\emph{An architected realization.} The construction can be built as an interpenetrating lattice, and contact-only lattices on dual polyhedra have already been printed \cite{white_interpenetrating_2021}. In a printed lattice the joints resist bending, and bending lifts the mechanisms by itself \cite{rocklin_transformable_2017}. A beam-frame model, with struts that stretch, bend and twist, reproduces the pin-joint values and shows how bending enters (Section~\ref{S-sec:meta}). If the projected contacts act as one scalar, $\Pi S\Pi=\tilde s\Pi$, the first-order band is $\tilde s$ plus the projected bending spectrum. The contacts then add no first-order width in this scalar sector, and the flatness is set by the ratio $k_b/\kc$. Here $k_b=12EI/L^3$ is the transverse beam stiffness, with Young's modulus $E$, second moment of area $I$ and strut length $L$. Bending itself can be made harmless. If the contacts lock the rotation of the floppy joints and every floppy strut has the same $k_b$, the projected bending becomes $k_b\,\Pi(\mathcal L\otimes\one)\Pi$, with $\mathcal L$ the Bloch graph Laplacian. Its spectrum is wavevector independent whenever the mechanisms are local normals and any two bonded normals are orthogonal, which includes the case that no edge joins two mechanism vertices. The first case holds in \net{bcu}$\otimes$\net{nbo} and the second in \net{nia}$\otimes$\net{nia-d} and \net{pcu}$\otimes$\net{urk}. Releasing the bending rotations at both ends of every strut instead removes bending from the mechanisms at first order in all ten pairs (Fig.~\ref{S-fig:s_meta_map} and Table~\ref{S-tab:meta_map}).

\emph{Residual width.} We evaluated the frame model with $k_b=\kc$ at \metaNq{} wavevectors where the stiffness gap of the floppy net is at least $0.05k_a$, with $k_a$ their axial stiffness, the $k_F$ of the bar model. There locking reduces the largest relative branch width of \net{nia}$\otimes$\net{nia-d} from \metaNiaFree\% to \metaNiaLock\%, and that of \net{bcu}$\otimes$\net{nbo} from \metaNboFree\% to \metaNboLock\%. In these two pairs the contacts map the mechanisms onto themselves, so with a fixed cage and ideal locking the remaining width enters only at second order, through the bending over the stiffness gap of the floppy net. A larger axial stiffness $k_a$ at fixed $k_b$ therefore reduces it further.

\emph{Physical scales and finite response.} For a polymer lattice with a 20 mm cell, the band of the locked \net{bcu}$\otimes$\net{nbo} design lies near \metaHzPoly{} Hz, below the first strut resonance. A separate finite specimen, with massless struts, a mobile cage and $k_b=\kc$, tests the response in time. Over the final 20 periods, a burst on its band has a mean energy-weighted root-mean-square (RMS) radius of \metaLocCells{} cells. Its mean participation number is \metaPRon{} nodes. A burst on the band of the same lattice with welded joints spreads over \metaPRfree{} nodes (Fig.~\ref{fig:meta} and Section~\ref{S-sec:meta}).

\emph{Electronic crystal fields.} The rule rests only on the form $R^\dagger R+S$. It therefore also governs crystal-field splittings of $\sigma$-bonded $p$ orbitals, once the bare on-site energies are tuned to make the Bloch Hamiltonian $R^\dagger R$ \cite{wu_flat_2007,wu_p_xy-orbital_2008,zeng_interorbital_2024}. With that tuning, ligands on the dual shift the \net{flu} band by $\tfrac{14}{9}\lambda$ to first order. Without it, the nonzero flat bands sit at $\pm\tfrac{4\sqrt2}{3}V_{pp\sigma}$ with equal norm on the two sublattices, and their first-order shift is $\tfrac53\lambda$ (Section~S9).

\emph{Parametric drive.} A flat lifted branch has one first-order frequency across generic wavevectors, which a single parametric drive can target. In undamped calculations at finite coupling, a drive at $2\omega_{\rm flat}$ makes \gainPosNbo, \gainPosFlu{} and \gainPosHcb{} of \gainNq{} sampled wavevectors grow in \net{bcu}$\otimes$\net{nbo}, \net{fcu}$\otimes$\net{flu} and the planar pair \net{hxl}$\otimes$\net{hcb}, the honeycomb net inside its triangular dual (Section~S6), against \gainPosDia{} for \net{dia}$\otimes$\net{dia} (Section~S8).

\emph{Outlook.} Reticular caging splits flat-band design into two tasks, a scaffold rigid at generic wavevectors and contacts whose projection onto its mechanisms has wavevector-independent scalar pieces. Tiling duality addresses both, the first through the coordination of the dual and the second through inherited site symmetry and the balance law, which fixes the flat value from the coordination geometry before any diagonalization. Growing a second, different net inside the pores of a first is already possible \cite{perl_hetero-interpenetrated_2023}, and printed interpenetrating lattices build the mechanical version \cite{white_interpenetrating_2021}. Once driven, a flat band raises the question of temporal order, since one drive makes nearly every sampled wavevector grow at once. For the planar pair, whose single lifted band is non-degenerate at the time-reversal-invariant M points, we take this up in a companion paper.

More broadly, this result joins a growing set of physical laws that follow from reticular design principles rather than from material-specific parameters, such as the cap that ordered patches place on the stability of disordered frameworks \cite{nam_ordered_2026} and antisymmetric thermopower without magnetism in three-phase composites \cite{nam_antisymmetric_2026}. Together they point to a reticular physical science, in which the topological vocabulary of reticular chemistry becomes a predictive language for physics.
\section*{Methods}
\textbf{Nets and duals.} Nets were taken from the Reticular Chemistry Structure Resource in their embedding of maximum symmetry, together with the dual assignments listed there \cite{okeeffe_reticular_2008,delgado-friedrichs_identification_2003}. We kept the non-self-dual pairs whose two nets share a space group, 76 in total, and expanded both nets in the conventional cell of the scaffold. The relative origin was optimized on an $8^3$ grid to maximize the clearance of the dual from scaffold vertices and edges, and the symmetry of the composite was assessed after placement.

\textbf{Mechanical model.} Each edge is a central-force spring of unit stiffness, $k_F=\ks=1$, with rest length equal to its length, so the framework is unstressed and in equilibrium, and every vertex carries unit mass \cite{lubensky_phonons_2015,kane_topological_2014}. Each floppy vertex is joined by a contact spring of stiffness $\kc$ to every scaffold vertex within 1.05 times its nearest contact distance. Bloch dynamical matrices $D(\mathbf q)=R(\mathbf q)^\dagger K R(\mathbf q)$ were assembled from the compatibility matrix $R$ of the composite and the diagonal matrix $K$ of spring stiffnesses \cite{calladine_buckminster_1978,lubensky_phonons_2015}.

\textbf{Mechanism counts and lifting.} Mechanisms at a wavevector were counted as the null space of $R(\mathbf q)$, at a singular-value threshold of $10^{-8}$ relative to the largest singular value. In the atlas screen, thresholds between $2\times10^{-9}$ and $2\times10^{-6}$ give the same ranks. The scaffold was called rigid when it has no mechanism at random wavevectors. The first-order lifted spectrum of Eq.~(\ref{eq:lift}) was evaluated at 40 random wavevectors per pair. In the screen a pair was called exactly flat when its relative spread rounds to zero at four decimals. The smallest nonzero spread of Table~\ref{S-tab:screen} is 0.004, in two pairs with unlifted mechanisms, and the smallest among pairs that lift every mechanism is 0.009. The second implementation gives the ten flat pairs relative spreads $s$ of $\xfFlatMin$ to $\xfFlatMax$ and the near-flat pairs at least $\xfNearMin$ (Table~\ref{S-tab:full_flat}). The choice of threshold therefore does not matter. The prospective tests used a separate, stated rule: full scaffold rank at all sampled wavevectors, every lifted value above $10^{-8}\|S\|_2$, and a largest branch range below $10^{-8}\|S\|_2$. High-symmetry lines and planes were scanned separately because measure-zero sets of scaffold mechanisms escape random sampling. The scans cover 13 low-index directions, low-index planes and the faces and edges of the primitive zone, with 11,965 to 21,793 distinct wavevectors for each of the 17 flat and near-flat pairs, 267,367 pair--wavevector samples in total.

\textbf{Full composite and second order.} The second-order bandwidth was obtained by diagonalizing the full composite at $\kc=10^{-3}$ to $3\times10^{-2}$ with $\ks=1$. Figure~\ref{fig:concept}b follows the stated high-symmetry path. Wavevectors were drawn uniformly at random for Fig.~\ref{S-fig:drive}b and Tables~\ref{S-tab:scaling} and \ref{S-tab:control}, and as 8,192 scrambled Sobol points for the zone fractions of Table~\ref{S-tab:zerovol}.

\textbf{Parametric gain.} The contact stiffness was modulated as $\kc(t)=\kc[1+h\cos\Omega t]$ in the full composite, with $\kc=10^{-3}$, $h=0.1$ and no damping. The linear gain at each wavevector was the largest Floquet exponent of the full monodromy matrix over one drive period \cite{kovacic_mathieus_2018}, on 256 wavevectors uniform in the primitive zone and 175 drive frequencies.

\textbf{Controls.} Non-dual controls cage each floppy net of the exactly flat dual pairs with 25 other nets of the same space group, using the same alignment and contact rule. The homo-interpenetrated control was built from the doubly interpenetrated \net{dia} entry of the same resource. The unbiased control draws 32 floppy nets uniformly from the 78 distinct nets with $\nu>0$ that appear in either role of a same-group dual pair, and cages each by its dual and by three uniformly drawn nets of the same space group.

\textbf{Beam-frame model.} For the architected realization, every vertex carries three translations and three rotations, and every edge of either net is an Euler--Bernoulli beam of length $L$. It has axial stiffness $k_a$, transverse bending stiffness $k_b=12EI/L^3$, with Young's modulus $E$ and second moment of area $I$, and torsion $GJ/L$ with $GJ=EI/(1+\nu_P)$ and Poisson ratio $\nu_P=0.3$. Struts are massless and nodes carry unit mass, so the rotations are condensed statically; a distributed-mass variant subdivides each strut into beam elements with consistent mass. Contacts are normal springs $\kc$, and rotation-locking contacts add rotational springs between a floppy node and the scaffold nodes it touches. The lifted band was identified by overlap with the pin-joint mechanism space, and widths are reported at wavevectors where the stiffness gap of the floppy net is at least $0.05k_a$ (Section~\ref{S-sec:meta}).

\textbf{Zn(CN)$_2$.} Force constants of ordered Zn(CN)$_2$ were computed with the MACE-MPA-0 machine-learned interatomic potential \cite{batatia_foundation_2025} and phonopy \cite{togo_implementation_2023} after a symmetry-constrained relaxation, with finite displacements of 0.01~\AA{} in a $2\times2\times2$ supercell (Section~S5.3).

\section*{Data availability}
All data supporting the findings are given in the Supplementary Information and archived at Zenodo (\url{https://doi.org/10.5281/zenodo.24043086}). The nets were taken from the Reticular Chemistry Structure Resource (\url{https://rcsr.net}).

\section*{Code availability}
The code and its results (JSON) are archived at Zenodo (\url{https://doi.org/10.5281/zenodo.24043086}) and are available on GitHub at \url{https://github.com/joohan-git/lifting-zero-modes}. The computational scripts used in this study were developed with assistance from Claude Code (Anthropic) and Codex (OpenAI). All AI-assisted code was reviewed and tested by the author.

\renewcommand\bibsection{\section*{References}}
\bibliography{refs_clean}
\clearpage
\onecolumngrid
\begin{center}{\large\bfseries Supplementary Information}\end{center}
\setcounter{section}{0}\setcounter{table}{0}\setcounter{figure}{0}\setcounter{equation}{0}
\setcounter{secnumdepth}{1}
\renewcommand{\thesection}{S\arabic{section}}\renewcommand{\thetable}{S\arabic{table}}
\renewcommand{\thefigure}{S\arabic{figure}}\renewcommand{\theequation}{S\arabic{equation}}
\section{Setting and the flat-band theorem}

This section proves the flatness criterion behind Eq.~(\ref{M-eq:lift}) of the main text. It fixes the model first and then states the result.

\textbf{S1.1 Model.} Each subnet is a central-force spring network with unit masses and the embedding of the Reticular Chemistry Structure Resource. The floppy net $F$ has $V$ vertices and $E$ edges per primitive cell in $d$ dimensions. Its compatibility matrix $R_F(\mathbf q)$ maps vertex displacements to edge extensions at Bloch wavevector $\mathbf q$. The Maxwell--Calladine index $\nu=dV-E$ equals $\dim\ker R_F(\mathbf q)-\dim\ker R_F^\dagger(\mathbf q)$ at every $\mathbf q$, because a rank $r(\mathbf q)$ leaves kernels of dimensions $dV-r$ and $E-r$, including where the rank drops. When $\nu>0$, the mechanism space $\mathcal K(\mathbf q)=\ker R_F(\mathbf q)$ is nonzero throughout the Brillouin zone. In general the generic number of mechanisms is $M_F=\dim\mathcal K(\mathbf q)=\nu+N_{\rm ss}$, with $N_{\rm ss}$ the states of self-stress per cell; \net{hxg} and \net{nia-d} have $\nu=0$ and are floppy through states of self-stress, and \net{she} carries eight mechanisms beyond $\nu$ per conventional cell. The theory of this section is stated per primitive cell. The counts $M_S$, $M_F$ and $\nu$ of the screen (Tables~\ref{tab:screen} and \ref{tab:split}) refer instead to the conventional cell of the scaffold, in which both nets are expanded (Section~S3.1). Each floppy vertex $i$ is joined to its nearest scaffold vertices by contacts of stiffness $\kc$ along unit vectors $\hat{\mathbf c}$. If the scaffold is held fixed, the contacts add the on-site tensor $S_i=\kc\sum_{c\ni i}\hat{\mathbf c}\hat{\mathbf c}^{\mathsf T}$, and $S=\bigoplus_i S_i=\kc\bar S$, with $\bar S$ independent of $\mathbf q$ and $\kc$. The bond stiffnesses of $F$ and of the scaffold are $k_F$ and $\ks$, and the screen uses $k_F=\ks=1$.

\textbf{S1.2 Conventions.} A Bloch vector $\mathbf u$ carries one displacement $\mathbf u_j$ per vertex of the cell, and vertex $j$ in cell $\mathbf R$ moves by $\mathbf u_je^{i\mathbf q\cdot\mathbf R}$. The released code writes $\mathbf q\cdot\mathbf R$ as $2\pi\mathbf q_{\rm red}\cdot\mathbf s$ in reduced coordinates. For an edge from $i$ in cell 0 to $j$ in cell $\mathbf R$, the compatibility row is $(R_F\mathbf u)_e=\hat{\mathbf b}_e^{\mathsf T}(e^{i\mathbf q\cdot\mathbf R}\mathbf u_j-\mathbf u_i)$, with a real unit bond vector $\hat{\mathbf b}_e$. Plane normals of vertices are written $\mathbf n_i$ and contact directions $\hat{\mathbf c}$. Norms and projectors use the Hermitian inner product, and bond projections use a transpose because bond vectors are real. The floppy net contributes $D_F=k_FR_F^\dagger R_F$. A generic domain $U$ is a set of wavevectors on which $m=\dim\mathcal K(\mathbf q)$ is constant and, for a movable scaffold, its stiffness matrix $D_S(\mathbf q)$ is positive definite. On $U$, $g_F(\mathbf q)$ is the smallest positive eigenvalue of $D_F$, and $g_S(\mathbf q)$ is the smallest eigenvalue of $D_S$. Wavevectors outside $U$ form sets of zero measure and are treated separately.

\textbf{S1.3 First-order lifting.}
\begin{lemma}\label{lem:first}
Let $\Pi(\mathbf q)$ be the orthogonal projector onto $\mathcal K(\mathbf q)$ and $Z(\mathbf q)$ an orthonormal basis of it, and hold the scaffold fixed. At $\mathbf q\in U$ with $\|S\|\ll g_F(\mathbf q)$, the $m$ squared frequencies that emerge from the mechanisms are the eigenvalues of $Z^\dagger SZ$, the matrix of $\Pi S\Pi$ on $\mathcal K(\mathbf q)$, up to corrections of order $\|S\|^2/g_F(\mathbf q)$.
\end{lemma}
\begin{proof}
With unit masses and the scaffold fixed, the dynamical matrix is $D(\mathbf q)=D_F(\mathbf q)+S$. The zero eigenspace of $D_F$ is $\mathcal K(\mathbf q)$, and the rest of its spectrum lies at or above $g_F(\mathbf q)>0$. Degenerate perturbation theory in $S$ gives the first-order eigenvalues as those of $Z^\dagger SZ$. The second-order effective matrix is $-Z^\dagger SD_F^+SZ$, and $\|D_F^+\|=1/g_F$ bounds it by $\|S\|^2/g_F(\mathbf q)$, of order $\kc^2/k_F$ at fixed geometry. The estimate is pointwise. It is uniform on any subset of $U$ with a common positive lower bound on $g_F$, and it fails near the sets of zero measure where $g_F$ closes.
\end{proof}
A movable scaffold couples to the floppy net through the off-diagonal contact block $\kc G(\mathbf q)$, whereas the floppy-floppy contact block is still $S$. Through second order it adds $-\kc^2Z^\dagger GD_S^{-1}G^\dagger Z$ to the effective matrix. Since $\|D_S^{-1}\|=1/g_S$, this correction is of order $\kc^2\|G\|^2/g_S(\mathbf q)$, with $g_S$ of order $\ks$ away from the soft sets of the scaffold (Section~\ref{sec:second}). The estimate needs $D_S(\mathbf q)>0$, that is, no scaffold mechanism at that wavevector. Both corrections are pointwise, and both grow near gap closings. Both second-order terms are negative semidefinite, and neither has to create width, because each may vanish or be independent of $\mathbf q$.

\textbf{S1.4 Flatness criterion.}
\begin{theorem}
The $m$ lifted bands are flat to first order on $U$ if and only if the spectrum of $Z^\dagger(\mathbf q)SZ(\mathbf q)$, counted with multiplicities, does not depend on $\mathbf q\in U$. Equivalently, $\mathcal K(\mathbf q)$ splits into orthogonal subspaces $\mathcal K_\sigma(\mathbf q)$ of $\mathbf q$-independent dimension, with projectors $\Pi_\sigma(\mathbf q)$, such that $\Pi_\sigma S\Pi_\tau=\delta_{\sigma\tau}\,\tilde s_\sigma\Pi_\sigma$ with $\mathbf q$-independent scalars $\tilde s_\sigma$. The scalars are eigenvalues of $S$ when $\Pi(\mathbf q)$ commutes with the eigenprojectors of $S$. In general they need not be.
\end{theorem}
\begin{proof}
The first statement restates Lemma~\ref{lem:first}. At each $\mathbf q\in U$, group the spectral projectors of $\Pi S\Pi$ on $\mathcal K(\mathbf q)$ by distinct eigenvalue. A constant spectrum with multiplicities gives these projectors constant ranks and constant eigenvalues, which is the decomposition. Conversely, $\Pi S\Pi=\sum_\sigma\tilde s_\sigma\Pi_\sigma$ has the spectrum $\{\tilde s_\sigma\}$ with the fixed multiplicities $\dim\mathcal K_\sigma$. The argument is pointwise and needs no globally chosen eigenvectors.
\end{proof}

\section{Selection rule and its three realizations}

The operator identity of Theorem~1 is not generic. This section gives three structural ways to satisfy it, which are the routes of the main text: single-site normals (L), an invariant sector on which $S$ is scalar (U), and fixed norm shares between vertex classes (C). Route U gives eigenvalues of $S$. Route L can give either eigenvalues or non-eigenvalue projections, and route C can give weighted values outside the spectrum of $S$.

\textbf{S2.1 Selection rule.}
\begin{lemma}[Inherited site symmetry]
Let the composite of $F$ and its scaffold keep the space group $G$ of $F$, and let the contacts of vertex $i$ join it to all scaffold vertices within a fixed distance. Then the contact star of $i$ is invariant under the site group $G_i$, and $S_i$ commutes with the action of $G_i$ on displacements.
\end{lemma}
\begin{proof}
An element $g\in G_i$ fixes vertex $i$, maps the scaffold onto itself and preserves distances. Its orthogonal linear part $O_g$ therefore permutes the contact directions of $i$, and $O_gS_iO_g^{\mathsf T}=\kc\sum_c(O_g\hat{\mathbf c})(O_g\hat{\mathbf c})^{\mathsf T}=S_i$.
\end{proof}
A dual built on the tile centres satisfies the hypothesis when the alignment of the two nets keeps the common space group. The clearance-maximizing alignment of Section~S3 can lower it, as for \net{pcu}$\otimes$\net{urk}, and the hypothesis must then be checked on the composite. Decompose the vector representation as $\Gamma_v|_{G_i}=\bigoplus_\alpha m_\alpha\Gamma_\alpha$ into real irreducible representations, with isotypic components $V_\alpha$. Because $S_i$ is real symmetric and commutes with $G_i$, it preserves each $V_\alpha$, and each of its eigenspaces in $V_\alpha$ is invariant under $G_i$. When $m_\alpha=1$, $V_\alpha$ is irreducible, so $S_i$ has a single eigenvalue there and acts as a scalar $s_\alpha$. When $m_\alpha>1$, the commuting symmetric operator may act as a non-scalar matrix on the multiplicity space.
\begin{theorem}[Selection rule]
Assume the hypotheses of the inherited-site-symmetry lemma, and let every multiplicity $m_\alpha$ at the vertices of $F$ equal one. Suppose that on $U$ the mechanism space is a direct sum $\mathcal K(\mathbf q)=\sum_\sigma\mathcal K_\sigma(\mathbf q)$, not assumed orthogonal, of subspaces of $\mathbf q$-independent dimension, such that the local components of every $\mathbf u\in\mathcal K_\sigma(\mathbf q)$ lie in sectors $V_{\alpha_i(\sigma)}$ with a common value $s_\sigma$. Then the first-order lifted spectrum is $\{s_\sigma\}$ with multiplicities $\dim\mathcal K_\sigma$ at every $\mathbf q\in U$.
\end{theorem}
\begin{proof}
For every $\mathbf u\in\mathcal K_\sigma$ the local hypotheses give $S\mathbf u=s_\sigma\mathbf u$. Thus $\mathcal K$ is invariant under the Hermitian operator $S$, and subspaces with distinct scalars are orthogonal. Subspaces that share a scalar combine into their direct sum, whose dimension is the sum of theirs and on which $S$ acts as that scalar. An orthonormal basis adapted to these sums diagonalizes $Z^\dagger SZ$ with the stated spectrum, and Theorem~1 applies. At wavevectors outside $U$ a larger kernel must be treated separately.
\end{proof}
For crystallographic point groups acting on three-dimensional displacements, all multiplicities equal one for groups with a principal axis of order three or more, for the cubic groups, and for $222$, $mm2$, $mmm$ and $\bar4$. They exceed one only for $1$, $\bar1$, $2$, $m$ and $2/m$, where the rule makes no prediction. Whether the mechanisms stay inside sectors is a property of the floppy net. The next three subsections give structural conditions that guarantee it, and the first of them needs no symmetry at all.

\textbf{S2.2 Local normal mechanisms (route L).}
\begin{lemma}
If a vertex $i$ of $F$ has planar coordination with real unit plane normal $\mathbf n_i$, then the Bloch vector $\mathbf e_i$ with component $\mathbf n_i$ on vertex $i$ and zero elsewhere lies in $\mathcal K(\mathbf q)$ for every $\mathbf q$.
\end{lemma}
\begin{proof}
Every edge at $i$, including an edge to a periodic image of $i$, lies in the coordination plane, so its direction is orthogonal to $\mathbf n_i$. Displacing every copy of $i$ along $\mathbf n_i$, with phase $e^{i\mathbf q\cdot\mathbf R}$ in cell $\mathbf R$, therefore changes no edge length to first order, whatever the neighbours are, and $R_F\mathbf e_i=0$ at every $\mathbf q$.
\end{proof}
The lemma needs no condition on the other vertices: for a set $P$ of planar vertices, each $\mathbf e_i$ with $i\in P$ is a mechanism.
\begin{theorem}[Local route]
Let the vectors $\mathbf e_i$ of a set $P$ of planar vertices span $\mathcal K(\mathbf q)$ on $U$, that is, $m=|P|$ per cell. Then the first-order lifted spectrum on $U$ is $\{\mathbf n_i^{\mathsf T}S_i\mathbf n_i\}_{i\in P}$, independent of $\mathbf q$ and of any symmetry of the contacts, and the band is lifted when every $\mathbf n_i^{\mathsf T}S_i\mathbf n_i>0$.
\end{theorem}
\begin{proof}
The vectors $\mathbf e_i$, $i\in P$, have disjoint vertex support and unit norm, so they form an orthonormal basis of $\mathcal K(\mathbf q)$. Because $S=\bigoplus_iS_i$ acts on each vertex separately, $\mathbf e_i^\dagger S\mathbf e_j=\delta_{ij}\,\mathbf n_i^{\mathsf T}S_i\mathbf n_i$, independent of $\mathbf q$.
\end{proof}
Symmetry is not needed for flatness here. It only makes $\mathbf n_i^{\mathsf T}S_i\mathbf n_i$ equal on each vertex orbit, so the number of distinct first-order frequencies is at most the number of orbits in $P$. The \net{nbo} net inside \net{bcu} realizes this case with two contacts along each normal, giving $\omega^2=2\kc$. In \net{lvt}, at a site $2/m$, the normal is not an eigenvector of $S_i$, and the band still sits at $\mathbf n^{\mathsf T}S\mathbf n=1.645\kc$, a value of the embedding. The fixed-cage statement needs only $\dim\mathcal K(\mathbf q)=|P|$. Read as the leading spectrum of the movable composite, it needs two conditions at each $\mathbf q$: the scaffold has no mechanism, and $\dim\mathcal K(\mathbf q)=|P|$. Both hold at generic wavevectors and can fail on sets of zero measure. In \net{bsn}$\otimes$\net{lvt} the floppy net gains a ninth mechanism on the lines $(t,t,2t)$, and one band moves from $1.645\kc$ to $\lvtSplit\kc$ there, although the scaffold stays rigid. In the other route-L pairs, including the local branch of \net{acs}$\otimes$\net{gra}, no floppy-side splitting with a rigid scaffold was found. Table~\ref{tab:split} lists, for every flat pair, the sets on which the band splits and whether the scaffold or the floppy net gains the mechanisms there.

\textbf{S2.3 Uniaxial sector (route U).} If every local component of every mechanism lies in an eigenspace of $S_i$ with the same eigenvalue $s$ at all occupied sites, then $S\mathbf u=s\mathbf u$, and all mechanisms see the same stiffness. Two separate facts are needed. Schur's lemma at the uniaxial site makes $S_i$ scalar on the in-plane sector. At generic $\mathbf q$ the little group is trivial, so the confinement of the mechanisms to that sector must come from bond geometry. Every edge of $F$ is either along or perpendicular to the axis, so $R_F$ splits into an axial and an in-plane block. The axial-chain block has no kernel at generic wavevectors, although it gains one where the axial Bloch phase is unity. The pair \net{hex}$\otimes$\net{bnn} is of this type, with $S_i$ eigenvalues $(\tfrac{12}{7},\tfrac{12}{7},\tfrac{18}{7})\kc$ and mechanisms in the degenerate plane. These values hold for the database embedding with $c=a$, in which axial edges are $\sqrt3$ times longer than in-plane edges.

\textbf{S2.4 Cubic site and fixed norm shares (route C).} At a site of cubic symmetry the vector representation is irreducible, so $S_i=s_i\one$. Isotropy flattens the band when, in addition, every mechanism distributes its norm over the vertex classes in fixed proportions. The following lemma supplies this property.
\begin{lemma}[Inter-class balance]\label{lem:balance}
Partition the vertices of $F$ into classes $A$ and $B$, and let $T'_i=\sum_{b}\hat{\mathbf b}\hat{\mathbf b}^{\mathsf T}$ be the inter-class bond tensor of vertex $i$, summed over its edges to the other class only. Then every mechanism $\mathbf u=(\mathbf u_A,\mathbf u_B)\in\mathcal K(\mathbf q)$ satisfies $\sum_{i\in A}\mathbf u_i^\dagger T'_i\mathbf u_i=\sum_{j\in B}\mathbf u_j^\dagger T'_j\mathbf u_j$ at every $\mathbf q$, whatever the intra-class edges do.
\end{lemma}
\begin{proof}
Keep only the rows of $R_F$ that belong to inter-class edges, $R'_F=[R'_A\,|\,R'_B]$. A mechanism annihilates every row of $R_F$, in particular these, so $R'_A\mathbf u_A=-R'_B\mathbf u_B$ and $\|R'_A\mathbf u_A\|^2=\|R'_B\mathbf u_B\|^2$. Each inter-class row touches one vertex of each class, so $R'^\dagger_AR'_A=\bigoplus_{i\in A}T'_i$, the Bloch phases cancelling in the squared norm. The same holds for $B$.
\end{proof}
Suppose the inter-class tensors act as positive scalars $t'_A$ and $t'_B$ on every local component of every vector of a mechanism subspace $\mathcal H$. The lemma then gives $\|\mathbf u_A\|^2=f_A\|\mathbf u\|^2$ for all $\mathbf u\in\mathcal H$, with $f_A=t'_B/(t'_A+t'_B)$, at every $\mathbf q$. Polarization of this quadratic-form identity gives $\Pi_{\mathcal H}P_A\Pi_{\mathcal H}=f_A\Pi_{\mathcal H}$, with $P_A$ the projector onto class $A$. For a bipartite net with isotropic stars of unit-stiffness bonds this is $f_A=z_B/(z_A+z_B)$, and edge counting, $N_Az_A=N_Bz_B$, then gives the two classes equal mean squared amplitude per vertex. If the contact tensors act as scalars $s_A$ and $s_B$ on the same local components, then $\Pi_{\mathcal H}S\Pi_{\mathcal H}=(f_As_A+f_Bs_B)\Pi_{\mathcal H}$, which for such a net is the vertex-weighted average of the contact scalars over a cell. Every bond of \net{flu} is inter-class, and $T'_4=\tfrac43\one$ and $T'_8=\tfrac83\one$ give $f_4=\tfrac23$. The band therefore sits at $\tfrac23\cdot\tfrac43+\tfrac13\cdot2=\tfrac{14}{9}$ in units of $\kc$. With $P_4$ the projector onto the displacements of the 4-coordinated class, direct evaluation gives $\Pi P_4\Pi=\tfrac23\Pi$ at all 200 sampled wavevectors. Applied within the in-plane sector, the lemma gives the branch of \net{gra} at $\tfrac34\kc$. Its two classes carry in-plane contact scalars 0 and $\tfrac32\kc$ and equal inter-class tensors, so $f=\tfrac12$. Its branch at $2\kc$ is made of local normals, and the pair is therefore labelled L + C. Both vertex classes of \net{she} are planar, and its two values, $2\kc$ and $\tfrac83\kc$, are the normal Rayleigh quotients of the two classes (route L). For a cluster restricted to rigid translations, sum its site tensors and keep its total mass in the norm. With nonunit masses the problem is $K\mathbf u=\omega^2M\mathbf u$, and the first-order operator is $Z^\dagger SZ$ in a mechanism basis with $Z^\dagger MZ=\one$. A translating cluster $C$ of total mass $m_C$ therefore enters with the tensors $\sum_{i\in C}S_i/m_C$ and $\sum_{i\in C}T'_i/m_C$. General rigid-unit motions need rotation coordinates as well, with a $6\times6$ generalized stiffness and mass. The translation reduction explains the first-order value met in Section~S3.7. Flatness at all orders in a fixed cage is a separate numerical result there.

\textbf{S2.5 Scope of the rule.} The contact star of a vertex inherits its site group whenever the composite structure keeps the common space group. Schur's lemma then makes $S_i$ scalar on every isotypic component of multiplicity one. For the site groups $1$, $\bar 1$, $2$, $m$ and $2/m$ some components have multiplicity two or three, and the rule makes no prediction there.

\textbf{S2.6 Exactness with a fixed cage.} With the scaffold held fixed, the dynamical matrix is $D=D_F+S$, and two of the conditions above make the lifted band exact at every order in $\kc$, not only at first order.
\begin{lemma}[Fixed-cage exactness]\label{lem:exact}
(i) If $\mathcal K(\mathbf q)=\bigoplus_a[\mathcal K(\mathbf q)\cap\mathcal E_a]$ for $\mathbf q\in U$, with $\mathcal E_a$ the eigenspaces of $S$, every lifted branch lies at exactly its eigenvalue $\omega^2=a$. Without the decomposition, each eigenvalue $a$ with $\mathcal K(\mathbf q)\cap\mathcal E_a\neq0$ still carries an exact branch of multiplicity at least $\dim(\mathcal K\cap\mathcal E_a)$. (ii) If every edge of $F$ joins classes $A$ and $B$, the bond tensors of $D_F$ are $t_A\one$ and $t_B\one$ with $t_A,t_B>0$, and the contact tensors are $s_A\one$ and $s_B\one$, the lifted band lies at the lower root of $(t_A+s_A-\omega^2)(t_B+s_B-\omega^2)=t_At_B$ at every $\mathbf q\in U$. The upper root gives a second exact flat band. Each root has multiplicity at least $m$ on $U$, and further degeneracies or crossings with other bands are not excluded. At $\kc=0$ the upper root is $t_A+t_B$, with eigenvector $(t_A\mathbf u_A,-t_B\mathbf u_B)$. At finite $\kc$ its components follow from the reduced $2\times2$ matrix.
\end{lemma}
\begin{proof}
(i) For $\mathbf u\in\mathcal K\cap\mathcal E_a$, $D\mathbf u=D_F\mathbf u+S\mathbf u=a\mathbf u$ identically, so no gap is needed; the gaps enter only to isolate the band and to bound the remainder. This identity does not use the decomposition, which gives the second statement. (ii) Write $D_F=\begin{pmatrix}t_A\one&M\\M^\dagger&t_B\one\end{pmatrix}$. A mechanism obeys $t_A\mathbf u_A=-M\mathbf u_B$ and $t_B\mathbf u_B=-M^\dagger\mathbf u_A$, so $M^\dagger M\mathbf u_B=t_At_B\mathbf u_B$. Each mechanism therefore selects a singular value $\sqrt{t_At_B}$ of $M$, and on the corresponding singular pair $D$ reduces to the wavevector-independent $2\times2$ matrix $\begin{pmatrix}t_A+s_A&\sqrt{t_At_B}\\\sqrt{t_At_B}&t_B+s_B\end{pmatrix}$.
\end{proof}
\begin{lemma}[Exact balance-law band]\label{lem:balance_exact}
Hold the cage fixed and take unit masses, so that $D=D_F+S$ with $S=\kc\bar S$. Every inter-class bond has the stiffness $k_F$, and intra-class bonds may have any stiffness. Partition the floppy vertices periodically into classes $A$ and $B$, allowing intra-class edges, and let $T'_i$ be the geometric inter-class bond tensor of Lemma~\ref{lem:balance}. Suppose that, for every $\mathbf q\in U$, a nonzero mechanism $\mathbf u=(\mathbf u_A,\mathbf u_B)\in\mathcal K(\mathbf q)$ satisfies
\begin{equation*}
T'_i\mathbf u_i=\tau_c\mathbf u_i,\qquad\bar S_i\mathbf u_i=\sigma_c\mathbf u_i\qquad(i\in c,\ c=A,B),
\end{equation*}
with constants $\tau_A,\tau_B>0$ and $\sigma_A,\sigma_B\ge0$ that do not depend on $i$ within a class, on $\mathbf q$ or on $\kc$. These equations allow $\mathbf u_i=0$ at some vertices. Put $t'_c=k_F\tau_c$ and $s_c=\kc\sigma_c$. Then $\mathbf u_A\neq0\neq\mathbf u_B$, $\mathrm{span}\{(\mathbf u_A,0),(0,\mathbf u_B)\}$ is invariant under $D$, and $D$ acts on it with the eigenvalues
\begin{equation}
\lambda_\pm=\tfrac12\Big[(t'_A+s_A)+(t'_B+s_B)\pm\sqrt{\big((t'_A+s_A)-(t'_B+s_B)\big)^2+4t'_At'_B}\,\Big],
\label{eq:balance_roots}
\end{equation}
with $\lambda_-\le\lambda_+$, the roots of $(t'_A+s_A-\lambda)(t'_B+s_B-\lambda)=t'_At'_B$. Both are therefore in the spectrum of $D$ at every $\mathbf q\in U$. If the mechanisms that satisfy these equations form a space of constant dimension $m'>0$ on $U$, each root has at least $m'$ independent eigenvectors there.
\end{lemma}
\begin{proof}
Split the rows of $R_F$ into intra-$A$, intra-$B$ and inter-class edges. An intra-$A$ row touches only class $A$, so $R_F\mathbf u=0$ gives $R_{AA}\mathbf u_A=0$; likewise $R_{BB}\mathbf u_B=0$. Hence $R_F(\mathbf u_A,0)$ is supported on the inter-class rows, where it equals $R'_A\mathbf u_A$, and
\begin{equation*}
D_F(\mathbf u_A,0)=k_FR'^\dagger R'_A\mathbf u_A=k_F(R'^\dagger_AR'_A\mathbf u_A,\,R'^\dagger_BR'_A\mathbf u_A).
\end{equation*}
 Each inter-class row touches one vertex of each class, so $R'^\dagger_AR'_A=\bigoplus_{i\in A}T'_i$, and the eigenvector condition gives $k_FR'^\dagger_AR'_A\mathbf u_A=t'_A\mathbf u_A$. The mechanism condition on the inter-class rows, $R'_A\mathbf u_A=-R'_B\mathbf u_B$, gives $k_FR'^\dagger_BR'_A\mathbf u_A=-k_FR'^\dagger_BR'_B\mathbf u_B=-t'_B\mathbf u_B$. Therefore $D_F(\mathbf u_A,0)=t'_A(\mathbf u_A,0)-t'_B(0,\mathbf u_B)$ and, by symmetry, $D_F(0,\mathbf u_B)=t'_B(0,\mathbf u_B)-t'_A(\mathbf u_A,0)$. The contact term acts as $s_A$ and $s_B$ on the two vectors. Lemma~\ref{lem:balance} gives $t'_A\|\mathbf u_A\|^2=t'_B\|\mathbf u_B\|^2$, so neither component vanishes. In the orthonormal basis $(\mathbf u_A,0)/\|\mathbf u_A\|$, $(0,\mathbf u_B)/\|\mathbf u_B\|$, $D$ is the $\mathbf q$-independent Hermitian matrix
\begin{equation*}\begin{pmatrix}t'_A+s_A&-\sqrt{t'_At'_B}\\-\sqrt{t'_At'_B}&t'_B+s_B\end{pmatrix},\end{equation*}
whose eigenvalues are Eq.~(\ref{eq:balance_roots}). By the same balance identity, a mechanism of this kind with $\mathbf u_A=0$ vanishes. A basis of $m'$ such mechanisms therefore has independent class components, and the reduction applied to it gives the multiplicity bound.
\end{proof}
Write $T=t'_A+t'_B$. The lower root is the lifted band,
\begin{equation*}
\lambda_-=\frac{t'_Bs_A+t'_As_B}{T}-\frac{t'_At'_B(s_A-s_B)^2}{T^3}+O\!\left(\frac{\kc^3}{k_F^2}\right),
\end{equation*}
and the upper root is its optical partner, which starts at $T$ when $\kc=0$. Both are exact eigenvalues for every $\kc\ge0$. The lower root is nonlinear in $\kc$ when $\sigma_A\neq\sigma_B$, and it equals $\kc\sigma$ exactly when $\sigma_A=\sigma_B=\sigma$. Without intra-class edges and with isotropic stars the lemma reduces to part (ii) of Lemma~\ref{lem:exact}, which is the singular-value reduction of a bipartite operator with scalar blocks \cite{calugaru_general_2022} (App.~I.3(a) therein). The extension adds two things: intra-class edges drop out because the mechanism already annihilates them, and the scalars need only hold on the local components that the mechanism occupies. The flat value then follows from the bond and contact geometry alone, before any diagonalization. In the released embedding of \net{pdb}$\otimes$\net{xav}, $(t'_A,t'_B)=(2,\tfrac12)k_F$ and $(s_A,s_B)=(3.999998,0.997199)\kc$. In \net{ith-e}$\otimes$\net{urg}, $(t'_A,t'_B)=(\tfrac32,2)k_F$ and $(s_A,s_B)=(1.866916,1.957543)\kc$. With the unrounded tensors, the lower root of Eq.~(\ref{eq:balance_roots}) for \net{pdb}$\otimes$\net{xav} is $1.2851$ at $\kc/k_F=1$ and $3.3973$ at $\kc/k_F=3$, in units of $k_F$. For \net{ith-e}$\otimes$\net{urg} it is $1.9052$ and $5.7122$ at the same two ratios. These equal the sampled fixed-cage band means to within $2\times10^{-14}$, here and at $\kc/k_F=0.1$. The lemma also covers the in-plane branch of \net{gra} at $\tfrac34\kc$. Both geometric inter-class tensors act as $\tfrac32$ on the in-plane sector, so $t'_A=t'_B=\tfrac32k_F$, and the contact scalars are $s_A=0$ and $s_B=\tfrac32\kc$. Eq.~(\ref{eq:balance_roots}) then gives $0.0731$, $0.5729$ and $1.0458$ in units of $k_F$ at $\kc/k_F=0.1$, 1 and 3, and the sampled fixed-cage widths are below $10^{-14}$. For \net{ttr}$\otimes$\net{omy}, in contrast, the lemma does not apply. Its translating tetrahedron occupies local directions that lie in no single eigenspace of the inter-class and contact tensors of each vertex. The two class components then do not span an invariant subspace, although their compression has fixed class averages, so the exactness of this band is a numerical finding.
Route L with normals that are eigenvectors of $S_i$ (\net{nbo} in \net{bcu}) and route U (\net{bnn} in \net{hex}) satisfy (i). Route C (\net{flu} in \net{fcu}) satisfies (ii); its energy is nonlinear in $\kc$ but independent of $\mathbf q$, with first-order term $f_As_A+f_Bs_B$. The hypothesis of (i) holds exactly when $\sum_a\dim(\mathcal E_a\cap\mathcal K)=\dim\mathcal K$. Table~\ref{tab:counting} shows the full decomposition for \net{bcu}$\otimes$\net{nbo}, \net{hxg-d}$\otimes$\net{hxg}, \net{nia}$\otimes$\net{nia-d}, \net{she-d}$\otimes$\net{she} and \net{hex}$\otimes$\net{bnn}. For \net{acs}$\otimes$\net{gra} it gives $\dim(\mathcal K\cap\mathcal E_{2\kc})=2$, so the branch at $2\kc$ is exact by the second statement of (i). Lemma~\ref{lem:balance_exact} covers its branch at $\tfrac34\kc$. Direct diagonalization with the scaffold fixed at $\kc/k_F=\exactKcs$, with $k_F$ the bond stiffness of the floppy net, checks \net{bcu}$\otimes$\net{nbo}, \net{hex}$\otimes$\net{bnn} and both roots of \net{fcu}$\otimes$\net{flu} (Table~\ref{tab:full_exact}). Their sampled squared-frequency widths are below $\exactWidth\,k_F$ on \exactNq{} wavevectors each. A second derivation, made without the first code, reaches the same lemma and the same widths. When a local normal is not an eigenvector of $S_i$, condition (i) fails and the guarantee is lost; flatness is not forbidden, but in \net{bsn}$\otimes$\net{lvt} and \net{qtz-x}$\otimes$\net{uom} the fixed-cage band does disperse at second order. On sampled wavevectors where the gap of the floppy net is at least $0.1k_F$, the largest branch widths are $\lvtCtwo\kc^2/k_F$ and $\uomCtwo\kc^2/k_F$, and the coefficients grow near gap closings. The same calculation for \net{pcu}$\otimes$\net{urk}, for $\kc/k_F$ from \urkKcMin{} to \urkKcMax, gives a largest branch width that scales with a log--log slope of \urkSlope{} in $\kc$. Its coefficient is $\urkCoef\kc^2/k_F$ on the \urkNused{} of \urkNq{} random wavevectors with a gap of at least $\urkGap k_F$, and $\urkCoefLarge\kc^2/k_F$ on \urkNusedLarge{} of \urkNqLarge. With a mobile scaffold the residual $\kc^2/\ks$ dispersion of Section~\ref{sec:second} remains; the lemma concerns the fixed cage.

\textbf{S2.7 Counting, resolved.} Three counts bear on the flat bands of this paper, and they must be kept apart. The first is the bound $\max(0,n_a-E)$, with $n_a$ the multiplicity of an eigenvalue $a$ of $S$ and $E$ all edges of $F$ per primitive cell; it is zero for all ten flat dual pairs (Table~\ref{tab:counting}). The second is the lower bound of Ref.~\cite{calugaru_general_2022} (App.~I.4 therein) restricted to the $E_{\rm eff}$ edges whose compatibility rows act on the eigenspace $\mathcal E_a$, $\dim(\mathcal E_a\cap\mathcal K)\ge\max(0,n_a-E_{\rm eff})$; equality needs those rows to have full rank. The third is the rank-resolved dimension $\dim(\mathcal E_a\cap\mathcal K)=n_a-\operatorname{rank}(R_FB_a)$, with $B_a$ an orthonormal basis of $\mathcal E_a$, which is an equality by rank--nullity and was evaluated at generic wavevectors. In the ten pairs, no row acts on the eigenspace of any eigenvalue band except at the level $\tfrac83\kc$ of \net{she-d}$\otimes$\net{she} and at $\tfrac{12}{7}\kc$ of \net{hex}$\otimes$\net{bnn}. The second bound is therefore sharp everywhere except at the \net{she} level, where $E_{\rm eff}=\xeSheE$ rows act on $n_a=\xeSheN$ coordinates per primitive cell and rank resolution gives $\xeSheDim$. For the pairs in Table~\ref{tab:counting}, the rank-resolved count equals the number of observed flat bands at eigenvalues of $S$. Each such band then consists of the $d_a=\dim(\mathcal E_a\cap\mathcal K)$ common eigenvectors of $S$ and $D_F$, and its multiplicity equals $d_a$. This equality is an observation for these pairs, not a general theorem, because a compression of $S$ to $\mathcal K$ can have eigenvalues equal to $a$ without common eigenvectors. Conversely, the effective-row bound can miss levels that rank resolution captures, because the rows that act on $\mathcal E_a$ need not be independent. The count explains \net{bcu}$\otimes$\net{nbo}, \net{hxg-d}$\otimes$\net{hxg}, \net{nia}$\otimes$\net{nia-d}, \net{hex}$\otimes$\net{bnn} and \net{she-d}$\otimes$\net{she} entirely and the branch at $2\kc$ of \net{acs}$\otimes$\net{gra}. No eigenspace count produces a flat value outside the spectrum of $S$. Five dual pairs have such values, and they fall into two classes. In \net{bsn}$\otimes$\net{lvt}, \net{qtz-x}$\otimes$\net{uom} and \net{pcu}$\otimes$\net{urk} the value is the single-site quotient $\mathbf n^{\mathsf T}S\mathbf n$ of a normal that is not an eigenvector of $S_i$. These bands are flat at first order only, and the computed fixed-cage bands of all three disperse at order $\kc^2$ (Section~S2.6). In \net{fcu}$\otimes$\net{flu} and in the branch at $\tfrac34\kc$ of \net{acs}$\otimes$\net{gra} the value is set by the inter-class balance of Section~S2.4, and the band is exact at every order in a fixed cage. Its algebra and what the balance lemma adds are given in Section~S2.6.

\section{Dual-pair screen}

This section reports the screen behind the atlas of the main text. It covers every non-self-dual pair in the Reticular Chemistry Structure Resource whose two nets share a space group.

\textbf{S3.1 Protocol.} Both nets were expanded in the conventional cell of the scaffold. The relative origin was optimized on an $8^3$ grid to maximize the clearance of the dual from scaffold vertices and edges. The symmetry of the composite was assessed after placement; for \net{pcu}$\otimes$\net{urk} it is lower than that of the nets. Each floppy vertex was joined to every scaffold vertex within 1.05 times its nearest contact distance. The scaffold role was assigned to the net with fewer mechanisms at generic wavevectors. The lifted spectrum of Lemma~\ref{lem:first} was evaluated at 40 random wavevectors.

\textbf{S3.2 Results.} Table~\ref{tab:screen} lists all 76 pairs. Seventy have a floppy net with generic mechanisms. In 49 of them the scaffold has none, and in 30 the contacts lift every mechanism. Seventeen pairs keep the spread below 0.05, and ten are exactly flat.

\textbf{Computational details.} Bloch matrices were built from the quotient graphs of the database embedding. Mechanisms were counted from the rank of $R_F(\mathbf q)$ at a relative singular-value threshold of $10^{-8}$; ranks are unchanged between $2\times10^{-9}$ and $2\times10^{-6}$. Scaffold mechanisms on sets of zero measure, such as the zone planes of \net{pcu}-type scaffolds and the $\langle110\rangle$ lines of \net{bcu}, were scanned separately on 11,965 to 21,793 wavevectors for each of the 17 flat and near-flat pairs. Table~\ref{tab:split} lists the sets on which the band of each flat pair splits. Every edge and contact has its rest length equal to its length, so all ten flat pairs are unstressed and in exact equilibrium.

\begin{longtable}{llrrrlrl}
\caption{RCSR dual pairs whose two nets share a space group. $M_S$, $M_F$: mechanisms of the scaffold and of the floppy net per conventional cell of the scaffold at a generic wavevector. $U$: median over wavevectors of the mechanisms left unlifted by the contacts. band: range of the first-order $\omega^2/k_c$ over all branches, unlifted ones included, clipped at 0. spread: at each of 40 random wavevectors of the conventional cell of the scaffold, with the scaffold fixed, the first-order eigenvalues are sorted into branches, unlifted ones included. The mean over branches of their range is divided by the mean of all values, and the result is stored to four decimals. Tables~\ref{tab:full_flat} and \ref{tab:full_retro} use the same definition. class F: exactly flat; N: spread below 0.05; blank: neither; skip: floppy net without generic mechanisms. The spread is listed for completeness but is not meaningful when the mean band value is near zero. $^a$In \net{xbp}$\otimes$\net{ttt} all but one mechanism stay unlifted and every first-order value lies below $10^{-4}\kc$, so the band rounds to zero while the spread divides a small width by a mean near zero.}\label{tab:screen} \\
\hline Scaffold & Floppy & $M_S$ & $M_F$ & $U$ & band & spread & class \\\hline\endfirsthead
\hline Scaffold & Floppy & $M_S$ & $M_F$ & $U$ & band & spread & class \\\hline\endhead
\net{acs} & \net{gra} & 0 & 4 & 0 & 0.750--2.000 & 0.000 & F \\
\net{bcu} & \net{nbo} & 0 & 6 & 0 & 2.000--2.000 & 0.000 & F \\
\net{bsn} & \net{lvt} & 0 & 8 & 0 & 1.645--1.645 & 0.000 & F \\
\net{fcu} & \net{flu} & 0 & 4 & 0 & 1.556--1.556 & 0.000 & F \\
\net{hex} & \net{bnn} & 0 & 1 & 0 & 1.714--1.714 & 0.000 & F \\
\net{hxg-d} & \net{hxg} & 0 & 4 & 0 & 2.000--2.000 & 0.000 & F \\
\net{nia} & \net{nia-d} & 0 & 2 & 0 & 2.000--2.000 & 0.000 & F \\
\net{pcu} & \net{urk} & 0 & 3 & 0 & 0.333--0.333 & 0.000 & F \\
\net{qtz-x} & \net{uom} & 0 & 6 & 0 & 1.471--1.471 & 0.000 & F \\
\net{she-d} & \net{she} & 0 & 20 & 0 & 2.000--2.667 & 0.000 & F \\
\net{bet} & \net{mep} & 0 & 46 & 0 & 0.435--1.393 & 0.031 & N \\
\net{bsn} & \net{lig} & 0 & 24 & 0 & 0.000--1.730 & 0.024 & N \\
\net{mgc-x} & \net{mtn} & 0 & 136 & 0 & 0.302--1.194 & 0.018 & N \\
\net{mgz-x} & \net{mgz-x-d} & 0 & 68 & 0 & 0.203--1.198 & 0.024 & N \\
\net{nas} & \net{nas-d} & 0 & 232 & 0 & 0.187--1.643 & 0.011 & N \\
\net{sve} & \net{cai} & 0 & 2 & 0 & 1.210--1.225 & 0.009 & N \\
\net{thp} & \net{ctn-x} & 0 & 20 & 0 & 1.317--1.557 & 0.021 & N \\
\net{alb-x} & \net{alb-x-d} & 0 & 16 & 3 & 0.000--1.411 & 0.117 &  \\
\net{ast-d} & \net{ast} & 0 & 40 & 20 & 0.000--1.713 & 0.103 &  \\
\net{bct} & \net{xbe} & 0 & 16 & 4 & 0.000--1.920 & 0.012 &  \\
\net{bcu-x} & \net{nbo-a} & 0 & 36 & 6 & 0.000--1.842 & 0.004 &  \\
\net{bcu-x} & \net{sod} & 0 & 12 & 0 & 0.802--1.600 & 0.052 &  \\
\net{bcu-x-a} & \net{xag} & 22 & 42 & 0 & 0.141--2.749 & 0.106 &  \\
\net{bsn} & \net{dia-f} & 0 & 24 & 8 & 0.000--0.879 & 0.112 &  \\
\net{cal} & \net{cal-d} & 0 & 68 & 0 & 0.000--1.083 & 0.081 &  \\
\net{ccu} & \net{doh} & 0 & 34 & 0 & 0.238--1.182 & 0.053 &  \\
\net{cla-d} & \net{cla} & 0 & 82 & 10 & 0.000--1.088 & 0.035 &  \\
\net{ddi} & \net{jph} & 0 & 28 & 0 & 0.000--2.000 & 0.161 &  \\
\net{deh-d} & \net{deh} & 0 & 48 & 16 & 0.000--0.993 & 0.185 &  \\
\net{deh-d} & \net{gis} & 0 & 16 & 0 & 0.019--1.194 & 0.262 &  \\
\net{dmp} & \net{utp} & 4 & 12 & 4 & 0.000--0.992 & 0.053 &  \\
\net{eca} & \net{xbd} & 0 & 10 & 2 & 0.000--1.427 & 0.000 &  \\
\net{fsc} & \net{mot} & 1 & 3 & 1 & 0.000--1.778 & 0.000 &  \\
\net{fse} & \net{xux} & 2 & 14 & 2 & 0.000--1.923 & 0.163 &  \\
\net{fsm} & \net{cnw} & 8 & 16 & 0 & 0.000--1.240 & 0.275 &  \\
\net{gan} & \net{lcs} & 8 & 24 & 0 & 0.667--2.650 & 0.229 &  \\
\net{gan} & \net{vab} & 8 & 72 & 0 & 0.667--0.667 & 0.000 &  \\
\net{gra} & \net{etc} & 4 & 12 & 6 & 0.000--1.000 & 0.111 &  \\
\net{hcp} & \net{hcp-d} & 0 & 2 & 0 & 0.677--1.000 & 0.305 &  \\
\net{ild} & \net{epv} & 0 & 18 & 3 & 0.000--0.412 & 0.026 &  \\
\net{jsm} & \net{uoc} & 0 & 8 & 0 & 0.027--1.350 & 0.569 &  \\
\net{lon} & \net{gra} & 4 & 4 & 0 & 1.171--2.000 & 0.000 &  \\
\net{mcf} & \net{mcd} & 8 & 8 & 2 & 0.000--1.092 & 0.226 &  \\
\net{mcf-d} & \net{mcf} & 4 & 8 & 0 & 0.000--0.906 & 0.678 &  \\
\net{odf-d} & \net{odf} & 0 & 260 & 26 & 0.000--1.026 & 0.014 &  \\
\net{qzd} & \net{bto} & 3 & 9 & 0 & 0.000--1.859 & 0.236 &  \\
\net{qzd} & \net{eta} & 3 & 9 & 0 & 0.203--1.901 & 0.235 &  \\
\net{qzd} & \net{qtz} & 3 & 3 & 0 & 0.075--2.000 & 0.577 &  \\
\net{qzd} & \net{qtz-h} & 3 & 18 & 6 & 0.000--1.000 & 0.035 &  \\
\net{qzd} & \net{uog} & 3 & 6 & 0 & 0.007--1.046 & 0.348 &  \\
\net{reo-d} & \net{cab} & 0 & 3 & 3 & 0.000--0.000 & 0.000 &  \\
\net{reo-d} & \net{nba} & 0 & 21 & 3 & 0.000--2.000 & 0.007 &  \\
\net{sta} & \net{ats} & 0 & 2 & 2 & 0.000--0.000 & 0.000 &  \\
\net{svn} & \net{svn-d} & 0 & 36 & 0 & 0.000--1.000 & 0.060 &  \\
\net{sxb} & \net{nod} & 0 & 24 & 0 & 0.000--1.438 & 0.162 &  \\
\net{sxb} & \net{nof} & 0 & 24 & 8 & 0.000--0.983 & 0.094 &  \\
\net{tsi} & \net{xbj} & 0 & 4 & 0 & 0.425--1.805 & 0.717 &  \\
\net{tsl} & \net{sgt} & 0 & 64 & 4 & 0.000--1.164 & 0.125 &  \\
\net{uni} & \net{uni-d} & 6 & 18 & 0 & 0.000--1.979 & 0.011 &  \\
\net{unj} & \net{etd} & 6 & 18 & 6 & 0.000--0.999 & 0.010 &  \\
\net{wzz} & \net{xaa} & 0 & 120 & 48 & 0.000--1.000 & 0.004 &  \\
\net{xaf} & \net{xag} & 6 & 42 & 24 & 0.000--2.549 & 0.149 &  \\
\net{xbk} & \net{ntv} & 24 & 60 & 0 & 0.635--1.372 & 0.035 &  \\
\net{xbp} & \net{ttt} & 0 & 12 & 11 & 0.000--0.000 & 2.605$^a$ &  \\
\net{xbp} & \net{xbz} & 0 & 44 & 0 & 0.000--0.999 & 0.073 &  \\
\net{xbr} & \net{spn} & 0 & 48 & 0 & 0.000--1.000 & 0.176 &  \\
\net{xbs} & \net{bod} & 0 & 15 & 5 & 0.000--1.561 & 0.203 &  \\
\net{xbs} & \net{bor} & 0 & 9 & 0 & 0.088--1.636 & 0.117 &  \\
\net{xbt} & \net{tof} & 2 & 14 & 0 & 0.427--1.048 & 0.108 &  \\
\net{zra} & \net{zra-d} & 0 & 40 & 1 & 0.000--1.236 & 0.039 &  \\
\net{crs} & \net{crs-d} & -- & 0 & -- & -- & -- & skip \\
\net{hcp} & \net{seh} & -- & 0 & -- & -- & -- & skip \\
\net{kag} & \net{kag-d} & -- & 0 & -- & -- & -- & skip \\
\net{keb} & \net{xak} & -- & 0 & -- & -- & -- & skip \\
\net{nbo-x} & \net{nbo-x-d} & -- & 0 & -- & -- & -- & skip \\
\net{reo} & \net{reo-d} & -- & 0 & -- & -- & -- & skip \\
\hline
\end{longtable}

\begingroup\small
\begin{longtable}{llllcl}
\caption{Measure-zero sets on which the generic flat band of the ten exactly flat pairs splits. The scans cover 13 low-index lines, low-index planes and the faces and edges of the primitive zone, a face $(hkl)$ being labelled by its outward normal and $\partial(hkl)$ denoting the set of its edges, with the scaffold free (joint first-order spectrum). Wavevectors are reduced coordinates of the conventional reciprocal cell. Side: scaffold, the scaffold has mechanisms on the set; floppy, the scaffold is rigid and the floppy net gains a mechanism. $M_S$, $M_F$: mechanisms of the scaffold and of the floppy net per conventional cell of the scaffold at the example wavevector; $M_F$ above its generic value means that the floppy net also gains mechanisms there. Spectrum: first-order $\omega^2/\kc$ there, with multiplicities; zeros are mechanisms left unlifted. A dash means that no split was found on the scanned sets. Extra floppy mechanisms do not always split the original flat band. When the scaffold remains rigid and the value and multiplicity of the band persist, the corresponding sets are omitted. For \net{bcu}$\otimes$\net{nbo}, this occurs at 1,133 of 2,909 scanned points. On its $\langle100\rangle$ lines, four extra mechanisms per conventional cell remain at zero frequency. There are also 14 such points each for \net{hxg-d}$\otimes$\net{hxg} and \net{she-d}$\otimes$\net{she}.}\label{tab:split} \\
\hline Pair & Side & Split set & Example $\mathbf q$ & $M_S$, $M_F$ & Spectrum \\\hline\endfirsthead
\hline Pair & Side & Split set & Example $\mathbf q$ & $M_S$, $M_F$ & Spectrum \\\hline\endhead
\net{bcu}$\otimes$\net{nbo} & scaffold & \parbox[t]{5.4cm}{\raggedright 6 lines: $[01\bar{1}]$, $[011]$, $[1\bar{1}0]$, $[10\bar{1}]$, $[101]$, $[110]$} & $(0,\,-\tfrac{1}{4},\,\tfrac{1}{4})$ & 1, 8 & \parbox[t]{2.9cm}{\raggedright $0^{\times2}$, $0.586$, $2^{\times5}$, $3.414$} \\
\net{hxg-d}$\otimes$\net{hxg} & scaffold & \parbox[t]{5.4cm}{\raggedright 3 lines: $[001]$, $[010]$, $[100]$} & $(0,\,0,\,-\tfrac{1}{4})$ & 2, 4 & \parbox[t]{2.9cm}{\raggedright $0.465^{\times2}$, $2^{\times2}$, $2.869^{\times2}$} \\
\net{nia}$\otimes$\net{nia-d} & scaffold & \parbox[t]{5.4cm}{\raggedright 20 isolated zone-boundary points, on $q_z=0$ and on the faces $q_z=\pm\tfrac12$} & $(-\tfrac{1}{2},\,-\tfrac{1}{2},\,0)$ & 3, 3 & \parbox[t]{2.9cm}{\raggedright $0$, $0.75^{\times2}$, $1.639$, $2$, $2.111$} \\
\net{she-d}$\otimes$\net{she} & -- & -- & -- & -- & -- \\
\net{bsn}$\otimes$\net{lvt} & scaffold & \parbox[t]{5.4cm}{\raggedright 9 lines: $[001]$, $[01\bar{1}]$, $[010]$, $[011]$, $[1\bar{1}0]$, $[10\bar{1}]$, $[100]$, $[101]$, $[110]$; 3 planes: $q_y=0$, $q_x=0$, $q_z=0$} & $(0,\,0,\,-\tfrac{1}{4})$ & 4, 8 & \parbox[t]{2.9cm}{\raggedright $0.414^{\times4}$, $1.645^{\times4}$, $2.48^{\times4}$} \\
 & floppy & \parbox[t]{5.4cm}{\raggedright line $[112]$ and parts of planes; the band splits at 171 of the 552 scanned points with extra floppy mechanisms} & $(\tfrac{1}{128},\,\tfrac{1}{128},\,\tfrac{1}{64})$ & 0, 9 & \parbox[t]{2.9cm}{\raggedright $0$, $1.645^{\times7}$, $1.801$} \\
\net{qtz-x}$\otimes$\net{uom} & scaffold & \parbox[t]{5.4cm}{\raggedright 11 lines: $[001]$, $[01\bar{1}]$, $[010]$, $[011]$, $[1\bar{1}\bar{1}]$, $[1\bar{1}0]$, $[1\bar{1}1]$, $[10\bar{1}]$, $[100]$, $[101]$, $[110]$; 5 planes: $q_y=0$, $q_x=0$, $q_x+q_y=0$, $q_z=0$, $q_z=\tfrac12$; 2 zone faces: $(00\bar{1})$, $(001)$; 2 zone-face edge sets: $\partial(001)$, $\partial(00\bar{1})$} & $(0,\,0,\,-\tfrac{1}{2})$ & 3, 9 & \parbox[t]{2.9cm}{\raggedright $0^{\times3}$, $0.496^{\times2}$, $0.5$, $1.471^{\times3}$, $2$, $2.378^{\times2}$} \\
\net{pcu}$\otimes$\net{urk} & scaffold & \parbox[t]{5.4cm}{\raggedright 9 lines: $[001]$, $[01\bar{1}]$, $[010]$, $[011]$, $[1\bar{1}0]$, $[10\bar{1}]$, $[100]$, $[101]$, $[110]$; 3 planes: $q_y=0$, $q_x=0$, $q_z=0$} & $(0,\,0,\,-\tfrac{1}{4})$ & 2, 9 & \parbox[t]{2.9cm}{\raggedright $0^{\times4}$, $0.057$, $0.239$, $0.414$, $0.627$, $0.777$, $2.466$, $3.086$} \\
\net{hex}$\otimes$\net{bnn} & scaffold & \parbox[t]{5.4cm}{\raggedright 1 line: $[001]$} & $(0,\,0,\,-\tfrac{1}{4})$ & 2, 2 & \parbox[t]{2.9cm}{\raggedright $0.655^{\times2}$, $4.488^{\times2}$} \\
\net{fcu}$\otimes$\net{flu} & -- & -- & -- & -- & -- \\
\net{acs}$\otimes$\net{gra} & scaffold & \parbox[t]{5.4cm}{\raggedright 4 lines: $[010]$, $[1\bar{1}0]$, $[100]$, $[110]$; 1 plane: $q_z=0$} & $(0,\,-\tfrac{1}{4},\,0)$ & 1, 5 & \parbox[t]{2.9cm}{\raggedright $0^{\times2}$, $0.75$, $2^{\times2}$, $2.25$} \\
\hline
\end{longtable}
\endgroup

\textbf{S3.3 Non-dual controls.} Duality could matter in two ways, through the rigidity of the scaffold or through the flatness itself. To separate them, each floppy net of the ten exactly flat pairs was caged by 25 other nets of its space group, with the roles held fixed. Only 21 of 250 non-dual scaffolds are rigid, against 49 of 70 dual scaffolds (Table~\ref{tab:nondual}). These controls are biased, because they start from floppy nets already known to flatten and their cells are matched to the scaffold cell. The unbiased control of Section~S3.4 replaces them for the comparison of rigidity rates.

\textbf{S3.4 Unbiased control.} From the 78 distinct nets with $\nu>0$ that appear in either role of a same-group dual pair, 32 were drawn uniformly without replacement (seed 20261111). Each was caged by its dual and by three nets of the same space group drawn uniformly, with no condition on size, rigidity or flatness. The dual was rigid for 21 of 32 nets (66\%) and the other nets for 18 of 96 (19\%). Resampling floppy nets with their four cages gives a 95\% interval of 31 to 63 percentage points for the difference. Among rigid and fully lifting cages, 4 of 12 dual cages and 1 of 15 others gave an exactly flat band. Rigidity requires a mean scaffold coordination of at least six, the Maxwell bound in three dimensions, and 49 of the 50 dual scaffolds that meet it are rigid; the exception is \net{xbt}$\otimes$\net{tof}. The same bound sorts the \xcRows{} cages of the unbiased control. Below six, none of the \xcLowDual{} dual and \xcLowOther{} other scaffolds is rigid. At or above six, \xcHighDualRigid{} of \xcHighDual{} dual and \xcHighOtherRigid{} of \xcHighOther{} other scaffolds are rigid (Fisher exact test, $p=\xcFisherP$), and \xcAtSix{} scaffolds sit exactly at six. Dual cages reach the bound far more often, \xcHighDual{} of 32 against \xcHighOther{} of 96 other cages. Most of the difference in rigidity therefore lies in how often each kind reaches the bound, and this sample resolves no difference above it. For a three-periodic tiling, Euler's relation per cell is $V-E+F-C=0$, with $F$ faces and $C$ tiles. The dual has $V^*=C$ vertices and $E^*=F$ edges, so $\nu(G^*)=3C-F$ and, with $E-F=V-C$, $\nu(G^*)=\nu(G)+2(C-V)$. A net whose tiles are few and many-faced relative to its vertices therefore has a highly coordinated dual. The relation holds only where the scaffold is the dual of a tiling of the floppy net in the scaffold cell, and two checks test this (Table~\ref{tab:full_atlas}). The first is the Euler residual of the composite cell, which is nonzero for four pairs (\net{ild}$\otimes$\net{epv}, \net{pcu}$\otimes$\net{urk}, \net{qzd}$\otimes$\net{qtz-h}, \net{qzd}$\otimes$\net{uog}). The second compares the mean face count of the catalogue natural tiling of the floppy net with the mean scaffold coordination, and it fails for seven further pairs whose residual vanishes. Five of them have the reverse catalogue assignment only: \net{acs}$\otimes$\net{gra}, \net{sta}$\otimes$\net{ats}, \net{bcu-x-a}$\otimes$\net{xag}, \net{xaf}$\otimes$\net{xag} and \net{jsm}$\otimes$\net{uoc}. In each, the catalogue lists the floppy net as the dual of the scaffold but not the scaffold as the dual of the floppy net. The listed natural duals are \net{lon} for \net{gra}, \net{ocu} for \net{xag} and \net{cds} for \net{uoc}. No dual is listed for \net{ats}. The scaffold is then the dual of a tiling of the floppy net other than its natural one, and the face count of the natural tiling does not apply. In the other two, \net{fse}$\otimes$\net{xux} and \net{xbp}$\otimes$\net{xbz}, the catalogue names the scaffold as the dual of the floppy net, yet the two means differ; the source of this mismatch was not resolved. The Euler relation is therefore applied only to the 59 pairs that pass both checks, whereas the coordination test above uses the computed scaffold graph and covers all 70 pairs. In two dimensions the dual can be realized as the reciprocal framework, with opposite Maxwell index at every wavevector \cite{zhou_topological_2019}. Table~\ref{tab:h5} lists the counts.

\begin{table}[h]\centering
\caption{Dual versus non-dual scaffolds. Non-dual controls pair each floppy net of the ten exactly flat dual pairs with 25 other nets of the same space group. Rigid: no scaffold mechanisms at a generic wavevector. Lifted: rigid and no unlifted mechanism. Near flat: lifted with spread below 0.05. Exact: lifted with zero spread.}\label{tab:nondual}
\begin{tabular}{lccccc}
\hline
Scaffold set & pairs & rigid & lifted & near flat & exact \\
\hline
dual (pairs with a floppy net) & 70 & 49 & 30 & 17 & 10 \\
non-dual controls & 250 & 21 & 13 & 11 & 11 \\
\hline
\end{tabular}
\end{table}

\begin{table}[h]\centering
\caption{Unbiased control: 32 floppy nets drawn uniformly from those with $\nu>0$ and a same-group dual, each caged by its dual and by three nets of the same space group drawn uniformly. Lifted: rigid and every mechanism lifted. Exactly flat: lifted with a first-order spread of zero.}\label{tab:h5}
\begin{tabular}{lcccc}
\hline
Cages & pairs & rigid & lifted & exactly flat \\
\hline
dual & 32 & 21 & 12 & 4 \\
same space group, non-dual & 96 & 18 & 15 & 1 \\
\hline
\end{tabular}
\end{table}

\textbf{S3.5 Flatness at finite stiffness ratio.} The first-order band is exact only in the weak-contact limit, with $\kc$ small against both $g_F(\mathbf q)$ and $g_S(\mathbf q)$. Taking $\kc/\ks\to0$ alone leaves the finite-$\kc/k_F$ corrections of Section~S2.6, and here $k_F=\ks$. We diagonalized the full composite at 8,192 wavevectors drawn uniformly from the zone (eight scrambled Sobol blocks) and assigned to each reference band the eigenstates with the largest overlap with its mechanism projector. A wavevector is counted when any assigned branch satisfies $|\omega_\alpha^2/(\kc c_\alpha)-1|>\delta$, where $\kc c_\alpha$ is its first-order squared frequency. At $\kc/\ks=10^{-3}$, the zone fractions counted for $\delta=0.01$ are 5.5\% for \net{bcu}$\otimes$\net{nbo}, 3.0\% for \net{hex}$\otimes$\net{bnn} and 4.2\% for \net{hxg-d}$\otimes$\net{hxg}. For $\delta=0.05$ they are 1.3\%, 0.6\% and 1.4\%. Five other pairs have much larger fractions, reaching 76\% and 41\% for \net{qtz-x}$\otimes$\net{uom}. Table~\ref{tab:zerovol} gives every pair and both thresholds. Counting eigenvalues inside the same windows instead of assigning branches gives the same count at every wavevector for \net{qtz-x}$\otimes$\net{uom}, so the large values are physical. They come from soft modes of the cage and of the floppy net across much of the zone. In \net{qtz-x}$\otimes$\net{uom} the median lowest scaffold eigenvalue is only 12 times the contact scale, and freezing the scaffold still leaves 58\% of the zone outside the 1\% window, against 0\% for \net{bcu}$\otimes$\net{nbo}. The relevant small parameter is therefore the ratio of the contact scale to the softest competing mode, not $\kc/\ks$ alone. Two independent implementations agree within sampling error. The two pairs whose band splits on no measure-zero set (Table~\ref{tab:split}), \net{fcu}$\otimes$\net{flu} and \net{she-d}$\otimes$\net{she}, were then computed with the released script on the same wavevectors, and they give the lowest fractions of the ten, \zvT{fcu}{flu}\% and \zvT{she-d}{she}\% outside the 1\% window at $\kc/\ks=10^{-3}$. At $\kc/\ks=10^{-2}$ the band of \net{she-d}$\otimes$\net{she} leaves the 1\% window almost everywhere but stays within 5\% on most of the zone, so its deviation there is a few per cent across the zone rather than a local excursion. Table~\ref{tab:full_zerovol} adds their sampling intervals and the share of each counted fraction that lies in the zone-centre pocket.

\begin{table}[h]\centering
\caption{Zone fractions (\%) where any assigned lifted branch deviates from its first-order squared frequency by more than 1\%; parentheses give the 5\% threshold. Full-composite spectra use 8,192 wavevectors in eight scrambled Sobol blocks, with $\ks=1$. \net{fcu}$\otimes$\net{flu} and \net{she-d}$\otimes$\net{she}, which split on no measure-zero set, were computed with the same script and wavevectors. Where the 1\% share is large but the 5\% share small, as for \net{she-d}$\otimes$\net{she} at $\kc=0.01$, the lifted branches deviate by between 1 and 5\% over most of the zone.}\label{tab:zerovol}
\begin{tabular}{lccc}
\hline
Pair & $\kc=0.001$ & $\kc=0.01$ & $\kc=0.03$ \\
\hline
\net{acs}$\otimes$\net{gra} & 37.4 (16.1) & 99.9 (51.2) & 100.0 (80.9) \\
\net{bcu}$\otimes$\net{nbo} & 5.5 (1.3) & 31.6 (9.3) & 94.1 (21.1) \\
\net{hex}$\otimes$\net{bnn} & 3.0 (0.6) & 17.1 (5.2) & 30.7 (11.6) \\
\net{bsn}$\otimes$\net{lvt} & 59.8 (33.6) & 99.7 (77.4) & 100.0 (96.4) \\
\net{hxg-d}$\otimes$\net{hxg} & 4.2 (1.4) & 27.2 (7.8) & 85.5 (18.3) \\
\net{nia}$\otimes$\net{nia-d} & 31.7 (14.0) & 99.5 (42.4) & 100.0 (71.1) \\
\net{pcu}$\otimes$\net{urk} & 34.0 (15.4) & 89.4 (45.3) & 100.0 (72.4) \\
\net{qtz-x}$\otimes$\net{uom} & 75.7 (41.0) & 100.0 (98.1) & 100.0 (100.0) \\
\net{fcu}$\otimes$\net{flu} & 1.5 (0.1) & 16.8 (3.5) & 32.8 (10.2) \\
\net{she-d}$\otimes$\net{she} & 0.7 (0.1) & 97.6 (2.5) & 100.0 (20.0) \\
\hline
\end{tabular}
\end{table}

\textbf{S3.6 Frozen classifier and prospective test.} The classifier takes as input only the floppy net, its contact stars and the rank of $R_F$ at generic wavevectors; it never sees the lifted spectrum. It predicts a flat band if every mechanism is a local normal with $\mathbf n_i^{\mathsf T}S_i\mathbf n_i>0$ (route L), or if the mechanisms lie in a uniaxial in-plane sector with a common contact value (route U). It also predicts a flat band if the floppy net is bipartite with isotropic bond and contact tensors at cubic sites (route C), or if the mechanisms split into such sectors, as in \net{gra}. Otherwise it abstains; it makes no negative predictions. Predictions apply to cages that are rigid and lift every mechanism, which requires $M_F>0$. The code was frozen with a recorded hash before the test. We drew 300 ordered pairs of distinct three-dimensional nets sharing a space-group setting, excluding all listed duals and all previously screened pairs (seed 20261121), recorded the predictions, and then computed the spectra. Five drawn pairs contained two-dimensional nets, IDs 46, 95, 153, 269 and 272 of the sample (\net{fwe}$\otimes$\net{dhd}, \net{hnd}$\otimes$\net{dhe}, \net{krf}$\otimes$\net{mth}, \net{mtd}$\otimes$\net{dhc} and \net{tts-a}$\otimes$\net{stz}). These five were replaced, under the same IDs, by uniform draws from the remaining three-dimensional pairs (\net{xag-a}$\otimes$\net{she}, \net{fti}$\otimes$\net{fit-a}, \net{mer}$\otimes$\net{tfj-a}, \net{ifi}$\otimes$\net{pmc} and \net{lrj}$\otimes$\net{nod}). The replacement followed an initial prediction run that was never scored, and it preceded both the sealing of the final predictions and the evaluation of any prospective spectrum. Prospective spectra were computed in the common primitive cell of each pair, with the eigenvalues checked against band folding of the conventional cell. Of the 20 pairs with a rigid, fully lifting cage, 5 were exactly flat. The classifier designated 3 of them by route L, \net{mob}$\otimes$\net{scp} at $1.092\kc$ and $1.333\kc$, \net{xfs}$\otimes$\net{nfc} at $0.134\kc$ and \net{npo-e}$\otimes$\net{ybh} at $5.708\kc$. It abstained on the other 2 and on all 15 dispersive pairs. Coverage is 3 of 5 (Wilson 95\% interval 23--88\%), and no pair was designated flat that is not (one-sided 95\% upper bound on the false-call rate among dispersive pairs, 18\%). With only three positive calls, the one-sided 95\% bound on the error rate among flat calls is \falseCallBound\%. A post hoc diagnosis, not added to the classifier, covers the two flat pairs it missed, \net{pdb}$\otimes$\net{xav} and \net{ith-e}$\otimes$\net{urg}. Their collective non-eigenvalue bands follow from the inter-class balance calculations of Section~S3.7, which were made after the predictions were scored. A second implementation reproduced the eligible set, the five flat pairs and the three designations pair by pair.

Routes U and C need high site symmetry, which uniform draws rarely reach. We therefore ran a second prospective test with the same frozen classifier on 1000 new pairs stratified by crystal system (seed 20261131). The sample holds 334 cubic (space groups 195--230), 333 hexagonal (168--194) and 333 tetragonal (75--142) pairs and excludes every pair screened before in either order. The sample and the 1000 predictions were sealed with recorded hashes before any spectrum was computed. One input-parser correction was made before sealing. The catalogue file writes \net{tep} in a syntax that the first parser did not read, so its pair, ID 327 (\net{edq}$\otimes$\net{tep}), first stopped with a geometry error and received no call. After the reader was corrected, with the pair, its roles and its coordinates unchanged, the classifier abstains on it, and the pair is not eligible because its scaffold keeps 36 mechanisms at generic wavevectors. Eligibility used the conservative definition of full scaffold rank at all 32 sampled generic wavevectors and every lifted value above $10^{-8}\|S\|_2$. Of 68 eligible pairs, 9 were exactly flat and 5 were designated by route L. The designated pairs were \net{kpl}$\otimes$\net{uro}, \net{ckq}$\otimes$\net{nfa}, \net{nia}$\otimes$\net{ats}, \net{svd}$\otimes$\net{knc} and \net{wfm}$\otimes$\net{hua}. The classifier abstained on the other 4 flat pairs and on all 59 dispersive pairs. Coverage was 5 of 9, with Wilson 95\% interval 27--81\%. The false-call rate among dispersive pairs was 0 of 59, with one-sided 95\% upper bound 5\%. The stratification did not reach its aim. The classifier made one route-U call and two route-C calls, all on scaffolds with mechanisms at generic wavevectors (\net{xly-a}$\otimes$\net{bnn}, \net{fnf}$\otimes$\net{dia}, \net{gng}$\otimes$\net{dia}). All route-U and route-C calls therefore fell outside the eligible set. A second implementation, written without reading the first, sealed its own predictions before evaluation and reproduced the nine flat pairs, five designations and four abstentions. It counted 70 eligible pairs, including two additional dispersive cases near the eligibility thresholds. For \net{mer-e}$\otimes$\net{xan}, the smallest scaffold singular value is about $4\times10^{-8}$. For \net{rkf}$\otimes$\net{wjf}, the lowest lifted branch approaches zero near a special plane. These differences leave the flat-pair and designation counts unchanged. The two tests are reported separately because their sampling differs; their sums, \protFlatCalled{} designated of \protFlat{} flat among \protEligible{} eligible pairs, are simple totals, not a pooled estimate. Table~\ref{tab:full_prospective} lists every eligible pair of both tests.

Retrospectively the classifier designates all 10 flat dual pairs, all 11 flat pairs among the rigid non-dual cages of Section~S3.3 and all 5 of the unbiased control, with no false designation. Counting each ordered pair once, this is 22 of 22 flat pairs against 36 dispersive ones (Fig.~\ref{M-fig:atlas}b of the main text).

\textbf{S3.7 Post hoc diagnosis of the prospective misses.} The frozen classifier abstained on six of the fourteen flat pairs of the two prospective tests, and their diagnosis shows where the frozen routes were too narrow. All six pairs, rebuilt independently for this diagnosis, remain flat and keep their total mechanism multiplicities. The diagnostic grouping merges the two nearby \net{tca}$\otimes$\net{hyb} levels into one eight-dimensional group at $\missGroup\kc$. Table~\ref{tab:full_prospective} resolves them as $1.93066\kc$ and $1.93073\kc$, with multiplicities two and six. Three of the misses have collective first-order values outside the spectrum of $S$. Their fixed-cage squared-frequency widths are at most $\missWidth\,k_F$ on ten sampled wavevectors at $\kc/k_F=0.1$, 1 and 3. Lemma~\ref{lem:balance_exact} proves this exactness for \net{ith-e}$\otimes$\net{urg} at every generic wavevector and for \net{pdb}$\otimes$\net{xav} wherever its sector mechanism exists, as shown below. For \net{ttr}$\otimes$\net{omy} it is a numerical finding. In \net{pdb}$\otimes$\net{xav} the band at $1.5978\kc$ lives in the in-plane sectors of two non-cubic classes. The geometric inter-class scalars $\tau_0=2$ and $\tau_1=\tfrac12$ fix the shares at $\tfrac15$ and $\tfrac45$. The full bond tensors would give the wrong shares $\tfrac37$ and $\tfrac47$, so this pair requires the inter-class form of the lemma. Its three further branches, at $4.6\times10^{-6}\kc$, lie far above the eligibility floor of $10^{-8}\|S\|_2$, so the pair counts as lifting every mechanism. At every sampled wavevector each local component of this mechanism lies in a common eigenspace of $T'_i$ and $S_i$, so Lemma~\ref{lem:balance_exact} applies there, and Eq.~(\ref{eq:balance_roots}) gives the band at every $\kc$. The active compatibility rows outnumber the coordinates of this sector space, so its mechanism away from the sampled wavevectors is inferred from the sampled ranks rather than from counting. In \net{ith-e}$\otimes$\net{urg} the band at $1.9058\kc$ follows in the same way from the geometric scalars $\tau_1=\tfrac32$ and $\tau_2=2$, with shares $\tfrac47$ and $\tfrac37$. The remaining class, at the 8-coordinated vertices, carries no motion, and its bonds run along the axes of their class-1 ends, so they drop out of the in-plane sector space. That space has 28 coordinates per cell, and only the 24 bonds between classes 1 and 2 act on it. At least four common-sector mechanisms therefore exist at every wavevector. Merging the resting 8-coordinated class into class 1, with $\mathbf u_i=0$ there as the lemma allows, turns its bonds into intra-class edges. Lemma~\ref{lem:balance_exact} then makes the band exact at every generic wavevector. In \net{ttr}$\otimes$\net{omy} a regular tetrahedron of four vertices translates rigidly. Treated as one effective vertex with summed tensors and its mass in the norm, it shares the norm with the central vertex in the ratio $\tfrac45:\tfrac15$. Its summed contact tensor is $4\kc\one$ over mass 4, and the central vertex has $\tfrac{16}{3}\kc\one$, so the first-order band sits at $\tfrac45\cdot1+\tfrac15\cdot\tfrac{16}{3}=\tfrac{28}{15}$ in units of $\kc$. The sum of the four uniaxial contact tensors of the tetrahedron is isotropic by its point group $T_d$. Rotations of the tetrahedron lie outside this translation reduction. The translating tetrahedron occupies local directions that lie in no single eigenspace of the inter-class and contact tensors of its vertices, so the hypothesis of Lemma~\ref{lem:balance_exact} fails, and flatness at all orders rests on the sampled widths above. The other three misses are of known types. \net{tca}$\otimes$\net{hyb} is an eigenvalue case carried by rigid triangle motions. \net{sta}$\otimes$\net{smv} and \net{wfm}$\otimes$\net{wih} carry compact multi-vertex mechanisms in which every vertex moves along a fixed line, and their bands are flat at first order only. These line directions are not eigenvectors of the local bond and contact tensors, so neither exactness lemma applies. The sampled fixed-cage bands leave their first-order values already at $\kc/k_F=0.1$. Together with \net{fcu}$\otimes$\net{flu} and \net{acs}$\otimes$\net{gra}, five collective bands outside the spectrum of $S$ are therefore flat beyond first order in a fixed cage, three of them from the prospective samples. The invariant two-class subspace explains four of them, and \net{ttr}$\otimes$\net{omy} is flat numerically. After scoring, we also tested whether the projected class-and-sector Gram operators commute and have a $\mathbf q$-independent joint spectrum. This spectral diagnosis, close to a direct evaluation of Eq.~(\ref{M-eq:lift}), reproduces all fourteen flat pairs and excludes the 74 dispersive ones. The prospective coverage of the frozen classifier remains \protFlatCalled{} of \protFlat.

\section{Second-order bandwidth}\label{sec:second}

For \net{nbo} the residual width of the band comes from scaffold motion; in other pairs soft modes of the floppy net also contribute (Section~S3.5). This section measures it for \net{nbo}.

The deviation in units of the contact stiffness, $|\omega^2-2\kc|/\kc$, was sampled at 1,500 random wavevectors for the \net{bcu} scaffold and for the \net{bcu} scaffold with added axial bonds. Its median, measured away from the zone centre, grows linearly in $\kc$ (Table~\ref{tab:scaling}), so the absolute deviation is of order $\kc^2/\ks$. The axial bonds halve the prefactor. Near the zone centre the band hybridizes with the scaffold acoustic branch, and this pocket shrinks as $\kc\to0$.

\begin{table}[h]\centering
\caption{Deviation of the \net{nbo} band from $2\kc$. The median reports $|\omega^2-2\kc|/\kc$ for wavevectors more than 0.25 reduced units from $\Gamma$. The column dev.$/\kc$ divides it by $\kc$. The last column gives the fraction of assigned modes within 1\% of $2\kc$, including the zone centre. Each sample contains 1,500 random wavevectors and six assigned modes per wavevector.}\label{tab:scaling}
\begin{tabular}{lcccc}
\hline
Scaffold & $k_c$ & median dev. & dev.$/k_c$ & within 1\% \\
\hline
\net{bcu} & 0.001 & 0.00089 & 0.890 & 0.981 \\
\net{bcu} & 0.003 & 0.00266 & 0.887 & 0.947 \\
\net{bcu} & 0.01 & 0.00887 & 0.887 & 0.821 \\
\net{bcu} & 0.03 & 0.02660 & 0.887 & 0.395 \\
\net{bcu} with axial bonds & 0.001 & 0.00041 & 0.410 & 0.998 \\
\net{bcu} with axial bonds & 0.003 & 0.00122 & 0.407 & 0.989 \\
\net{bcu} with axial bonds & 0.01 & 0.00408 & 0.408 & 0.924 \\
\net{bcu} with axial bonds & 0.03 & 0.01224 & 0.408 & 0.589 \\
\hline
\end{tabular}
\end{table}

\section{Controls}

Two controls separate the role of the scaffold from that of caging alone.

\textbf{S5.1 Homo-interpenetration.} Two \net{dia} nets interpenetrated as in cuprite lift every mechanism through 32 contacts per cubic cell. The band, however, stays dispersive, and its quantiles do not narrow with $\kc$ (Table~\ref{tab:control}). Neither subnet is rigid, so mechanisms couple to mechanisms at first order.

\textbf{S5.2 Planar dual pair.} The honeycomb net inside its triangular dual has three contacts at $120^\circ$ per vertex. The tensor is isotropic, $S_i=\tfrac32\kc\one$, and the single-band median, taken from the mode with the largest mechanism overlap, is $\hcbMedian\,\kc$ at $\kc=0.003$ (Table~\ref{tab:control}).

\textbf{S5.3 Zn(CN)$_2$ as a real homo-interpenetrated crystal.} The bar model of S5.1 can be checked in a real crystal of the same topology. We built ordered Zn(CN)$_2$ (two interpenetrated \net{dia} nets, head-to-tail cyanide order) in the face-centred cell of four formula units, which holds both nets, starting from the cubic cell $a=5.902$~\AA. We relaxed it to $a=5.943$~\AA{} with the MACE-MPA-0 machine-learned interatomic potential, and computed harmonic force constants by finite displacements in a $2\times2\times2$ supercell. The mechanism space of one net was taken as its rigid-unit modes, with ZnC$_4$ and ZnN$_4$ tetrahedra held rigid and joined by C--N bars; it has four modes per tetrahedron at generic wavevectors, the count known for this structure \cite{goodwin_negative_2005,tan_rigid_2024}. From the full-crystal force constants, so that no atom had to be removed, we define $A_i=-\sum_p\Phi_{ip}$ over the atoms $p$ of the second net and their images, and $S_i=(A_i+A_i^{\mathsf T})/2$. We divide each $S_i$ by the atomic mass $m_i$ and project $\bigoplus_i(S_i/m_i)$ onto an orthonormal mass-weighted rigid-unit basis $Z$. The eigenvalues of $S_i$ at the threefold C and N sites differ between the axial and transverse directions. The on-site projection nevertheless has wavevector-independent branches, each of width below $\znOnsiteWidth$ THz$^2$, with a mean of \znContactMean{} THz$^2$ and a pooled 5--95\% range of \znContactRange{} THz$^2$. This range measures the separation between flat branches, not dispersion. The rigid-unit band of a single net has a mean of \znIntrinsicMean{} THz$^2$, about \znRatio{} times larger, and each of its \znNbranch{} branches disperses by \znBranchWidthMin{} to \znBranchWidthMax{} THz$^2$ across the zone. A separate correction, $Z^\dagger[D_{FF}^{\rm filled}(\mathbf q)-D^{\rm isolated}(\mathbf q)]Z$, with $D^{\rm filled}_{FF}$ the block of one net in the full crystal and $D^{\rm isolated}$ that net alone, is dispersive, with a mean of \znOperatorMean{} THz$^2$ and a pooled range \znOperatorRel{} times its mean. The calculated intrinsic rigid-unit spectrum is thus already dispersive, and its mean is about \znRatio{} times the mean on-site contact contribution. Within this weak contact action the balance law still holds at the level of rigid units. Treat the ZnC$_4$ and ZnN$_4$ tetrahedra of a cyanide bridge as two classes of translating clusters, with the reduction of Section~S2.4. Each unit then has the inter-class tensor $\tfrac43\one$ over its mass, 113.4 and 121.4~u, so the mass-weighted shares are $0.483$ and $0.517$. The translational contact scalars, $\operatorname{tr}(\sum_{i}S_i)/(3m_{\rm unit})$ over each unit, are $0.455$ and $-0.548$ THz$^2$. Their share-weighted mean predicts the on-site translation doublet at $-0.0636$ THz$^2$, and the computed doublet agrees to $10^{-12}$ THz$^2$ without fitting. The value is an eigenvalue of the contact contribution alone. It is a curvature of the interaction between the two nets, negative along some local directions, and not a squared frequency of the crystal, whose rigid-unit modes also carry the intrinsic stiffness of the net.

\begin{table}[h]\centering
\caption{Lifted bands of the homo-interpenetrated self-dual pair \net{dia}$\otimes$\net{dia} and of the planar dual pair \net{hxl}$\otimes$\net{hcb}: 5\%, 50\% and 95\% quantiles of $\omega^2/k_c$, and unlifted mechanisms. For \net{dia}$\otimes$\net{dia}, which has several lifted branches, the quantiles pool all lifted modes over the sampled wavevectors. For \net{hxl}$\otimes$\net{hcb}, these are single-band quantiles over 400 random wavevectors. At each wavevector we select the eigenvalue with the largest overlap with the \net{hcb} mechanism, as for the markers of Fig.~\ref{fig:planar}b.}\label{tab:control}
\begin{tabular}{lccc}
\hline
Pair & $k_c$ & quantiles of $\omega^2/k_c$ & unlifted \\
\hline
\net{dia}$\otimes$\net{dia} & 0.003 & 0.142, 1.340, 2.536 & 0 \\
\net{dia}$\otimes$\net{dia} & 0.01 & 0.169, 1.354, 2.570 & 0 \\
\net{dia}$\otimes$\net{dia} & 0.03 & 0.155, 1.351, 2.602 & 0 \\
\net{hxl}$\otimes$\net{hcb} & 0.003 & 1.459, 1.4989, 1.5000 & 0 \\
\net{hxl}$\otimes$\net{hcb} & 0.01 & 1.369, 1.4964, 1.4999 & 0 \\
\net{hxl}$\otimes$\net{hcb} & 0.03 & 1.153, 1.4894, 1.4998 & 0 \\
\hline
\end{tabular}
\end{table}

\section{Closed form for the planar pair}\label{sec:hcb}

\begin{figure*}[t]\centering
\includegraphics[width=0.66\textwidth]{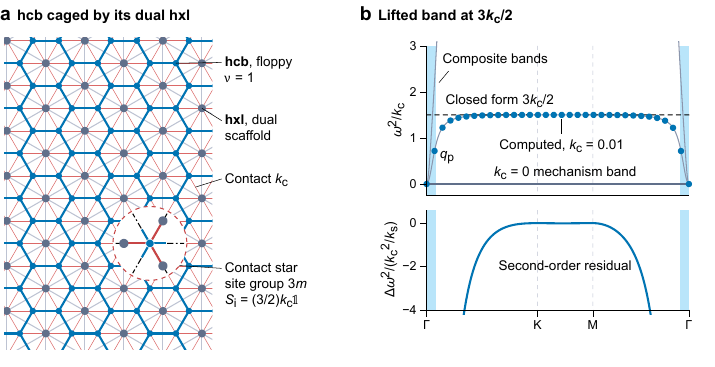}
\caption{\textbf{The planar pair \net{hxl}$\otimes$\net{hcb}.} \textbf{a}, The honeycomb net \net{hcb} ($\nu=1$) inside its dual, the triangular net \net{hxl}, whose vertices sit at the hexagon centres. Each \net{hcb} vertex has three contacts at $120^\circ$ along the bisectors of the angles between its edges. The highlighted contact star, marked by the red dashed circle, has site group $3m$ (grey dash-dot mirror lines), so $S_i=\tfrac32\kc\one$. \textbf{b}, Band structure of the full \net{hxl}$\otimes$\net{hcb} composite along $\Gamma$--K--M--$\Gamma$ at $\kc=0.01$, with $\ks=1$, unit masses and unit \net{hcb} edge length. Markers show the lifted band from direct diagonalization (the eigenvector with the largest overlap with the \net{hcb} mechanism), and the dashed line shows the closed form $\omega^2=\tfrac32\kc$. The line at zero is the mechanism band at $\kc=0$, and the thin lines are the other composite bands. The shaded zone-centre pocket $|\mathbf q|<q_p$ marks the hybridization with the transverse acoustic branch, on the scale $q_p=\sqrt{3\kc/2}/c_T=1.99\sqrt{\kc/\ks}$ obtained by extrapolating the small-wavevector sound speed $c_T$ of the composite to $\tfrac32\kc$. The lower strip shows the deviation of the lifted band from $\tfrac32\kc$ in units of $\kc^2/\ks$. It vanishes at K and M and grows towards the pocket. With the same rule at 2,000 random wavevectors, the zone median is $\hcbFigMedian\,\kc$ at $\kc=0.01$.}
\label{fig:planar}
\end{figure*}

The honeycomb net inside its triangular dual is solvable by hand. This section derives its mechanism, its lifted band and the origin of its residual width. Figure~\ref{fig:planar} shows the pair and its band.

\textbf{S6.1 Mechanism.} Let the three edges from vertex $A$ reach vertex $B$ through lattice vectors $\mathbf R_m$ along unit bond vectors $\hat{\mathbf b}_m$ at $120^\circ$, with Bloch phases $\phi_m=e^{i\mathbf q\cdot\mathbf R_m}$. A mechanism obeys $\hat{\mathbf b}_m\cdot\mathbf u_A=\phi_m\,\hat{\mathbf b}_m\cdot\mathbf u_B$ for $m=1,2,3$. Because $\sum_m\hat{\mathbf b}_m=0$, these three conditions are consistent only if $\mathbf w\cdot\mathbf u_B=0$ with $\mathbf w=\sum_m\phi_m\hat{\mathbf b}_m$. Hence
\begin{equation}
\mathbf u_B=J\mathbf w,\qquad \mathbf u_A=\frac23\sum_{m=1}^{3}\phi_m\,\hat{\mathbf b}_m(\hat{\mathbf b}_m\cdot\mathbf u_B),
\end{equation}
where $J$ rotates by $90^\circ$ and $\sum_m\hat{\mathbf b}_m\hat{\mathbf b}_m^{\mathsf T}=\tfrac32\one$ was used. The vector $\mathbf w$ vanishes only at $\Gamma$, so away from the zone centre the mechanism band is isolated, and the inter-class balance lemma of Section~S2.4 gives $\|\mathbf u_A\|=\|\mathbf u_B\|$ identically.

\textbf{S6.2 First order.} Each honeycomb vertex has three contacts along the bisectors of the angles between its edges, so $S_i=\tfrac32\kc\one$ by the threefold site symmetry. With the scaffold fixed, the mechanism band lies at $\omega^2=\tfrac32\kc$ at every $\mathbf q$. In the movable composite this is the first-order value away from $\Gamma$.

\textbf{S6.3 Second order and the zone-centre pocket.} With the scaffold free, let $\mathbf u(\mathbf q)$ be the normalized honeycomb mechanism, write the off-diagonal contact block as $\kc G(\mathbf q)$, and put $\bar D_S=D_S/\ks$. At each fixed $\mathbf q\neq\Gamma$, with the geometry and $k_F$ and $\ks$ fixed,
\begin{equation*}\omega^2(\mathbf q)=\tfrac32\kc-\frac{\kc^2}{\ks}\,\tilde g(\mathbf q)+O\!\left(\frac{\kc^3}{g(\mathbf q)^2}\right),\qquad\tilde g=\mathbf u^\dagger G\bar D_S^{-1}G^\dagger\mathbf u\ge0,\end{equation*}
with $g(\mathbf q)$ the smaller of the two unperturbed gaps. The floppy-net channel of Lemma~\ref{lem:first} vanishes, because $S$ acts as a scalar on the mechanism. The coefficient uses the scaffold inverse at zero energy, the $\kc\to0$ limit of the scaffold resolvent at the flat-band frequency. Numerically $\tilde g$ vanishes at K and M and grows toward the zone centre roughly as $1/q^2$. Along $\Gamma$--M it is \gtOne, \gtSix{} and \gtThree{} at $|\mathbf q|=1.0$, 0.6 and 0.3 for unit edge length, in the limit $\kc\to0$. Direct diagonalization at $\kc=10^{-4}$ and $10^{-5}$ gives these values to better than 1\%. A fit $\tilde g\approx C/q^2$ for $\gtFitLo\le|\mathbf q|\le\gtFitHi$ gives $C=\gtC$, with a log--log slope of $-\gtSlope$ and residuals below \gtFitRes\%. Divided by the first-order value $\tfrac32\kc$, the correction gives the relative deviation $\tfrac23(\kc/\ks)\tilde g$. Along $\Gamma$--M, a relative deviation larger than $\delta$ therefore sets in at the asymptotic scale $|\mathbf q|\simeq\sqrt{2C\kc/(3\ks\delta)}$, which shrinks as $\sqrt{\kc/\ks}$; it is a scale, not a sharp boundary at finite coupling. Closer to the zone centre the flat band hybridizes with the transverse acoustic branch of the composite, in a smooth crossover. That branch has the small-wavevector sound speed $c_T=0.617$ at $\kc/\ks=0.01$, and extrapolating its slope to $\tfrac32\kc$ gives the pocket scale $q_p=\sqrt{3\kc/2}/c_T\approx2.0\sqrt{\kc/\ks}$. This marks the size of the crossover region, which shrinks as $\sqrt{\kc}$, rather than an exact crossing. 

\section{Joint bending}\label{sec:bending}

The selection rule is a statement about pin-jointed frameworks. This section measures how joint bending degrades it.

Bending was added at first order as angular springs between edges at each vertex, or as beam-like straightening of collinear edges, with stiffness $k_b$. Bending lifts the mechanisms of the floppy net on its own, and its stiffness does not share the site structure of the contact tensor, so the band spreads (Table~\ref{tab:bending}). At $k_b/\kc=0.01$ the spread is a few per cent to a few tens of per cent, and at $k_b/\kc=1$ it is between 0.1 and 2. In these angular and collinear-straightening models, bending therefore broadens nine pairs already at $k_b/\kc=0.01$, the smallest bending stiffness tested. The pair \net{nia}$\otimes$\net{nia-d} is the exception. Its two mechanisms live on the threefold vertices of \net{nia-d}, which are isolated from each other by ninefold hubs that carry no mechanism. The angular term then acts on site and does not depend on $\mathbf q$, and the beam term vanishes on the mechanisms. The projected bending spectra of this pair therefore remain flat throughout the tested range. These two models carry no hub rotation. In the beam frame of Section~\ref{sec:meta}, welded joints couple the threefold vertices through the condensed rotations of the shared ninefold hubs, and the band of this pair then spreads (Table~\ref{tab:meta_map}). The full beam-frame operator and the rotation-locking condition are treated in Section~\ref{sec:meta}.

\begin{table}[h]\centering
\caption{Relative band spread of the ten exactly flat pairs with joint bending added at first order, for angular springs and for beam-like straightening of collinear edges, at bending stiffness $k_b/k_c=0.01$, $0.1$ and $1$. Zero spread at $k_b=0$ for every pair.}\label{tab:bending}
\begin{tabular}{lcccccc}
\hline
Pair & \multicolumn{3}{c}{angular} & \multicolumn{3}{c}{beam} \\
 & 0.01 & 0.1 & 1.0 & 0.01 & 0.1 & 1.0 \\
\hline
\net{acs}$\otimes$\net{gra} & 0.07 & 0.31 & 0.81 & 0.04 & 0.33 & 1.02 \\
\net{bcu}$\otimes$\net{nbo} & 0.06 & 0.32 & 0.51 & 0.07 & 0.39 & 0.51 \\
\net{hex}$\otimes$\net{bnn} & 0.19 & 0.67 & 1.44 & 0.09 & 0.70 & 1.83 \\
\net{bsn}$\otimes$\net{lvt} & 0.02 & 0.14 & 0.44 & 0.05 & 0.25 & 0.41 \\
\net{fcu}$\otimes$\net{flu} & 0.15 & 0.57 & 0.78 & 0.07 & 0.52 & 1.02 \\
\net{hxg-d}$\otimes$\net{hxg} & 0.04 & 0.17 & 0.24 & 0.06 & 0.28 & 0.41 \\
\net{nia}$\otimes$\net{nia-d} & 0.00 & 0.00 & 0.00 & 0.00 & 0.00 & 0.00 \\
\net{pcu}$\otimes$\net{urk} & 0.22 & 0.51 & 0.47 & 0.27 & 0.52 & 0.59 \\
\net{qtz-x}$\otimes$\net{uom} & 0.03 & 0.17 & 0.35 & 0.12 & 0.55 & 0.78 \\
\net{she-d}$\otimes$\net{she} & 0.03 & 0.14 & 0.20 & 0.02 & 0.08 & 0.13 \\
\hline
\end{tabular}
\end{table}

\section{Parametric drive}\label{sec:drive}

At first order, a flat band lets one drive frequency address every mechanism. This section defines the pump and reports its linear gain.

\textbf{S8.1 Pump operator.} The contacts are modulated as $\kc(t)=\kc[1+h\cos\Omega t]$. The pump then acts on the composite as $P=\partial D/\partial\kc$, including the cross terms between floppy and scaffold vertices. For the unstressed contacts studied here, the contact term is $\kc C^\dagger C$, so $P=C^\dagger C$, and the geometric stiffness of the contacts vanishes. On a flat branch with $\Pi_\sigma S\Pi_\sigma=\tilde s_\sigma\Pi_\sigma$, the projected pump is $\Pi_\sigma P\Pi_\sigma=(\tilde s_\sigma/\kc)\Pi_\sigma$, so at leading order every mode of that branch sees the same parametric coupling. At eligible generic wavevectors these modes therefore share the principal resonance at $\Omega=2\omega_\sigma$, and for weak modulation at exact resonance the threshold is $h_c\simeq4\zeta$ for damping ratio $\zeta$.

\begin{figure}[h]\centering
\includegraphics[width=\columnwidth,height=0.7\textheight,keepaspectratio]{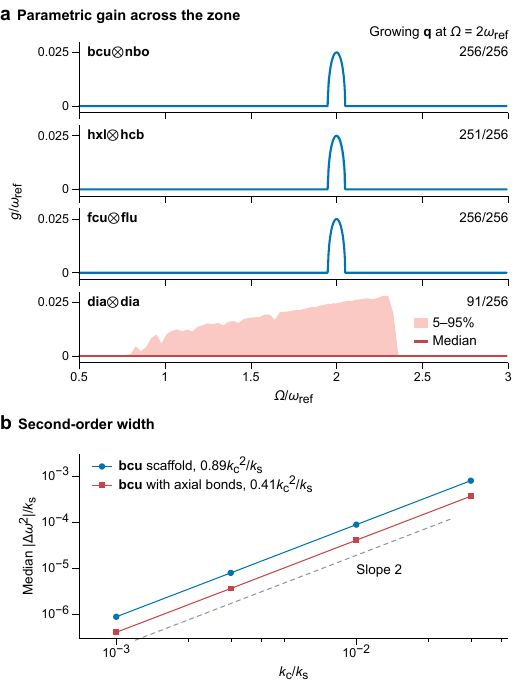}
\caption{\textbf{Drive and contrast.} \textbf{a}, Linear Floquet gain $g$ for contacts modulated as $\kc(t)=\kc[1+h\cos\Omega t]$ with $\kc=0.001$, $h=0.1$ and no damping (Section~S8). Lines show the median over \gainNq{} wavevectors and bands the 5--95\% range, which for the dual pairs is narrower than the line. The reference $\omega_{\rm ref}$ is the flat-band frequency, and $\sqrt{2\kc}$ for \net{dia}$\otimes$\net{dia}, which has no flat band. Counts give the growing wavevectors at $\Omega=2\omega_{\rm ref}$. The dual pairs gain at one drive frequency across the zone, whereas the \net{dia} pair spreads its gain over a wide range of $\Omega$. \textbf{b}, Median deviation of the \net{nbo} band from $2\kc$ away from the zone centre, for the \net{bcu} scaffold with and without added axial bonds (Section~S4). Both scale as $\kc^2/\ks$, with the coefficients of Table~\ref{tab:scaling}.}
\label{fig:drive}
\end{figure}

\textbf{S8.2 Linear gain spectrum.} The full monodromy matrix over one drive period was computed at 256 wavevectors uniform in the primitive zone and 175 drive frequencies, with $\kc=10^{-3}$, $h=0.1$, $\ks=1$, unit masses and no damping. At $\Omega=2\omega_{\rm flat}$ the median normalized rates $g_{\max}/\omega_{\rm flat}$ are 0.02499, 0.02498 and 0.02499 for \net{bcu}$\otimes$\net{nbo}, \net{hxl}$\otimes$\net{hcb} and \net{fcu}$\otimes$\net{flu}, against $h/4=0.025$ for a single Mathieu mode. A positive $g_{\max}$ shows that at least one direction grows at that wavevector; it does not establish growth of every lifted mode. A wavevector is counted as growing when $g_{\max}/\omega_{\rm flat}>10^{-5}$, the threshold of Fig.~\ref{fig:drive}a. Of the \gainNq{} wavevectors, \gainPosNbo, \gainPosHcb{} and \gainPosFlu{} grow, and \gainWinNbo, \gainWinHcb{} and \gainWinFlu{} grow within 4\% of $h/4$. For \net{dia}$\otimes$\net{dia}, referenced to $\omega^2=2\kc$, 91 of 256 grow and the median rate is zero. At the zone centre the lifted band of \net{bcu}$\otimes$\net{nbo} and \net{hxl}$\otimes$\net{hcb} merges with the acoustic branch and does not grow. In \net{fcu}$\otimes$\net{flu} the zone centre still grows, at $\gainGammaFlu\,\omega_{\rm flat}$, about \gainGammaFluPct\% of the single-mode rate. A second implementation reproduces \net{bcu}$\otimes$\net{nbo} on 52 primitive wavevectors, all growing with median rate $0.02499\,\omega_{\rm flat}$. The same drive applied to \otherPairs{} further flat pairs at $\kc=10^{-3}$, on \otherNq{} scrambled Sobol wavevectors per pair, makes \otherFracMin{} to \otherFracMax\% of wavevectors grow. Their mean largest exponent is \otherMeanMin{} to \otherMeanMax{} of the single-mode rate, lowest for \otherLow.

\textbf{S8.3 Temporal order.} A flat band gives nearly every wavevector the same gain, which is what a single drive needs and also what makes a uniform period-doubled pattern fragile. Whether a driven lifted band can sustain a discrete period-doubled pattern, and why its band edge must then sit at a time-reversal-invariant momentum, is analysed in a companion paper.

\section{Electronic crystal fields}\label{sec:orbital}

The flatness criterion of Theorem~1 applies to any on-site perturbation of a band of zero modes of an operator $R^\dagger R$. This section tests it on $p$ orbitals in a crystal field, a problem that becomes mathematically equivalent to the mechanical one once the hopping and on-site energies of S9.1 are imposed.

\textbf{S9.1 Model.} Each site carries spinless $p_x$, $p_y$ and $p_z$ orbitals $\mathbf c_i$. Nearest neighbours are joined by $\sigma$ hopping only, of magnitude $V_{pp\sigma}$, the two-centre integral between $p$ orbitals pointing along the bond, which sets the unit of energy. The on-site energies are chosen so that
\begin{equation}
H_{\rm e,0}=\sum_i\mathbf c_i^\dagger T_i\mathbf c_i-\sum_{\langle ij\rangle}\big[\mathbf c_i^\dagger\hat{\mathbf b}_{ij}\hat{\mathbf b}_{ij}^{\mathsf T}\mathbf c_j+{\rm h.c.}\big],\qquad T_i=\sum_{b\ni i}\hat{\mathbf b}\hat{\mathbf b}^{\mathsf T}.
\end{equation}
The minus sign of the hopping term corresponds to $V_{pp\sigma}=-1$ in this unit. Its Bloch matrix is $R^\dagger R$, the operator of Section~S1 with unit stiffness. The correspondence holds for these on-site energies and vanishing $\pi$ hopping. For the honeycomb net $T_i=\tfrac32\one$ in the $p_{x,y}$ block, so the model is the $p_{x,y}$-orbital honeycomb model with its flat bands \cite{wu_flat_2007,wu_p_xy-orbital_2008}, shifted by $\tfrac32$. The dual vertices act as a static ligand environment, with field $\lambda\bar S_i$ from the nearest ligands. Here $\bar S_i=\sum_\ell\hat{\mathbf c}_{i\ell}\hat{\mathbf c}_{i\ell}^{\mathsf T}=S_i/\kc$ is the dimensionless contact tensor, and $\bar S=\bigoplus_i\bar S_i$. For an isolated flat band of energy $E_0$ with orthonormal basis $Z(\mathbf q)$, the first-order energies are $E_0+\lambda\operatorname{eig}[Z^\dagger\bar SZ]$, the electronic form of Lemma~\ref{lem:first}.

\textbf{S9.2 Results.} Table~\ref{tab:orbital} lists the cases. On \net{hcb} the triangular dual gives $\bar S_i=\mathrm{diag}(\tfrac32,\tfrac32,0)$. The in-plane sector $E'$ of $D_{3h}$ occurs once, so route U shifts the collective in-plane band rigidly by $\tfrac32\lambda$, and the full diagonalization keeps it flat at finite $\lambda$. The anisotropic field $\mathrm{diag}(2.1,0.9,0)$ breaks the threefold axis and spreads the first-order shift from $\orbAnisoLo\lambda$ to $\orbAnisoHi\lambda$. On \net{flu} the fcc dual gives $\bar S_4=\tfrac43\one$ and $\bar S_8=2\one$, and route C fixes the norm shares at $\tfrac23$ and $\tfrac13$, so the band shifts by $\tfrac{14}{9}\lambda$. In this two-class model the band stays exactly flat at finite field, at
\begin{equation}
E_-(\lambda)=\tfrac12\Big[t_A+t_B+\lambda(s_A+s_B)-\sqrt{[t_A-t_B+\lambda(s_A-s_B)]^2+4t_At_B}\Big],
\end{equation}
with $(t_A,t_B,s_A,s_B)=(\tfrac43,\tfrac83,\tfrac43,2)$, because every zero mode selects the singular value $\sqrt{t_At_B}$ of the interclass hopping block. Adding $0.4\,\mathrm{diag}(1,-1,0)$ on every site disperses the first-order branches. On the $12^3$ grid of Table~\ref{tab:orbital} the widest branch spans $\orbFluWidthBroken\lambda$; explicit generic wavevectors widen this to at least $0.60\lambda$ per branch and $0.80\lambda$ for the whole bundle, so the grid value is a lower bound, not the zone-wide width. The two decoupled $p_z$ bands of \net{hcb}, placed at energy 4, test route L. The tilted field $\bar S_i=\orbSL$, in the basis $(p_x,p_y,p_z)$ and the same on every site, does not have $p_z$ as an eigenvector, yet the site-supported basis gives $Z^\dagger\bar SZ=\orbLvalue\,\one$, and the bands stay flat to first order. As a scalar control, a common potential on the two majority sites of the Lieb lattice shifts its zero band rigidly, whereas unequal potentials disperse it. The scalar case is the counting rule \cite{calugaru_general_2022} (App.~I.4 therein); the tensor cases follow from the routes of Section~S2. A separate implementation, written without the first code, reproduces the \net{hcb} route-U shift, the \net{flu} first-order value and its finite-field energy to $10^{-12}$, and the dispersion of both anisotropic controls.

\begin{table}[h]\centering
\caption{Electronic crystal-field test of the rule. The \net{hcb} in-plane and \net{flu} rows use spinless $p$ orbitals with $\sigma$ hopping and the on-site energies that make the Bloch matrix $R^\dagger R$. The Lieb rows use the zero band of the Lieb lattice, and the \net{hcb} $p_z$ row uses the two decoupled $p_z$ bands placed at energy 4 (Section~S9.2). The field is $\lambda\bar S$, with $\bar S$ the dimensionless contact tensor. Route: the condition of Section~S2 that applies (none: no condition holds). Shift: first-order energy shift per unit $\lambda$ over the sampled wavevectors. Width: spread of the first-order shift, and bandwidth of the full diagonalization at $\lambda=0.03$ (largest over branches). Energies in units of the $\sigma$ hopping.}\label{tab:orbital}
\begin{tabular}{lllccc}
\hline
Net & Field & Route & Shift & First-order width & Full width \\
\hline
\net{hcb} & common scalar, $\one$ on every site & scalar & 1.000 & $<10^{-12}$ & $<10^{-12}$ \\
\net{hcb} & triangular dual star, $\mathrm{diag}(\tfrac32,\tfrac32,0)$ & U & 1.500 & $<10^{-12}$ & $<10^{-12}$ \\
\net{hcb} & anisotropic, $\mathrm{diag}(2.1,0.9,0)$ & none & 0.900--2.100 & 1.200 & 0.036 \\
\net{flu} & fcc dual stars, $\tfrac43\one$ and $2\one$ & C & 1.556 & $<10^{-12}$ & $<10^{-12}$ \\
\net{flu} & plus $0.4\,\mathrm{diag}(1,-1,0)$ on every site & none & 1.293--1.818 & 0.431 & 0.015 \\
Lieb & common scalar on both majority sites & scalar & 1.000 & $<10^{-12}$ & $<10^{-12}$ \\
Lieb & unequal scalars 1.4 and 0.6 & none & 0.600--1.400 & 0.800 & 0.024 \\
\net{hcb} $p_z$ & tilted, $p_z$ not an eigenvector & L & 0.800 & $<10^{-12}$ & $9.9\times10^{-5}$ \\
\hline
\end{tabular}
\end{table}

\textbf{Computational details.} Honeycomb and Lieb bands were sampled on an $81\times81$ reciprocal grid, and \net{flu} on a $12^3$ grid of the conventional reciprocal cell, of which 1,680 points are generic; the 48 points with extra zero modes were diagonalized separately. Second-order behaviour was checked on a gapped honeycomb K--M--K$'$ path for $10^{-4}\le\lambda\le10^{-2}$: the error of the anisotropic first-order prediction and the width of the route-L bands scale with exponents \orbSlopeU{} and \orbSlopeL. In the pure $\sigma$-hopping model of \net{flu}, without the on-site energies $T_i$, the same structure gives exactly flat bands at $\pm\tfrac{4\sqrt2}{3}V_{pp\sigma}$, with sampled widths below $10^{-13}$. Because this Bloch matrix only couples the two classes, each of these nonzero bands carries equal norm on both, so the first-order shift is the plain mean $\tfrac12(\tfrac43+2)\lambda=\tfrac53\lambda$, not $\tfrac{14}{9}\lambda$. The on-site energies $T_4=\tfrac43\one$ and $T_8=\tfrac83\one$ of S9.1 are therefore what fix the shares at $\tfrac23$ and $\tfrac13$. The additional zero-energy bands that the sublattice imbalance of \net{flu} forces in this model are not covered by this statement. The two-class reduction requires scalar class tensors and a zero mode at every wavevector. The calculations use nearest-neighbour $\sigma$ hopping, with zero $\pi$ hopping and no spin--orbit coupling.

\clearpage
\section{Metamaterial realization}\label{sec:meta}

Section~\ref{sec:bending} showed that joint bending spreads the band of a pin-jointed pair. In three pairs the first-order spread comes from the rotation of the joints. Holding the joint rotations removes it there, because no bond joins two mechanism vertices with non-orthogonal normals. The other seven pairs stay dispersive under ideal locking, whereas released strut ends restore the first-order pin-joint spectrum in all ten. The section then follows that band through distributed strut mass, a finite specimen driven in time and fabrication errors, all within linear elastic models of an architected lattice.

\textbf{\ref{sec:meta}.1 Beam frame.} The frame keeps the screened placement of each pair and replaces every edge by an Euler--Bernoulli beam. Each vertex carries three translations and three rotations. A strut of length $L$ along the unit bond vector $\hat{\mathbf b}$ has axial stiffness $k_a$ and transverse stiffness $k_b=12EI/L^3$ in both bending planes. Here $E$ is Young's modulus and $I$ the second moment of area, and $k_b$ is the same for every floppy strut. As for a solid circular rod, the torsional stiffness is $GJ/L=k_bL^2/[12(1+\nu_P)]$, with shear modulus $G$, polar second moment $J$ and Poisson ratio $\nu_P=0.3$. Contacts are springs of stiffness $\kc$ along their unit vectors. Nodes carry unit mass and no rotary inertia, so the rotations are condensed statically and the translational stiffness $K(\mathbf q)$ enters $K\mathbf u=\omega^2\mathbf u$. Units are the floppy axial stiffness, $k_a=1$, and the node mass. The cage is thick and moves, with axial stiffness $k_{s,a}=10\ks$ and bending stiffness $k_{s,b}=\ks$ per strut. The calculations use $\ks=1$ in these units, and the estimates below scale both cage stiffnesses together with $\ks$. The lifted band is the set of $M=\dim\ker R_F(\mathbf q)$ eigenmodes at generic wavevectors with the largest overlap with the pin-joint mechanism space $\ker R_F(\mathbf q)$, and its width is the largest relative width of a sorted branch, $\max_b(\max_{\mathbf q}\omega_b^2-\min_{\mathbf q}\omega_b^2)/\langle\omega_b^2\rangle_{\mathbf q}$.

\begin{figure*}[!htbp]\centering
\includegraphics[width=\textwidth]{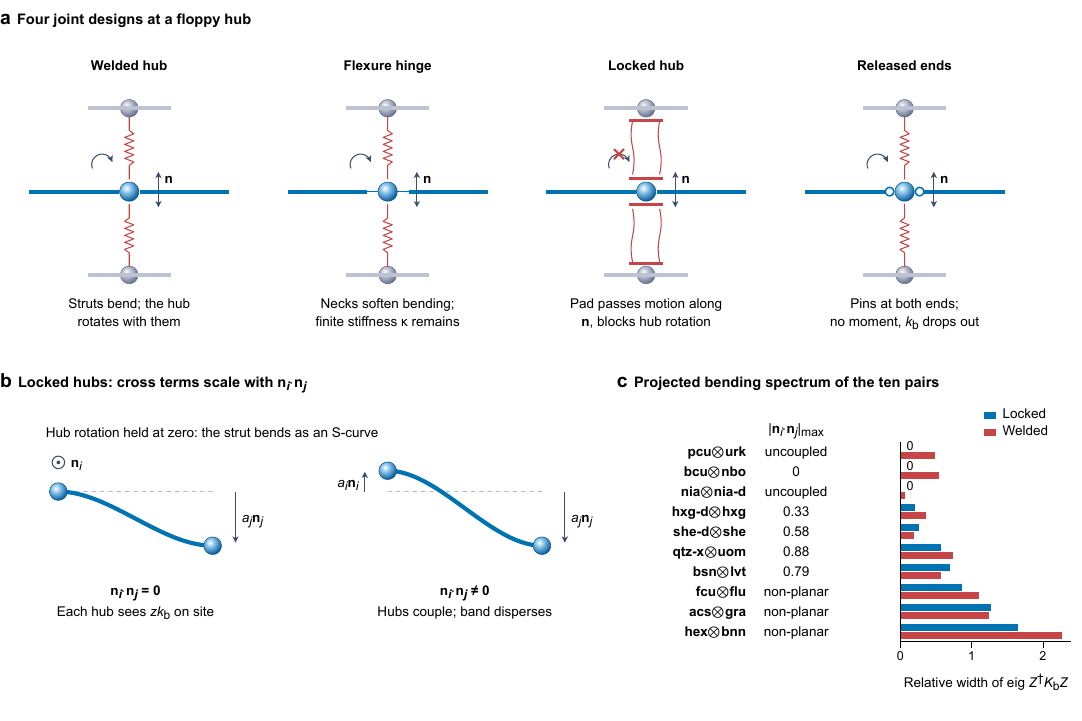}
\caption{\textbf{Holding hub rotation turns joint bending into an on-site term when neighbouring mechanism normals are orthogonal.} \textbf{a}, The three joint designs at a floppy hub and, for comparison, a flexure hinge, each between two cage nodes (grey) joined to the hub by contact springs (red) along the contact normal $\mathbf n$. A welded hub rotates with its struts, so strut bending couples neighbouring hubs. A flexure hinge softens the struts near the hub but keeps a finite rotational stiffness. A locked hub is guided by a parallel-leaf pad that passes motion along $\mathbf n$ and blocks rotation. Released struts end in pins and carry no moment, so $k_b$ drops out of the mechanism sector. \textbf{b}, With hub rotations held at zero, a strut bends as an S-curve with zero end slopes. When the mechanism of one hub lies along $\mathbf n_i$ and that of its neighbour along an orthogonal $\mathbf n_j$ (left, as in \net{bcu}$\otimes$\net{nbo}), the bending energy has no cross term and each hub sees $zk_b$ on site. Otherwise (right), the energy contains the cross term $-k_ba_ia_j(\mathbf n_i\cdot\mathbf n_j)$ for real amplitudes, which couples the hubs, and the band disperses. \textbf{c}, Relative width over 30 wavevectors of the projected bending spectrum $\operatorname{eig}Z^\dagger K_bZ$ for the ten exactly flat pairs, locked ($\theta=0$, Eq.~(\ref{eq:meta_lap})) and welded (rotations condensed), with the largest $|\mathbf n_i\cdot\mathbf n_j|$ over edges joining two planar mechanism vertices. Only \net{bcu}$\otimes$\net{nbo}, \net{nia}$\otimes$\net{nia-d} and \net{pcu}$\otimes$\net{urk} are flat when locked (Table~\ref{tab:meta_normals}).}
\label{fig:s_meta_concept}
\end{figure*}

The frame admits three joint designs at the floppy hubs (Fig.~\ref{fig:s_meta_concept}a). Welded hubs share one rotation among all their struts. Locked hubs are tied to the cage node across each contact by a rotational spring $k_{cr}=100$ about all three axes. This spring models a pad that passes motion along the contact direction $\hat{\mathbf c}$ and blocks rotation, such as a parallel-leaf or double-blade flexure. Its leaves bend easily along $\hat{\mathbf c}$, which sets the contact stiffness $\kc$. A rotation of the pad instead stretches one leaf and compresses the other, so its rotational stiffness is of order $k_\ell h^2$, with $k_\ell$ the axial stiffness of a leaf and $h$ the leaf separation. Locking needs this stiffness to dominate the bending of the struts at the hub, $k_{cr}\gg k_bL^2$, which in model units holds with $k_{cr}=100$ against $k_bL^2$ of order $10^{-3}$ to $10^{-2}$. Released struts have moment-free ends: each beam block is condensed over its own end rotations before assembly, which leaves the axial bar to rounding error. A flexure hinge lies between the welded and released designs and is not treated separately.

The width depends on the wavevector sample, because the dimensionless stiffness gap of the floppy net, $g(\mathbf q)=g_F(\mathbf q)/k_a$, the $(M+1)$th eigenvalue of $R_F^\dagger R_F$ at unit stiffness, closes on planes of the zone. For \net{bcu}$\otimes$\net{nbo} it closes where a component of $\mathbf q$ is an integer, and \metaNboGapOne{} of \metaNq{} uniform draws have $g<0.01$ and \metaNboGapFive{} have $g<0.05$. Near these planes the lifted modes mix with soft floppy modes at an order that the band width does not control. All design numbers therefore use \metaNq{} random wavevectors with $g\ge\metaGapMin$, drawn from the stream that also supplies the unfiltered sample, and Table~\ref{tab:meta_map} reports both.

\begin{figure*}[!htbp]\centering
\includegraphics[width=\textwidth]{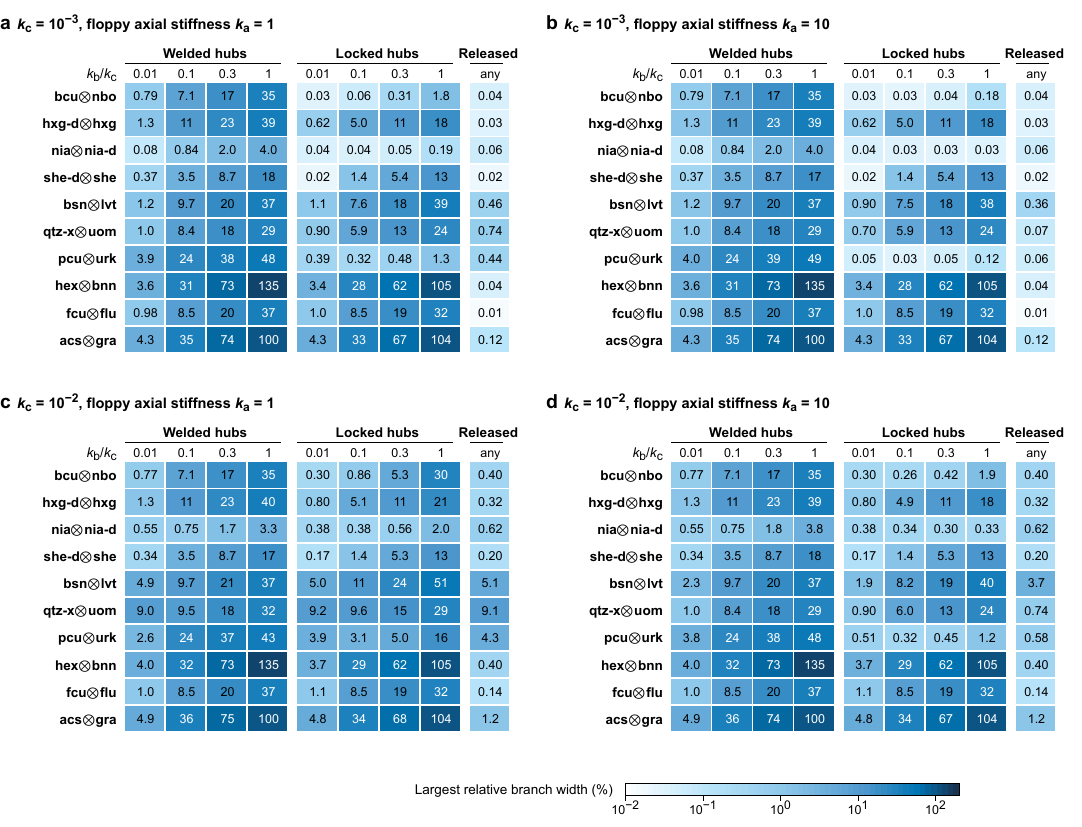}
\caption{\textbf{Locking removes first-order bending dispersion in three pairs, and released ends restore the pin-joint limit in all ten.} Largest relative branch width (\%) of the lifted band for the ten exactly flat pairs in the beam frame with a mobile thick cage (axial stiffness 10, bending stiffness 1, in units of the floppy axial stiffness), on \metaNq{} random wavevectors with floppy-net gap $g\ge\metaGapMin$. Panel a reuses the \net{bcu}$\otimes$\net{nbo} and \net{nia}$\otimes$\net{nia-d} measurements of main Fig.~\ref{M-fig:meta}b, and its other rows extend that comparison to eight further pairs. Columns give welded and locked hubs at $k_b/\kc=0.01$, 0.1, 0.3 and 1, and released strut ends, for which $k_b$ drops out. \textbf{a}, $\kc=10^{-3}$, floppy axial stiffness $k_a=1$. \textbf{b}, $\kc=10^{-3}$, $k_a=10$. \textbf{c}, $\kc=10^{-2}$, $k_a=1$. \textbf{d}, $\kc=10^{-2}$, $k_a=10$. At $\kc=10^{-2}$ the lifted level approaches the stiffness of the soft floppy modes, and the widths of all designs grow; a stiffer floppy net restores the locked widths of the three exact pairs. Released \net{bsn}$\otimes$\net{lvt}, \net{qtz-x}$\otimes$\net{uom} and \net{pcu}$\otimes$\net{urk} keep a finite width, the order-$\kc^2$ dispersion of their pin-jointed bands, and mobile-cage corrections remain in all pairs. Table~\ref{tab:meta_map} adds the unfiltered sample.}
\label{fig:s_meta_map}
\end{figure*}

\textbf{\ref{sec:meta}.2 First order.} At first order in $\kc$ and $k_b$, the lifted frequencies are $\omega^2=\operatorname{eig}[Z^\dagger(S+K_b)Z]$. Here $S$ is the contact operator of Section~S1, which already contains $\kc$. The matrix $Z$ is an orthonormal basis of $\ker R_F$, and $K_b$ is the condensed bending stiffness of the floppy net. The projected contact spectrum is flat for all ten pairs. Its eigenvalues add directly to the bending eigenvalues only within a common scalar sector that bending preserves. In other cases we diagonalize the combined projected operator. With welded hubs the bending term couples neighbouring hubs through their shared rotations. At $k_b=\kc$ the width is then \metaFreeMin{} to \metaFreeMax\% across the ten pairs (Fig.~\ref{fig:s_meta_map}a), and \metaNboFree\% for \net{bcu}$\otimes$\net{nbo}.

Released ends remove this term exactly. A mechanism keeps every strut at its length, so each strut moves as a rigid chord whose ends rotate by $\boldsymbol\alpha=\hat{\mathbf b}\times\mathbf d/L$, with $\mathbf d$ the transverse relative displacement. Free end rotations can follow $\boldsymbol\alpha$ at no energy, so $K_bZ=0$ for any $EI$, and the band returns to its pin-joint value at first order. The remaining width, \metaNboRel\% for \net{bcu}$\otimes$\net{nbo} at $\kc=10^{-3}$, comes from the mobile cage. The pairs \net{bsn}$\otimes$\net{lvt}, \net{qtz-x}$\otimes$\net{uom} and \net{pcu}$\otimes$\net{urk}, whose normals are not eigenvectors of $S_i$, also keep their fixed-cage dispersion of order $\kc^2$ (Table~\ref{tab:meta_map}). The released design therefore restores the pin-joint limit of every flat pair, but it needs ideal moment releases at both ends of every strut.

\textbf{\ref{sec:meta}.3 Exact condition for locked bending.} In the ideal locking limit, the hub rotations vanish. This limit gives the first-order condition below, whereas the finite-frame calculation uses rotational springs $k_{cr}$ to the mobile cage. Assume unit translational masses, unstressed struts and the common transverse stiffness $k_b$ of Section~\ref{sec:meta}.1. A strut then bends as an S-curve with zero end slopes, and its transverse energy is $\tfrac{k_b}{2}|P_\perp\boldsymbol\Delta_e|^2=\tfrac{k_b}{2}(|\boldsymbol\Delta_e|^2-|\hat{\mathbf b}\cdot\boldsymbol\Delta_e|^2)$, with $\boldsymbol\Delta_e$ the relative displacement of its ends. With the Bloch convention of Section~S1.2 for both $R_F$ and the scalar Bloch graph Laplacian $\mathcal L(\mathbf q)$ of the floppy net, the sum over struts is
\begin{equation*}
K_b^{(\theta=0)}=k_b\big[\mathcal L(\mathbf q)\otimes I_3-R_F^\dagger R_F\big],
\end{equation*}
Here each strut contributes $k_b(I_3-\hat{\mathbf b}\hat{\mathbf b}^{\mathsf T})$ on its relative displacement, and the two terms on the right contribute $k_bI_3$ and $k_b\hat{\mathbf b}\hat{\mathbf b}^{\mathsf T}$. Since $R_FZ=0$, this gives
\begin{equation}
Z^\dagger K_b^{(\theta=0)}Z=k_b\,Z^\dagger\left[\mathcal L(\mathbf q)\otimes I_3\right]Z .
\label{eq:meta_lap}
\end{equation}
For unequal strut stiffnesses $k_{b,e}$, $k_b\mathcal L$ is replaced by the weighted Laplacian $B^\dagger\operatorname{diag}(k_{b,e})B$, with $B$ the scalar Bloch incidence matrix. Locked bending is therefore flat at first order on $U$ if and only if the spectrum of $Z^\dagger(\mathcal L\otimes I_3)Z$ does not depend on $\mathbf q$. A sufficient condition follows when the mechanism space is spanned by the site-supported unit normals $\mathbf e_i$ of a fixed set $P$ of planar vertices (Section~S2.2). Writing $\mathcal L=D-A$, the degree term is then diagonal in this basis, with entries $z_ik_b$. Both normals of such an edge are orthogonal to it, so an edge from $i$ to $j$ in cell $\mathbf R_e$ contributes $-k_b\operatorname{Re}[a_i^*e^{i\mathbf q\cdot\mathbf R_e}a_j]\,(\mathbf n_i\cdot\mathbf n_j)$ to the bending energy (Fig.~\ref{fig:s_meta_concept}b). These cross terms vanish when those normals are orthogonal or when no edge joins two vertices of $P$, and the locked first-order levels are then $z_ik_b$, independent of $\mathbf q$. The condition is sufficient, not necessary.

Equation~(\ref{eq:meta_lap}) holds to $\metaIdentity$ in every matrix element for all ten pairs, and Table~\ref{tab:meta_normals} applies the condition to each. Three pairs meet it. In \net{bcu}$\otimes$\net{nbo} every hub is planar and fourfold, and neighbouring normals are orthogonal, so the locked level is $4k_b$. In \net{nia}$\otimes$\net{nia-d} and \net{pcu}$\otimes$\net{urk} the mechanism sits on threefold and fourfold planar vertices, respectively, that no edge joins, at $3k_b$ and $4k_b$. The other four pairs with normal mechanisms have neighbouring normals at $|\mathbf n_i\cdot\mathbf n_j|$ from \metaDotMin{} to \metaDotMax. In the three pairs with non-planar mechanism vertices, \net{hex}$\otimes$\net{bnn}, \net{fcu}$\otimes$\net{flu} and \net{acs}$\otimes$\net{gra}, the mechanism space is not spanned by plane normals, so the condition does not apply. Locking does not flatten these seven pairs, whose locked width at $k_b=\kc$ stays between \metaOtherLockMin{} and \metaOtherLockMax\% (Fig.~\ref{fig:s_meta_concept}c and Table~\ref{tab:meta_map}).

\begin{figure*}[!htbp]\centering
\includegraphics[width=\textwidth]{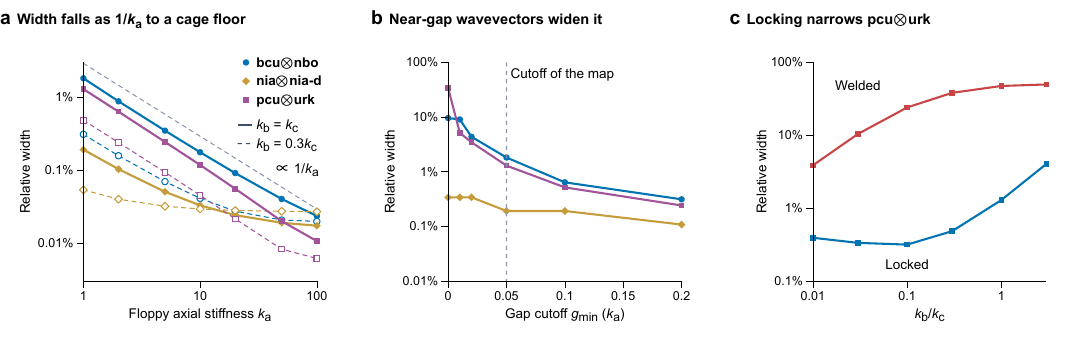}
\caption{\textbf{The residual width of a locked band comes from mixing with stretching modes and from cage motion, and it grows where the floppy-net gap closes.} \textbf{a}, Largest relative branch width of the locked band against the floppy axial stiffness $k_a$, at $\kc=10^{-3}$ and $k_b/\kc=1$ (filled, solid) or 0.3 (open, dashed), for the three pairs whose locked bending is exactly flat at first order. The width first falls as $1/k_a$ (dashed guide) and then approaches the cage floor of order $\kc/\ks$ near $k_a=100$. Equation~(\ref{eq:meta_res}) gives the bending contribution for \net{bcu}$\otimes$\net{nbo} and \net{nia}$\otimes$\net{nia-d}, and \net{pcu}$\otimes$\net{urk} also has contact-only and mixed terms. \textbf{b}, The same width at $k_b=\kc$ and $k_a=1$ against the gap cutoff of the \metaNq-point sample, with 0 for unfiltered draws; the dashed line marks the cutoff used throughout. Wavevectors near the planes where the floppy-net gap closes inflate the width of \net{bcu}$\otimes$\net{nbo} and \net{pcu}$\otimes$\net{urk} by more than an order of magnitude between $g_{\min}=0.2$ and no cutoff. \textbf{c}, Locked and welded widths of \net{pcu}$\otimes$\net{urk} against $k_b/\kc$; Fig.~4b of the main text shows the same comparison for \net{bcu}$\otimes$\net{nbo} and \net{nia}$\otimes$\net{nia-d}.}
\label{fig:s_meta_residual}
\end{figure*}

\textbf{\ref{sec:meta}.4 Residual width.} In the three exact pairs the first-order band is flat, so the residual width comes from higher orders. For a fixed cage and ideal locking, let $D_0=k_aR_F^\dagger R_F$, $Q=\one-ZZ^\dagger$ and $V=S+K_b$. At a gapped wavevector the effective operator through second order is
\begin{equation*}
H_{\rm eff}=Z^\dagger VZ-Z^\dagger VD_0^+VZ+O\!\left(\frac{\|V\|^3}{k_a^2g(\mathbf q)^2}\right).
\end{equation*}
The second term exists because $K_bZ$ is not confined to the mechanism space: the locked bending mixes each mechanism with the stretching modes of the floppy net, whose stiffness is $k_ag(\mathbf q)$. In \net{bcu}$\otimes$\net{nbo} each contact pair lies along the normal of its hub, and in \net{nia}$\otimes$\net{nia-d} the mechanisms lie in an eigenspace of $S$. In both, $QSZ=0$, so only bending enters this term. On a sample with $g\ge g_{\min}$, and with the geometry held fixed, this gives the upper scale
\begin{equation}
\frac{\Delta\omega^2}{\omega^2}=O\!\left(\frac{k_b^2}{(s\kc+zk_b)\,k_a\,g_{\min}}\right),
\label{eq:meta_res}
\end{equation}
with $s\kc$ the contact level. Its coefficient, and whether the band broadens at all, depend on the projected resolvent. In \net{pcu}$\otimes$\net{urk}, $QSZ\neq0$, and contact and mixed terms of orders $\kc^2$ and $\kc k_b$ enter as well. The finite rotational springs $k_{cr}$ and the mobile cage add corrections that this estimate omits. The frame follows this scale. At $k_b=\kc$ the locked width of \net{bcu}$\otimes$\net{nbo} falls from \metaNboLock\% at $k_a=1$ to \metaNboLockTen\% at $k_a=10$ and \metaNboLockHundred\% at $k_a=100$, where the cage floor of order $\kc/\ks$ takes over (Fig.~\ref{fig:s_meta_residual}a). It also grows as the sample admits wavevectors closer to the gap planes, from \metaNboCutTwo\% at $g_{\min}=0.2$ to \metaNboCutZero\% without a cutoff (Fig.~\ref{fig:s_meta_residual}b), so a width quoted without a cutoff characterizes the sample rather than the design. The gap of \net{nia-d} closes less often, and filtering changes its locked width only from \metaNiaLockRaw{} to \metaNiaLock\% (Table~\ref{tab:meta_map}).

\begin{figure*}[!htbp]\centering
\includegraphics[width=\textwidth]{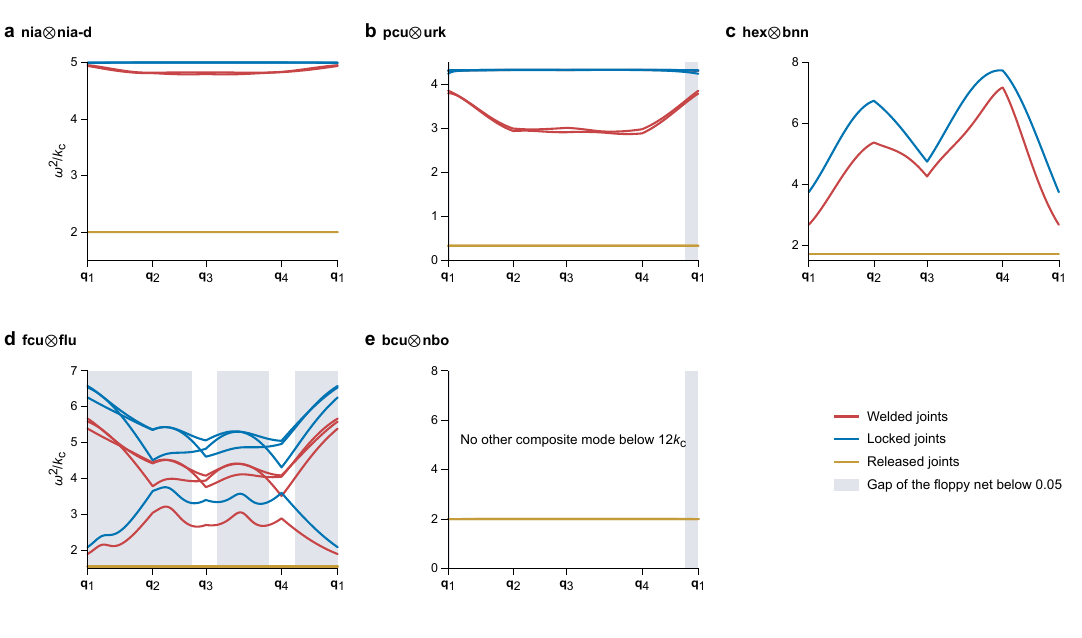}
\caption{\textbf{Released ends restore the first-order pin-joint spectrum of all pairs, whereas locking preserves first-order flatness only where mechanism normals are orthogonal or uncoupled.} Lifted bands of the beam frame, $\omega^2/\kc$, at $\kc=10^{-3}$, $k_b=\kc$, mobile thick cage and $k_a=1$, along a closed path through four generic wavevectors $\mathbf q_1$ to $\mathbf q_4$ of the conventional cell (Section~\ref{sec:meta}), for welded, locked and released joints. Grey bands mark segments where the gap of the floppy net falls below 0.05. \textbf{a}, \net{nia}$\otimes$\net{nia-d}: the locked band sits at the contact level plus $3k_b$. \textbf{b}, \net{pcu}$\otimes$\net{urk}: locked at the contact level plus $4k_b$. \textbf{c}, \net{hex}$\otimes$\net{bnn} and \textbf{d}, \net{fcu}$\otimes$\net{flu}: non-planar mechanism vertices leave both welded and locked bands dispersive, while the released bands lie near $\tfrac{12}{7}\kc$ and $\tfrac{14}{9}\kc$. \textbf{e}, \net{bcu}$\otimes$\net{nbo} with released ends, at $2\kc$ with no other composite mode below $12\kc$; its locked and welded bands are Fig.~4c of the main text.}
\label{fig:s_meta_bands}
\end{figure*}

\begin{figure*}[!htbp]\centering
\includegraphics[width=\textwidth]{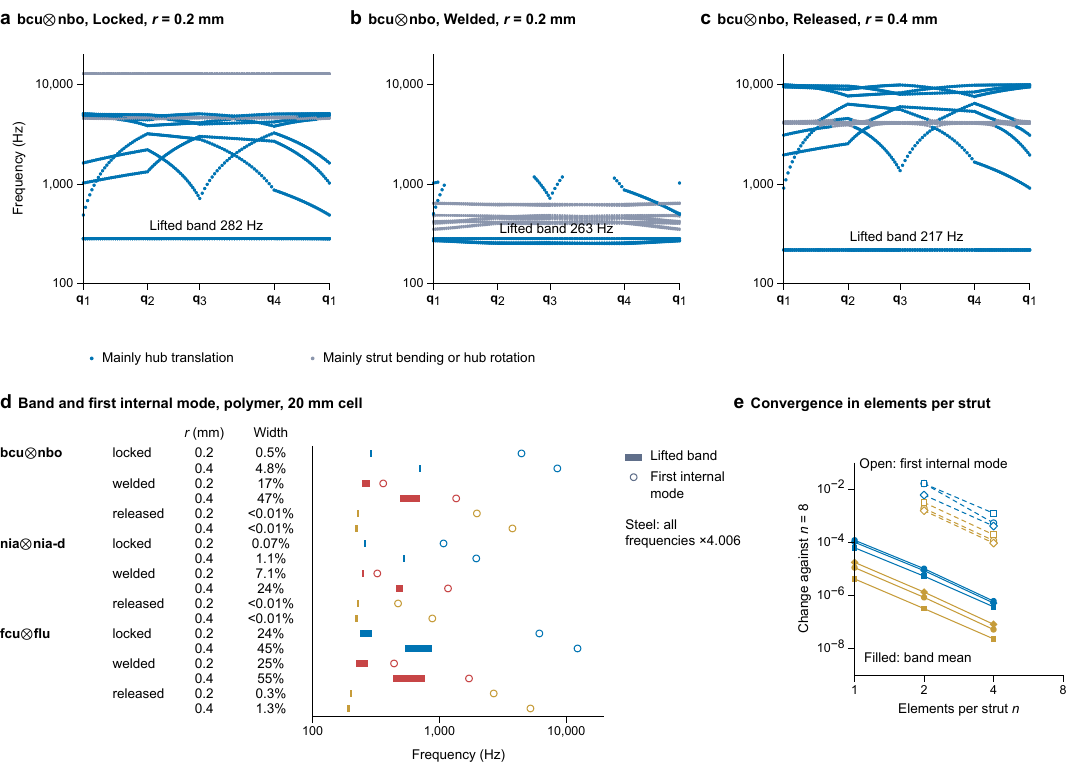}
\caption{\textbf{With distributed strut mass, the locked band of a 20~mm polymer cell stays nearly flat and well below the first internal resonance.} Generalized Bloch spectra with six consistent-mass Euler--Bernoulli elements per floppy strut and all six degrees of freedom per node, cage fixed. The material is polymer ($E=2$~GPa, $\rho=1200$~kg\,m$^{-3}$), with a conventional cell of 20~mm, node mass 100~mg and normal contact stiffness 100~N\,m$^{-1}$. \textbf{a}--\textbf{c}, \net{bcu}$\otimes$\net{nbo} along the path of Fig.~\ref{fig:s_meta_bands}, locked with strut radius 0.2~mm (\textbf{a}), welded with radius 0.2~mm (\textbf{b}) and released with radius 0.4~mm (\textbf{c}). Blue points are modes whose kinetic energy lies mainly on hub translations, grey points modes dominated by strut bending or hub rotation. The locked band at \metaHzPoly~Hz is isolated by a factor \metaMainRatio{} from the first strut mode, whereas welded hubs place rotation modes just above a dispersive band. \textbf{d}, Range of the lifted band (bars) and first internal mode (open circles) for three pairs, three joint designs and two strut radii, with the relative band width in $\omega^2$; steel at the same geometry scales every frequency by \metaSteelFactor. \textbf{e}, Relative change of the band mean (filled) and of the first internal mode (open) against eight elements per strut, for locked (blue) and released (gold) joints of the three pairs at radius 0.4~mm. Table~\ref{tab:meta_phys} lists the values.}
\label{fig:s_meta_mass}
\end{figure*}

\textbf{\ref{sec:meta}.5 Design rules.} Equations~(\ref{eq:meta_lap}) and (\ref{eq:meta_res}) set the design in sequence. Locking removes first-order bending dispersion when the mechanism normals meet the condition of Section~\ref{sec:meta}.3, as in \net{bcu}$\otimes$\net{nbo}, \net{nia}$\otimes$\net{nia-d} and \net{pcu}$\otimes$\net{urk}. Higher-order mixing then sets the residual width, which grows with $k_b/\kc$. At $k_b/\kc=0.3$ and $k_a=1$, locked \net{bcu}$\otimes$\net{nbo} has a width of \metaNboLockMid\%, against \metaNboFreeMid\% for welded hubs. Axially stiffer floppy struts reduce this residual as $1/k_a$, until cage motion sets a floor that falls with $\kc/\ks$. Ideal moment releases instead remove the bending contribution and recover the first-order pin-joint spectrum in all ten pairs.

\textbf{\ref{sec:meta}.6 Physical dimensions and distributed mass.} The rules can be met at printable dimensions. We took a conventional cell of 20~mm, circular floppy struts of radius 0.2 or 0.4~mm, nodes of 83.3~mm$^3$ (100~mg of polymer) and normal contact springs of 100~N\,m$^{-1}$ at $E=2$~GPa, as in leaf-spring cartridges. For steel the contact stiffness scales with $E$. At a fixed radius, struts of different length have different $k_b$, so the locked condition of Section~\ref{sec:meta}.3 uses the weighted Laplacian; in \net{nia}$\otimes$\net{nia-d} the mechanism vertices share no edge, so the locked level stays on site. Each floppy strut was split into six Euler--Bernoulli elements with the consistent mass matrix, including torsional inertia, and all six degrees of freedom of every node were kept in the generalized problem $K(\mathbf q)\mathbf x=\omega^2M(\mathbf q)\mathbf x$. Welded hubs carry the rotary inertia of a sphere of the node volume, locked hubs have their rotation held, and released struts have their own end rotations. The cage was held fixed to isolate the effect of mass. Band means change by less than $\metaConvBand$ and the first internal mode by less than \metaConvStrut\% between four and eight elements per strut (Fig.~\ref{fig:s_meta_mass}e).

The locked \net{bcu}$\otimes$\net{nbo} design with strut radius 0.2~mm, $k_a=\metaMainKa$~kN\,m$^{-1}$ and $k_b=\metaMainKb$~N\,m$^{-1}$, places the band at \metaHzPoly~Hz in polymer and \metaHzSteel~Hz in steel, with a width of \metaMainWidth\% in $\omega^2$ (Table~\ref{tab:meta_phys}). Its first internal mode, a strut bending mode, lies \metaMainRatio{} times higher. This separation supports a low-frequency picture in which strut inertia mainly renormalizes the projected hub mass, and the reported frequencies come from the full generalized eigenproblem. Thicker struts raise $k_b/\kc$ and with it the residual width of Section~\ref{sec:meta}.5: at radius 0.4~mm, $k_b/\kc=\metaLockFourRatio$, the band moves to \metaLockFourHz~Hz and widens to \metaLockFourWidth\%. Welded hubs fail differently. With strut radius 0.2~mm, they give a band centred at \metaWeldHz~Hz with \metaWeldWidth\% width in $\omega^2$. The first hub-rotation mode lies at \metaWeldRot~Hz, only \metaWeldRatio{} times the upper frequency of the band (Fig.~\ref{fig:s_meta_mass}b). These rotation modes must therefore be retained dynamically. The locked \net{nia}$\otimes$\net{nia-d} design gives \metaNiaHz~Hz at \metaNiaWidthPhys\% width. With struts of radius 0.4~mm, released \net{fcu}$\otimes$\net{flu} acquires a width of \metaFluRelWidth\% from strut mass alone, because its projected mass depends on $\mathbf q$. In \net{nbo}, by contrast, the orthogonal normals cancel the cross terms of the strut mass and keep its released band at \metaRelHz~Hz with a width of $\metaRelWidthPhys$\%. Steel scales every frequency by \metaSteelFactor{} at the same geometry.

\begin{table}[h]\centering
\caption{Physical realization with distributed strut mass in a 20~mm conventional cell. Nodes have volume 83.3~mm$^3$, which is 100~mg in polymer. Polymer has $E=2$~GPa and $\rho=1200$~kg\,m$^{-3}$; steel has $E=210$~GPa and $\rho=7850$~kg\,m$^{-3}$. The normal contact stiffness is 100~N\,m$^{-1}$ in polymer and scales with $E$ for steel. Each floppy strut has six consistent-mass elements, the cage is fixed, and spectra use 24 wavevectors with $g\ge0.05$. The strut columns give the radius $r$, the lengths $L$ and the bending stiffnesses $k_b=12EI/L^3$ in corresponding order, with $I=\pi r^4/4$. $f$ is the mean frequency of the lifted band, and width is its relative branch width in $\omega^2$. The polymer frequency $f_1$ is that of the first mode whose kinetic energy is mainly strut bending or hub rotation. Ratio is $f_1$ over the top of the band.}\label{tab:meta_phys}
\begin{tabular}{llcccccccc}
\hline
Pair & joints & $r$ (mm) & $L$ (mm) & $k_b$ (N\,m$^{-1}$) & $f$ polymer (Hz) & $f$ steel (Hz) & width (\%) & $f_1$ (Hz) & ratio \\
\hline
\net{bcu}$\otimes$\net{nbo} & locked & 0.2 & 10.0 & 30 & 281.8 & 1129 & 0.48 & 4434 & 15.7 \\
\net{bcu}$\otimes$\net{nbo} & locked & 0.4 & 10.0 & 483 & 701.9 & 2812 & 4.8 & 8500 & 12.1 \\
\net{bcu}$\otimes$\net{nbo} & welded & 0.2 & 10.0 & 30 & 263.0 & 1054 & 17 & 360 & 1.3 \\
\net{bcu}$\otimes$\net{nbo} & welded & 0.4 & 10.0 & 483 & 603.3 & 2417 & 47 & 1354 & 1.9 \\
\net{bcu}$\otimes$\net{nbo} & released & 0.2 & 10.0 & 30 & 222.8 & 893 & 0.00020 & 1972 & 8.9 \\
\net{bcu}$\otimes$\net{nbo} & released & 0.4 & 10.0 & 483 & 216.5 & 867 & 0.00065 & 3761 & 17.4 \\
\net{nia}$\otimes$\net{nia-d} & locked & 0.2 & 11.5/20.0 & 20/4 & 253.5 & 1016 & 0.072 & 1077 & 4.2 \\
\net{nia}$\otimes$\net{nia-d} & locked & 0.4 & 11.5/20.0 & 313/60 & 517.1 & 2072 & 1.1 & 1950 & 3.8 \\
\net{nia}$\otimes$\net{nia-d} & welded & 0.2 & 11.5/20.0 & 20/4 & 246.8 & 989 & 7.1 & 323 & 1.3 \\
\net{nia}$\otimes$\net{nia-d} & welded & 0.4 & 11.5/20.0 & 313/60 & 474.0 & 1899 & 24 & 1168 & 2.3 \\
\net{nia}$\otimes$\net{nia-d} & released & 0.2 & 11.5/20.0 & 20/4 & 223.1 & 894 & 0.000079 & 471 & 2.1 \\
\net{nia}$\otimes$\net{nia-d} & released & 0.4 & 11.5/20.0 & 313/60 & 217.6 & 872 & 0.00027 & 878 & 4.0 \\
\net{fcu}$\otimes$\net{flu} & locked & 0.2 & 8.7 & 46 & 268.1 & 1074 & 24 & 6131 & 20.7 \\
\net{fcu}$\otimes$\net{flu} & locked & 0.4 & 8.7 & 743 & 725.3 & 2906 & 45 & 12262 & 14.0 \\
\net{fcu}$\otimes$\net{flu} & welded & 0.2 & 8.7 & 46 & 245.8 & 985 & 25 & 439 & 1.6 \\
\net{fcu}$\otimes$\net{flu} & welded & 0.4 & 8.7 & 743 & 611.7 & 2451 & 55 & 1710 & 2.2 \\
\net{fcu}$\otimes$\net{flu} & released & 0.2 & 8.7 & 46 & 195.7 & 784 & 0.34 & 2677 & 13.7 \\
\net{fcu}$\otimes$\net{flu} & released & 0.4 & 8.7 & 743 & 188.1 & 754 & 1.3 & 5214 & 27.6 \\
\hline
\end{tabular}
\end{table}

\textbf{\ref{sec:meta}.7 Finite specimen in time.} An excitation confined to an exactly flat band has no relative phase evolution between its Bloch components and does not spread. A local force also excites other bands, so a finite specimen measures the resulting spatial response directly. We quantify the energy distribution over the finite simulation by its participation number and energy-weighted RMS radius. These specimens use the massless-beam frame of Section~\ref{sec:meta}.1 with a mobile cage, not the distributed-mass model of Section~\ref{sec:meta}.6. Specimens of $6\times6\times6$ conventional cells were built from the beam frame at $\kc=10^{-3}$ and $k_b=\kc$, with the cage clamped outside the block and floppy edges that leave it removed; \net{bcu}$\otimes$\net{nbo} has \metaNdof{} translational degrees of freedom after condensation. The condensed stiffness was diagonalized once and the response to a point force on the floppy hub nearest the centre, along its mechanism polarization, was integrated exactly mode by mode, with modal damping ratio $\zeta=1/(2Q)$ equal for every mode. The force is a sine under a Gaussian envelope of four carrier periods, whose relative bandwidth of \metaBurstBW\% covers the whole locked band. The on-band carrier is the mean frequency of the low-lying modes weighted by their amplitude at the driven hub, $\omega^2=\metaWonNbo\,\kc$ for locked \net{bcu}$\otimes$\net{nbo}, and the off-band carrier is the peak of the same weight among the stiff branches, $\omega^2=\metaWoffNbo\,\kc$. Energy per node gives the participation number $N$ and the energy-weighted RMS distance $R$ from the driven hub.

\begin{figure*}[!htbp]\centering
\includegraphics[width=\textwidth]{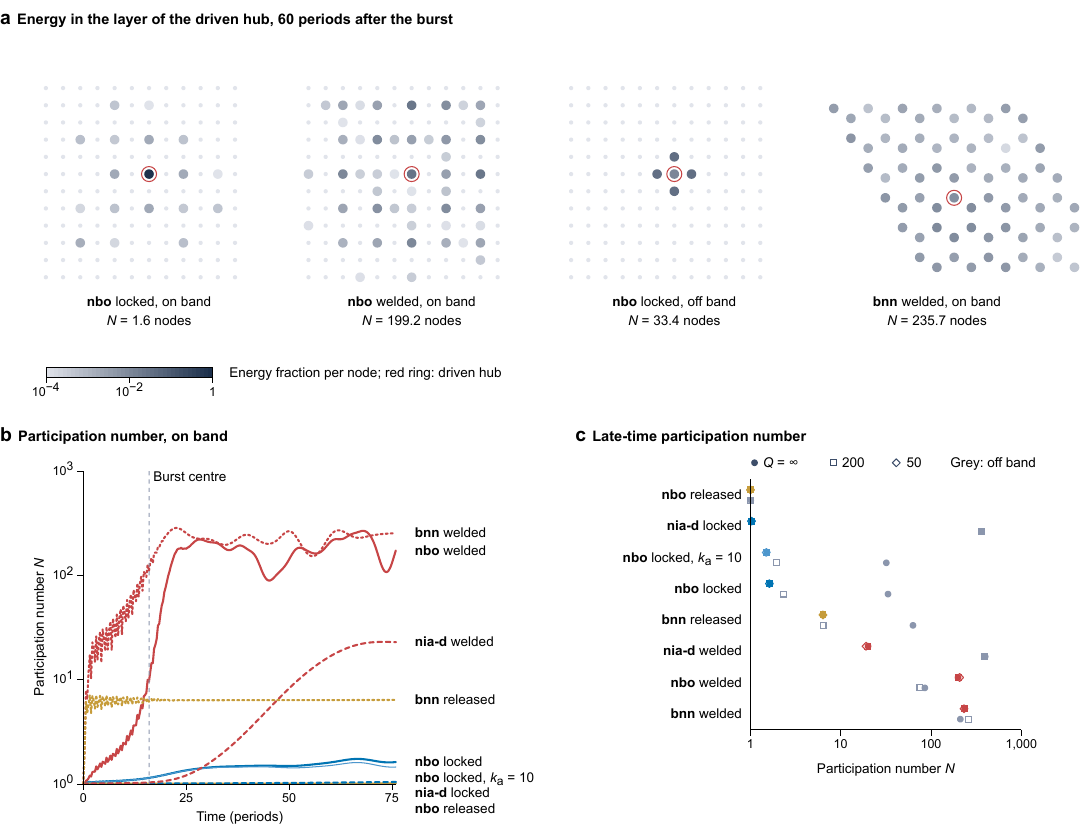}
\caption{\textbf{An on-band burst stays near the driven hub of a locked finite specimen and spreads in its welded counterpart.} Specimens of $6\times6\times6$ conventional cells, $\kc=10^{-3}$, $k_b=\kc$, mobile thick cage clamped outside the block, driven at one floppy hub by a Gaussian-windowed sine of four carrier periods. \textbf{a}, Energy fraction per node in the layer of the driven hub (red ring), \metaAfter{} periods after the burst, with the participation number $N$ averaged over the last 20 periods. \textbf{b}, Participation number against time for every on-band case without damping; the dashed line marks the centre of the burst. Locked \net{nbo} and \net{nia-d} and released \net{nbo} hold the energy on one or two nodes; welded joints spread it over tens to hundreds of nodes. \textbf{c}, Late-time participation number on band (coloured) and off band (grey) for $Q=\infty$ (filled circles), 200 (squares) and 50 (diamonds; on band only). Damping leaves the locked on-band participation unchanged to the reported precision. The RMS radius of \net{bcu}$\otimes$\net{nbo} is Fig.~4d of the main text, and Table~\ref{tab:meta_time} lists all cases.}
\label{fig:s_meta_time}
\end{figure*}

On the band, the energy of the locked specimen stays near the driven hub over the simulated time, with participation number $N=\metaPRon$ and mean RMS radius $R=\metaLocCells$ cells, with \metaFracOne\% of the energy within one cell. These are means over the last 20 periods of the \metaAfter{} periods after the burst (Fig.~4d of the main text, Fig.~\ref{fig:s_meta_time} and Table~\ref{tab:meta_time}). The welded specimen, driven at the carrier of its own band, has $N=\metaPRfree$ and $R=\metaRfree$ cells. Off the band, the locked specimen stores only about $10^{\metaEoffExp}$ of the on-band energy after the burst, and what it keeps spreads over $N=\metaPRoff$ nodes. Released \net{bcu}$\otimes$\net{nbo} is the exception among the off-band runs: its carrier, $\omega^2=\metaWoffRel\,\kc$, falls on stiff modes that are themselves confined to the driven hub, and the response stays on $N=\metaPRnboRelOff$ nodes. Locked \net{nia}$\otimes$\net{nia-d} holds its energy on $N=\metaPRnia$ nodes, whereas its welded version spreads over \metaPRniaFree{} nodes. Welded \net{hex}$\otimes$\net{bnn}, whose mechanism is collective, reaches \metaPRbnn{} nodes, and its released version stays on \metaPRbnnRel{} nodes. For locked \net{bcu}$\otimes$\net{nbo}, the participation number remains \metaPRonQfifty{} to the reported precision at $Q=50$. Its narrow frequency range gives similar modal decay rates. The corresponding welded value changes from \metaPRfreeOne{} to \metaPRfreeQfifty{}.

\begin{table}[h]\centering
\caption{Finite specimens of $6\times6\times6$ conventional cells after a point burst on one floppy hub, $\kc=10^{-3}$, $k_b=\kc$. Here $\omega_{\rm on}^2$ is the squared on-band carrier frequency and $N$ the participation number of the nodal energy. The energy-weighted RMS radius $R$ is measured from the driven hub in conventional cells. The next column gives the energy share within one cell of the hub. All observables are averaged over the final 20 periods, without damping unless marked.}\label{tab:meta_time}
\begin{tabular}{llccccccc}
\hline
Pair & joints & $\omega_{\rm on}^2/\kc$ & $N$ & $R$ & within one cell (\%) & $N$, $Q=50$ & $N$, off band & $R$, off band \\
\hline
\net{bcu}$\otimes$\net{nbo} & locked & 6.00 & 1.6 & 1.02 & 85 & 1.6 &33 & 1.93 \\
\net{bcu}$\otimes$\net{nbo} & locked, $k_a=10$ & 6.00 & 1.5 & 0.99 & 86 & 1.5 &32 & 1.93 \\
\net{bcu}$\otimes$\net{nbo} & welded & 5.18 & 199.2 & 2.68 & 9 & 207.4 &85 & 2.39 \\
\net{bcu}$\otimes$\net{nbo} & released & 2.00 & 1.0 & 0.11 & 100 & 1.0 &1 & 0.16 \\
\net{nia}$\otimes$\net{nia-d} & locked & 5.00 & 1.0 & 0.30 & 99 & 1.0 &363 & 3.70 \\
\net{nia}$\otimes$\net{nia-d} & welded & 4.86 & 20.1 & 1.63 & 30 & 19.1 &398 & 3.70 \\
\net{hex}$\otimes$\net{bnn} & welded & 4.44 & 235.7 & 2.86 & 4 & 232.0 &211 & 2.44 \\
\net{hex}$\otimes$\net{bnn} & released & 1.71 & 6.4 & 0.84 & 86 & 6.4 &63 & 2.15 \\
\hline
\end{tabular}
\end{table}

\begin{figure*}[!htbp]\centering
\includegraphics[width=\textwidth]{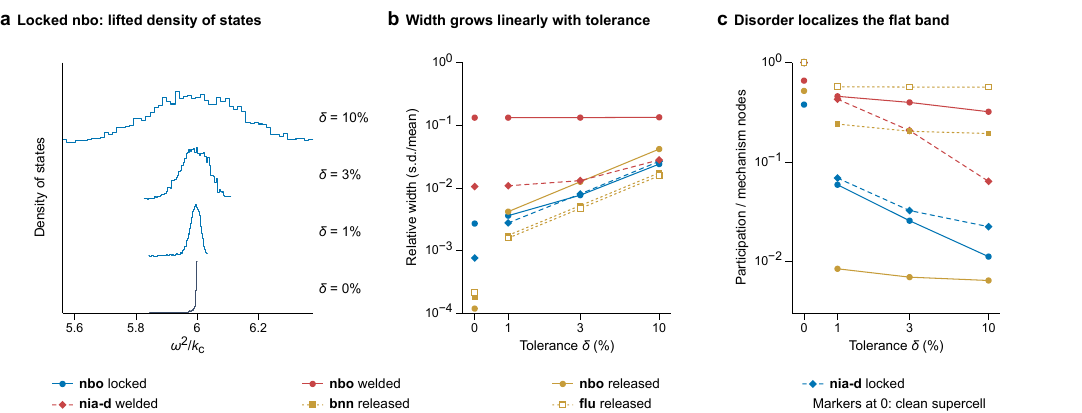}
\caption{\textbf{Fabrication errors widen the flat bands linearly and concentrate their states on one to ten hubs.} Periodic $3\times3\times3$ supercells at $\kc=10^{-3}$, $k_b=\kc$, mobile thick cage, with every contact and floppy-strut bending stiffness multiplied by $1+\delta\xi$, $\xi$ uniform on $[-1,1]$; \metaTolSeeds{} seeds and two supercell wavevectors per seed and tolerance. \textbf{a}, Density of lifted states of locked \net{bcu}$\otimes$\net{nbo}, pooled over seeds and normalized to its peak, for $\delta=0$, 1, 3 and 10\%. \textbf{b}, Standard deviation of the lifted density of states over its mean against $\delta$, with the clean supercell at 0. Locked and released flat bands widen in proportion to $\delta$, whereas the welded \net{nbo} band keeps its clean dispersion. \textbf{c}, Mean participation number of the lifted eigenvectors over the hubs that carry mechanisms, divided by their number. The flat bands of locked \net{nbo} and \net{nia-d} and released \net{nbo} localize on one to ten hubs from $\delta=1\%$ on; welded \net{nbo} and the collective mechanisms of released \net{bnn} and \net{flu} stay largely extended.}
\label{fig:s_meta_tolerance}
\end{figure*}

\textbf{\ref{sec:meta}.8 Fabrication tolerance.} A flat band has no dispersion to compete with disorder, so fabrication errors set both its width and the extent of its states. Every contact stiffness and every floppy-strut bending stiffness of a $3\times3\times3$ supercell was multiplied by $1+\delta\xi$, with $\xi$ uniform on $[-1,1]$ and $\delta=1$, 3 and 10\%, over \metaTolSeeds{} seeds. Each seed used two supercell wavevectors, chosen so that every folded wavevector has $g\ge0.05$ wherever such a choice exists (Fig.~\ref{fig:s_meta_tolerance}). For \net{fcu}$\otimes$\net{flu} no such choice exists, and the best draws reach $g=\metaTolFluGap$. The width here is the standard deviation of the lifted density of states over its mean, which is smaller than the branch width of Table~\ref{tab:meta_map}. For locked \net{bcu}$\otimes$\net{nbo} it rises from \metaTolLockZero\% in the clean supercell to \metaTolLockOne, \metaTolLockThree{} and \metaTolLockTen\%, so that at large $\delta$ it approaches \metaTolSlopeLock$\,\delta$. Locked \net{nia}$\otimes$\net{nia-d} follows \metaTolSlopeNia$\,\delta$, and released \net{bcu}$\otimes$\net{nbo} follows \metaTolSlopeRel$\,\delta$, more steeply because the contact errors then carry the whole level. Welded \net{bcu}$\otimes$\net{nbo} keeps its clean width of \metaTolFreeZero\% at every $\delta$.

The same errors localize the states of the flat bands. In the clean finite-stiffness supercell, the lifted band of locked \net{bcu}$\otimes$\net{nbo} has \metaTolLockZero\% relative standard deviation. Its computed eigenvectors have a mean participation of \metaTolPRLockZero\% of the \metaTolNmc{} mechanism hubs. At exact degeneracy, participation would depend on the chosen eigenbasis. At $\delta=1\%$ the participation number falls to \metaTolNodesOneLock{} hubs, and at $\delta=10\%$ to \metaTolNodesLock{}, an equivalent length of \metaTolXiLockTen{} cells. Released \net{bcu}$\otimes$\net{nbo} and locked \net{nia}$\otimes$\net{nia-d} localize on \metaTolNodesOneRel{} and \metaTolNodesOneNia{} hubs already at 1\%. Welded \net{bcu}$\otimes$\net{nbo} stays extended, with \metaTolPRFreeTen\% of its hubs at $\delta=10\%$, because its clean dispersion exceeds the disorder. The collective mechanisms of released \net{hex}$\otimes$\net{bnn} and \net{fcu}$\otimes$\net{flu} localize only partly, to \metaTolPRBnnTen{} and \metaTolPRFluTen\% of their hubs. A fabricated locked lattice with errors of a few per cent therefore carries a band of hub-local resonances whose spread grows linearly with the tolerance, the disordered counterpart of the localization seen in the time domain.

\textbf{Computational details.} Geometry, placement and contacts come from the screened records, with the RCSR embedding. Design-map wavevectors are uniform in the conventional reduced zone (seed 20261107); the filtered sample keeps draws with $g\ge0.05$ from the same stream. Band paths join the reduced points $(0.06,0.13,0.21)$, $(0.44,0.17,0.29)$, $(0.38,0.41,0.09)$ and $(0.11,0.36,0.45)$ of the conventional cell. The time integration uses the exact two-by-two propagator of each damped mode with the force held over steps of 1/64 carrier period. Supercell spectra use the same element assembly with per-element stiffness multipliers; a $2\times2\times2$ supercell reproduces the folded unit-cell spectrum to rounding error.

\begin{table}[h]\centering
\caption{Largest relative branch width (\%) of the lifted band in the beam frame, $\kc=10^{-3}$, cage $k_{s,a}=10$, $k_{s,b}=1$, on 48 random wavevectors with floppy-net gap $g\ge0.05$ unless marked. $M$, mechanisms per conventional cell. Welded and locked columns at $k_b/\kc=0.1$ and 1; $k_a$, floppy axial stiffness; raw, the same number of unfiltered draws; released, moment-free strut ends at any $k_b$.}\label{tab:meta_map}
\begin{tabular}{lcccccccc}
\hline
Pair & $M$ & \multicolumn{2}{c}{welded} & \multicolumn{2}{c}{locked} & locked, $k_a=10$ & locked, raw & released \\
 & & 0.1 & 1 & 0.1 & 1 & 1 & 1 & \\
\hline
\net{bcu}$\otimes$\net{nbo} & 6 & 7.1 & 35 & 0.060 & 1.8 & 0.18 & 9.6 & 0.040 \\
\net{hxg-d}$\otimes$\net{hxg} & 4 & 11 & 39 & 5.0 & 18 & 18 & 18 & 0.032 \\
\net{nia}$\otimes$\net{nia-d} & 2 & 0.84 & 4.0 & 0.037 & 0.19 & 0.033 & 0.34 & 0.061 \\
\net{she-d}$\otimes$\net{she} & 20 & 3.5 & 18 & 1.4 & 13 & 13 & 17 & 0.020 \\
\net{bsn}$\otimes$\net{lvt} & 8 & 9.7 & 37 & 7.6 & 39 & 38 & 67 & 0.46 \\
\net{qtz-x}$\otimes$\net{uom} & 6 & 8.4 & 29 & 5.9 & 24 & 24 & 74 & 0.74 \\
\net{pcu}$\otimes$\net{urk} & 3 & 24 & 48 & 0.32 & 1.3 & 0.12 & 34 & 0.44 \\
\net{hex}$\otimes$\net{bnn} & 1 & 31 & 135 & 28 & 105 & 105 & 106 & 0.041 \\
\net{fcu}$\otimes$\net{flu} & 4 & 8.5 & 37 & 8.5 & 32 & 32 & 63 & 0.014 \\
\net{acs}$\otimes$\net{gra} & 4 & 35 & 100 & 33 & 104 & 104 & 105 & 0.12 \\
\hline
\end{tabular}
\end{table}
\begin{table}[h]\centering
\caption{Exact condition for flat locked bending. Mechanism vertices: vertices carrying nonzero mechanism weight, out of the $n$ floppy vertices per conventional cell. The next columns give their coordination and planar count. Normal purity is the fraction of the mechanism norm on the plane normals. The dot-product column gives the largest $|\mathbf n_i\cdot\mathbf n_j|$ over bonds between planar mechanism vertices; ``none'' denotes no such bond. Locked spectra are eigenvalues of $Z^\dagger(\mathcal L\otimes I_3)Z$, in units of $k_b$. Welded spectra use the condensed bending operator. Both use 30 wavevectors with $g\ge0.02$. The identity $Z^\dagger K_b^{(\theta=0)}Z=k_bZ^\dagger(\mathcal L\otimes I_3)Z$ holds to $1\times10^{-15}$ for every pair.}\label{tab:meta_normals}
\begin{tabular}{lcccccccc}
\hline
Pair & $M$ & mech.\ vertices & coordination & planar & purity & $|\mathbf n_i\cdot\mathbf n_j|$ & locked & welded \\
\hline
\net{bcu}$\otimes$\net{nbo} & 6 & 6/6 & 4 & 6 & 1.000 & 0.000 & 4.000 & 2.02--4.00 \\
\net{hxg-d}$\otimes$\net{hxg} & 4 & 4/4 & 6 & 4 & 1.000 & 0.333 & 4.31--7.59 & 3.75--7.43 \\
\net{nia}$\otimes$\net{nia-d} & 2 & 2/4 & 3 & 2 & 1.000 & none & 3.000 & 2.80--2.99 \\
\net{she-d}$\otimes$\net{she} & 20 & 20/20 & 4,6 & 20 & 1.000 & 0.577 & 2.15--7.85 & 2.05--7.71 \\
\net{bsn}$\otimes$\net{lvt} & 8 & 8/8 & 4 & 8 & 1.000 & 0.789 & 1.05--6.95 & 0.84--6.63 \\
\net{qtz-x}$\otimes$\net{uom} & 6 & 6/6 & 4 & 6 & 1.000 & 0.882 & 1.34--6.66 & 0.58--5.62 \\
\net{pcu}$\otimes$\net{urk} & 3 & 3/7 & 4 & 3 & 1.000 & none & 4.000 & 2.20--4.00 \\
\net{hex}$\otimes$\net{bnn} & 1 & 2/2 & 5 & 0 & -- & -- & 0.79--6.73 & 0.39--6.51 \\
\net{fcu}$\otimes$\net{flu} & 4 & 12/12 & 4,8 & 0 & -- & -- & 0.94--4.86 & 0.52--3.93 \\
\net{acs}$\otimes$\net{gra} & 4 & 4/4 & 3,5 & 2 & -- & -- & 0.39--4.03 & 0.24--3.84 \\
\hline
\end{tabular}
\end{table}

\clearpage
\section{Figures of the atlas}\label{sec:figs}

These figures resolve the geometry and spectra behind the atlas. They show individual contact stars, selected mechanism displacements, band paths and pair-resolved controls. The numerical definitions and full records are in Section~\ref{sec:data}.

\begin{figure*}[t]\centering
\includegraphics[width=\textwidth]{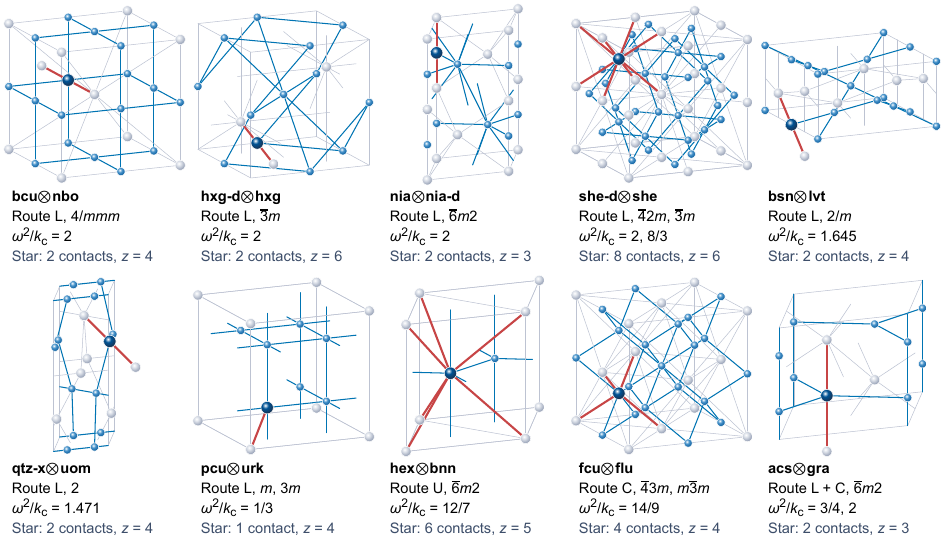}
\caption{\textbf{The ten exactly flat pairs cage a floppy net inside a rigid catalogued dual scaffold.} Each panel shows one conventional cell of the dual scaffold (grey) with the floppy net (blue) inside it. Edges that leave the cell are cut at its faces. The dark vertex and the red lines mark the contact star of one floppy vertex. It is taken from the vertex class that carries the stated value. For route L this is a planar vertex with $\mathbf n_i^{\mathsf T}S_i\mathbf n_i$ equal to a flat value. In \net{she-d}$\otimes$\net{she} the marked vertex is 6-c, and it is planar as well, carrying the branch at $\tfrac83$ with an isotropic $S_i$, while the 4-c vertices carry the branch at 2. For route U it is a vertex whose degenerate in-plane eigenvalue of $S_i$ equals the flat value, and for route C it is a 4-c vertex of \net{flu}. For \net{acs}$\otimes$\net{gra} it is the planar 3-c vertex of the local sector. Below each cell are the pair (scaffold $\otimes$ floppy net), the route, the site groups of the floppy vertices in the composite and the first-order flat values $\omega^2/\kc$. The last line gives the number of contacts and the bond degree $z$ of the marked vertex. The values 1.645, 1.471 and 12/7 are values of the embedding, and in \net{pcu}$\otimes$\net{urk} the screened alignment lowers the site group (Table~\ref{M-tab:atlas}). The pairs follow the order of that table. Placement and contacts are those of the screen: the database embedding, the clearance-optimized origin and contacts within 1.05 times the nearest contact distance.}
\label{fig:s_gallery}
\end{figure*}

\begin{figure*}[t]\centering
\includegraphics[width=\textwidth]{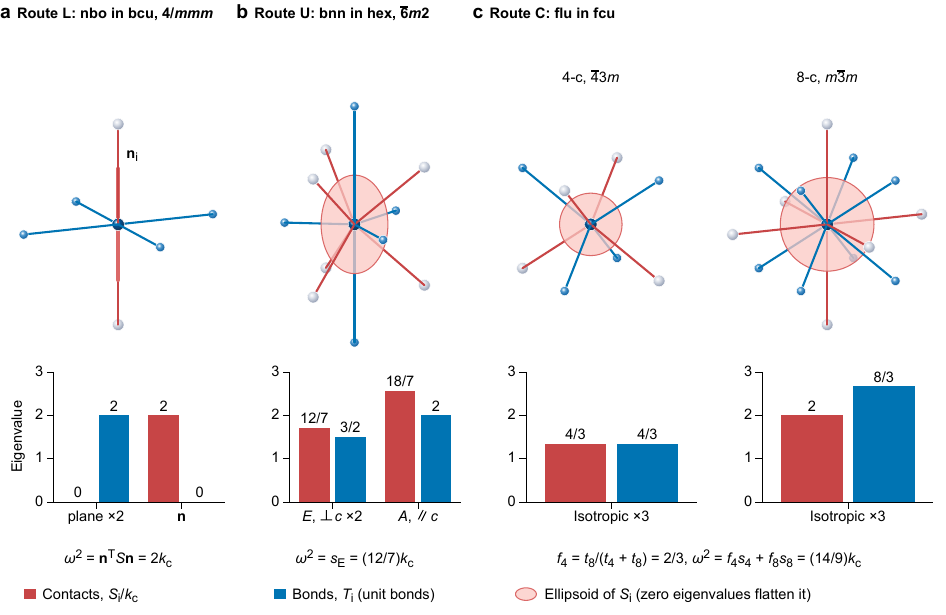}
\caption{\textbf{Site symmetry makes the contact tensor a scalar on each sector of the three route prototypes.} Top, one floppy vertex in its database geometry, with its bonds (blue, true lengths) and its contacts (red) to scaffold vertices (grey). The shaded glyph is the ellipsoid of $S_i$, with semi-axes proportional to its eigenvalues. A zero eigenvalue flattens it, so in \textbf{a} it collapses to a bar along $\mathbf n_i$. Bottom, eigenvalues of $S_i/\kc$ (red) and of the bond tensor $T_i=\sum\hat{\mathbf b}\hat{\mathbf b}^{\mathsf T}$ over bonds of unit stiffness (blue) on each symmetry sector, with its multiplicity. Both tensors share the sectors at every site shown. \textbf{a}, Route L, \net{nbo} in \net{bcu}. Two contacts lie along the normal of the square-planar vertex, so $S_i=2\kc\,\mathbf n_i\mathbf n_i^{\mathsf T}$ and $\omega^2=\mathbf n_i^{\mathsf T}S_i\mathbf n_i=2\kc$. \textbf{b}, Route U, \net{bnn} in \net{hex}. The six contacts give $s_E=\tfrac{12}{7}\kc$ in the plane normal to $c$ and $s_A=\tfrac{18}{7}\kc$ along it, and the mechanisms see only $s_E$. The axial bonds are drawn at their true length, $\sqrt3$ times the in-plane bonds. \textbf{c}, Route C, \net{flu} in \net{fcu}, drawn at one scale for both classes. Both tensors are isotropic. The contact scalars are $s_4=\tfrac43\kc$ and $s_8=2\kc$, and the bond scalars of unit-stiffness bonds are $t_4=\tfrac43$ and $t_8=\tfrac83$. The bond scalars fix the norm share $f_4=t_8/(t_4+t_8)=\tfrac23$ (Section~S2.4), which places the band at $\tfrac{14}{9}\kc$ to first order. These are the route prototypes shown schematically in Fig.~\ref{M-fig:rule}c. The present panels add the embedding and the bond tensors.}
\label{fig:s_stars}
\end{figure*}

\begin{figure*}[t]\centering
\includegraphics[width=\textwidth]{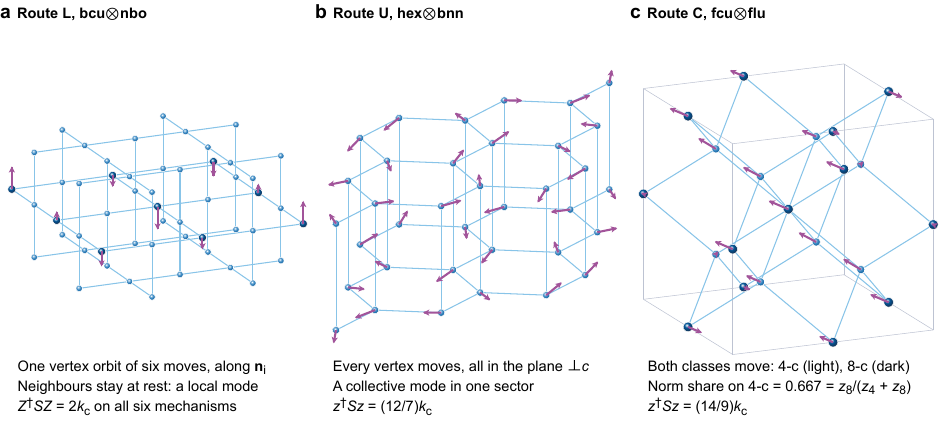}
\caption{\textbf{The lifted mechanism is local in route L and collective in routes U and C.} Arrows show the displacement $\mathrm{Re}[z_k e^{2\pi i\mathbf q\cdot\mathbf t}]$ of one lifted mechanism at the generic wavevector $\mathbf q=(0.31,0.17,0.43)$, in reduced coordinates of the conventional cell of each scaffold. Here $z$ is the normalized mechanism vector, $z_k$ its displacement at vertex $k$ and $\mathbf t$ the lattice translation of the drawn image. The global phase maximizes the drawn amplitude. The scaffold is held fixed and is not drawn; it has no mechanism at this $\mathbf q$. Arrows are scaled per panel. \textbf{a}, \net{bcu}$\otimes$\net{nbo}, $2\times2\times1$ cells. The kernel at $\mathbf q$ is six-dimensional, and $Z^\dagger SZ=2\kc\one$ on it. The drawn mechanism moves one of the six vertices of the conventional cell, with all its lattice images, along its normal (dark vertices), while all neighbours stay at rest. \textbf{b}, \net{hex}$\otimes$\net{bnn}, $3\times3\times2$ cells. The single mechanism per cell moves every vertex, always in the plane normal to $c$. This plane is the degenerate sector of $S_i$, so $z^\dagger Sz=\tfrac{12}{7}\kc$. \textbf{c}, \net{fcu}$\otimes$\net{flu}, one conventional cubic cell. The single mechanism of the primitive cell moves both classes, 4-c (light) and 8-c (dark). The 4-c class carries the norm share $8/(4+8)=\tfrac23$, set by the two coordinations, at this $\mathbf q$ and at 20 further random wavevectors, so $z^\dagger Sz=\tfrac{14}{9}\kc$. Because the 4-c vertices are twice as many, both classes carry the same mean squared amplitude per vertex.}
\label{fig:s_modes}
\end{figure*}

\begin{figure*}[t]\centering
\includegraphics[width=\textwidth]{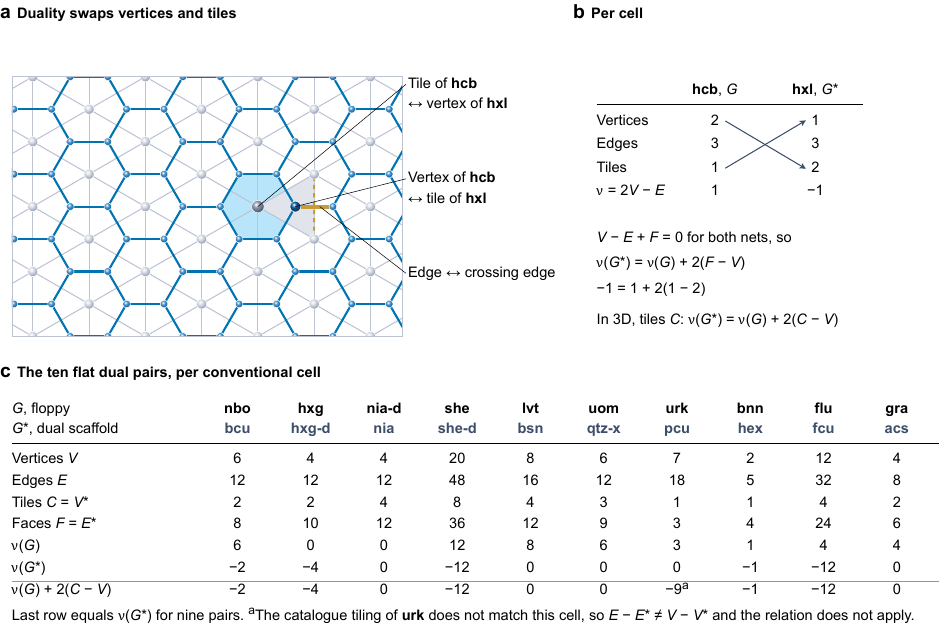}
\caption{\textbf{Tiling duality exchanges vertices and tiles, so Euler's relation fixes the Maxwell index of the dual.} \textbf{a}, The honeycomb net \net{hcb} (blue) and its dual, the triangular net \net{hxl} (grey). The shaded hexagonal tile of \net{hcb} holds one \net{hxl} vertex (dark grey). The shaded triangular tile of \net{hxl} holds one \net{hcb} vertex (dark blue), and every \net{hcb} edge crosses one \net{hxl} edge (gold). \textbf{b}, Counts per primitive cell. A periodic planar net obeys $V-E+F=0$, and its Maxwell index is $\nu=2V-E$. Duality swaps vertices and tiles and keeps the number of edges, so $\nu(G^*)=\nu(G)+2(F-V)$, and in two dimensions $\nu(G^*)=-\nu(G)$ always. For \net{hcb} this gives $\nu(G^*)=-1$, the opposite of $\nu(G)=1$. This is the planar analogue of $\nu(G^*)=\nu(G)+2(C-V)$ for three-periodic tilings (Section~S3.4), where the tiles $C$ take the place of the faces $F$. \textbf{c}, The same ledger for the ten exactly flat pairs, per conventional cell of the scaffold, with $\nu=3V-E$, $C=V^*$ and $F=E^*$. The counts come from the graphs used in the screen. For nine pairs, $\nu(G)+2(C-V)$ reproduces $\nu(G^*)$; the identity needs only a vanishing Euler residual, whereas the face-count check of Section~S3.4 tests whether the scaffold is the dual of the natural tiling. For \net{pcu}$\otimes$\net{urk} the catalogue tiling does not match the cell, $E-E^*\neq V-V^*$, and the relation is not applied, as in Section~S3.4.}
\label{fig:s_euler}
\end{figure*}
\begin{figure*}[t]\centering
\includegraphics[width=\textwidth]{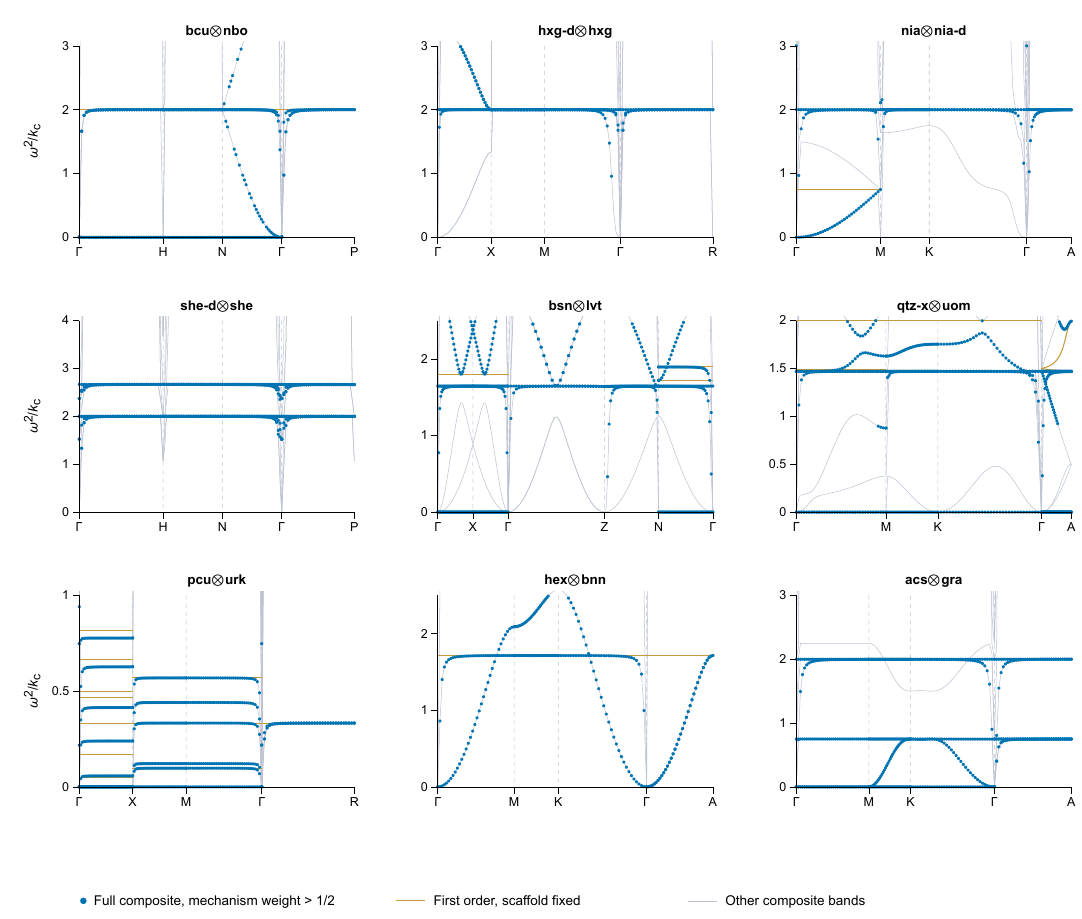}
\caption{\textbf{High-symmetry paths expose mode mixing in pairs whose lifted bands are flat at first order.} Band structures of the full composite at $\kc=10^{-3}$, with $\ks=1$, unit masses and the screened placement, in the primitive cell, for nine of the ten exactly flat pairs; \net{fcu}$\otimes$\net{flu} is shown in Fig.~\ref{M-fig:concept}b. Markers show modes with more than half their weight in the mechanism space of the floppy net, gold lines the first-order spectrum $\operatorname{eig}[\Pi S\Pi]$ at the same wavevector with the scaffold fixed, and thin grey lines all other composite bands. Special points are given in reduced coordinates of the conventional reciprocal cell. For primitive cubic lattices they are X $(0,\tfrac12,0)$, M $(\tfrac12,\tfrac12,0)$ and R $(\tfrac12,\tfrac12,\tfrac12)$, and for body-centred cubic lattices H $(0,0,1)$, N $(0,\tfrac12,\tfrac12)$ and P $(\tfrac12,\tfrac12,\tfrac12)$. For hexagonal lattices they are M $(\tfrac12,0,0)$, K $(\tfrac13,\tfrac13,0)$ and A $(0,0,\tfrac12)$, and for the body-centred tetragonal \net{bsn} cell X $(\tfrac12,\tfrac12,0)$, Z $(0,0,1)$ and N $(\tfrac12,0,\tfrac12)$. The full-composite modes approach the first-order values of Table~\ref{M-tab:atlas} where both uncoupled gaps are large compared with the contact scale. Finite-contact corrections remain away from the zone centre. Near $\Gamma$ the band meets the acoustic branches. High-symmetry lines often belong to the measure-zero sets of Table~\ref{tab:split}. There the floppy net or the scaffold gains mechanisms. These appear as marked modes at $\omega^2=0$, as on $\Gamma$--H of \net{bcu}$\otimes$\net{nbo}, where four extra mechanisms per conventional cell stay unlifted while the band at $2\kc$ keeps its multiplicity. They also appear as split values, as on $\Gamma$--X of \net{pcu}$\otimes$\net{urk}. For \net{hex}$\otimes$\net{bnn} the whole path lies on such sets, because the \net{hex} scaffold has mechanisms both in the plane $q_z=0$, which holds $\Gamma$--M--K--$\Gamma$, and on the hexagonal axis $\Gamma$--A. Its marked modes therefore hybridize with scaffold branches along this path, although its first-order band is flat at generic wavevectors.}
\label{fig:s_bands}
\end{figure*}

\begin{figure*}[t]\centering
\includegraphics[width=\textwidth]{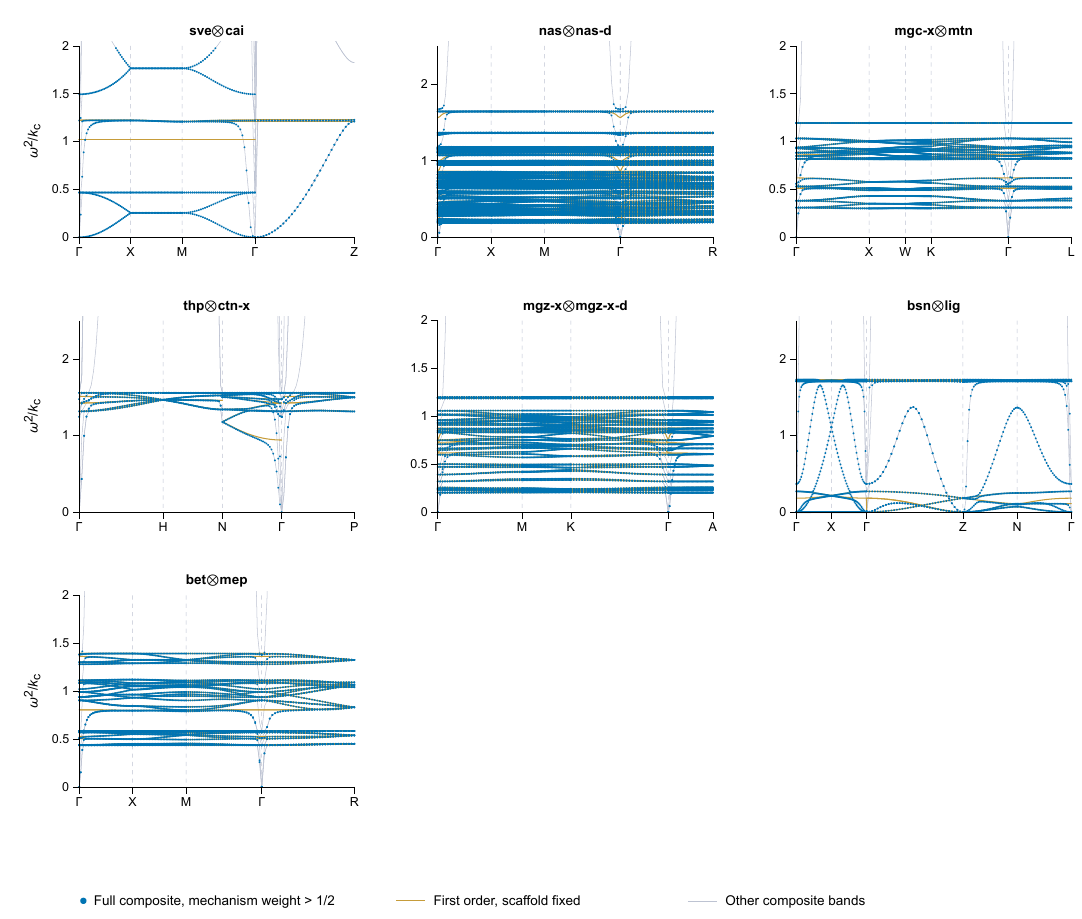}
\caption{\textbf{The seven near-flat pairs lift their mechanisms into narrow but distinct branches.} Band structures along high-symmetry paths, drawn as in Fig.~\ref{fig:s_bands}, for the pairs whose first-order spread is below 5\% without vanishing (Table~\ref{tab:screen}). Special points of the face-centred cubic lattice are X $(0,1,0)$, W $(\tfrac12,1,0)$, K $(\tfrac34,\tfrac34,0)$ and L $(\tfrac12,\tfrac12,\tfrac12)$, and of the primitive tetragonal \net{sve} cell X $(0,\tfrac12,0)$, M $(\tfrac12,\tfrac12,0)$ and Z $(0,0,\tfrac12)$. Each pair carries several lifted branches with different values. The spread of Table~\ref{tab:screen} takes, for each sorted branch, its range over 40 random wavevectors of the conventional cell of the scaffold and divides the mean of these ranges by the mean branch value. It therefore measures the widths of the branches, not their separation. A few wide branches among many narrow ones raise it only a little, and the worst branch, listed in Table~\ref{tab:full_flat}, can be several times wider than the spread suggests. This is the case in \net{bsn}$\otimes$\net{lig}, where some of the 24 branches disperse strongly along the path.}
\label{fig:s_bands_near}
\end{figure*}

\begin{figure*}[t]\centering
\includegraphics[width=\textwidth]{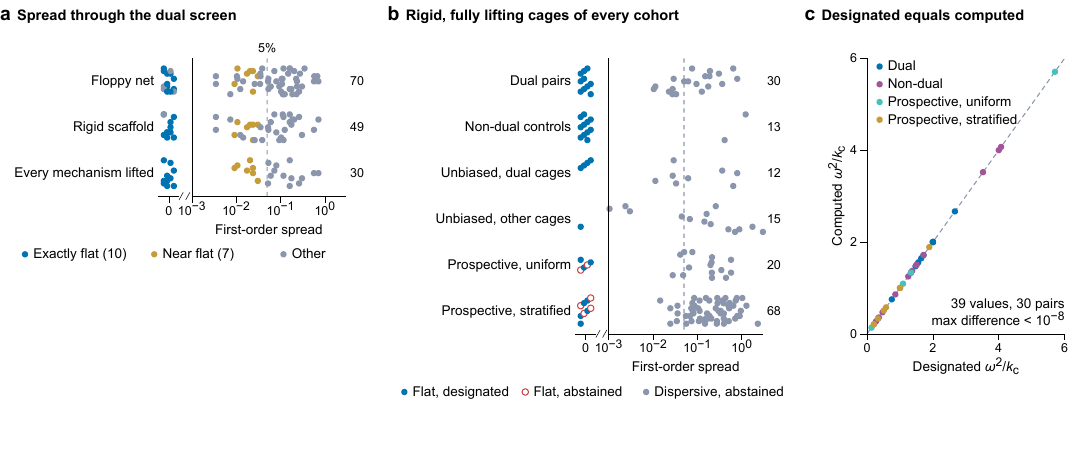}
\caption{\textbf{Exact flatness separates cleanly from dispersion, and every designated flat value equals the computed one.} \textbf{a}, First-order spread, the mean branch width over the mean band value at 40 random wavevectors with the scaffold fixed, for the dual pairs at each stage of the screen of Fig.~\ref{M-fig:atlas}a. The stages are a floppy net with generic mechanisms (70 pairs), a rigid scaffold (49) and every mechanism lifted (30). Exact zeros are plotted in the separate column at 0. Three pairs at each of the first two stages have no lifted band and are omitted, because their spread is undefined. Colours follow the final class: exactly flat (10), near flat (7) and other; the dashed line marks 5\%. \textbf{b}, The same spread, on 32 wavevectors of the common primitive cell, for the cages that are rigid and lift every mechanism in each cohort. The cohorts are the dual pairs (30), the non-dual controls of Section~S3.3 (13), the unbiased control with dual (12) and other cages (15), the uniform prospective test (20) and the stratified prospective test (68). Colours give the call of the frozen classifier. All 40 flat cohort entries, representing 36 distinct ordered pairs, lie at numerical zero, below $2\times10^{-13}$, and the nearest dispersive cage has a spread of $1.0\times10^{-3}$, nearly ten orders of magnitude higher. \textbf{c}, Designated value of the classifier, computed from the floppy net and its contact stars alone, against the computed first-order branch mean, for the 39 distinct values of the 30 designated flat pairs. They agree to better than $10^{-8}$, including the values that symmetry does not fix, such as 1.645 for \net{bsn}$\otimes$\net{lvt}, 1.471 for \net{qtz-x}$\otimes$\net{uom} and 5.708 for \net{npo-e}$\otimes$\net{ybh}.}
\label{fig:s_flatness}
\end{figure*}

\begin{figure*}[t]\centering
\includegraphics[width=\textwidth]{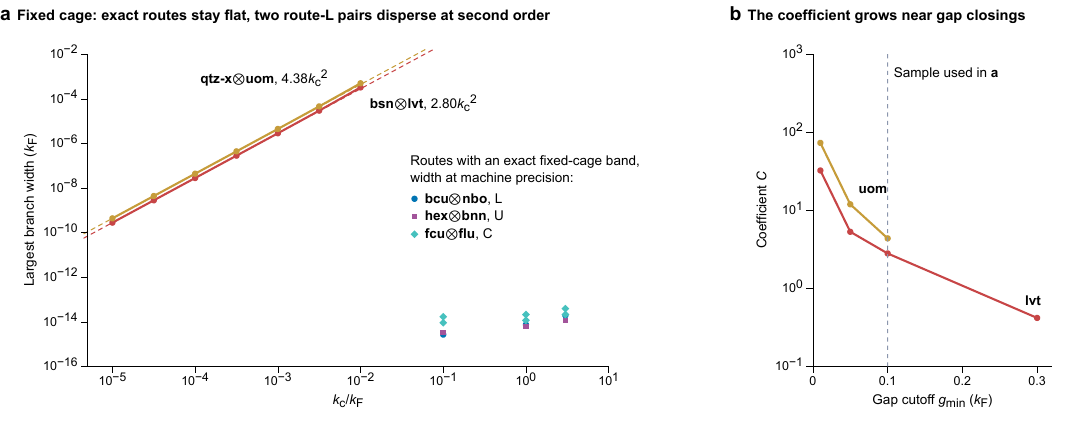}
\caption{\textbf{In a fixed cage, routes with eigenvector normals stay flat at all orders, whereas \net{bsn}$\otimes$\net{lvt} and \net{qtz-x}$\otimes$\net{uom} disperse at order $\kc^2$.} \textbf{a}, Largest sorted-branch width of the lifted band with the scaffold held fixed, in units of the floppy-net stiffness $k_F=1$, against $\kc/k_F$. For \net{bsn}$\otimes$\net{lvt} (147 wavevectors) and \net{qtz-x}$\otimes$\net{uom} (161 wavevectors), the sample is restricted to wavevectors where the gap of the floppy net is at least $0.1k_F$. The widths follow $\lvtCtwo\kc^2/k_F$ and $\uomCtwo\kc^2/k_F$ (dashed lines), with fitted exponents of 2.005 for both; the coefficient rises to 3.2 and 4.9 at $\kc/k_F=10^{-2}$ through higher orders. Route L with normals that are eigenvectors of $S_i$ (\net{bcu}$\otimes$\net{nbo}), route U (\net{hex}$\otimes$\net{bnn}) and route C (\net{fcu}$\otimes$\net{flu}, both roots) give widths below $\exactWidth$ on \exactNq{} wavevectors at $\kc/k_F=0.1$, 1 and 3. These widths are at the level of rounding error. \textbf{b}, The second-order coefficient of the two dispersing pairs against the gap cutoff of the sample. It grows as wavevectors closer to gap closings are admitted and exceeds $10^{8}$ for \net{bsn}$\otimes$\net{lvt} without a cutoff, so it characterizes the sample and is not a zone-wide constant. The mobile-scaffold width of order $\kc^2/\ks$ is a separate effect, shown for \net{nbo} in Fig.~\ref{fig:drive}b.}
\label{fig:s_order}
\end{figure*}

\begin{figure*}[t]\centering
\includegraphics[width=\textwidth]{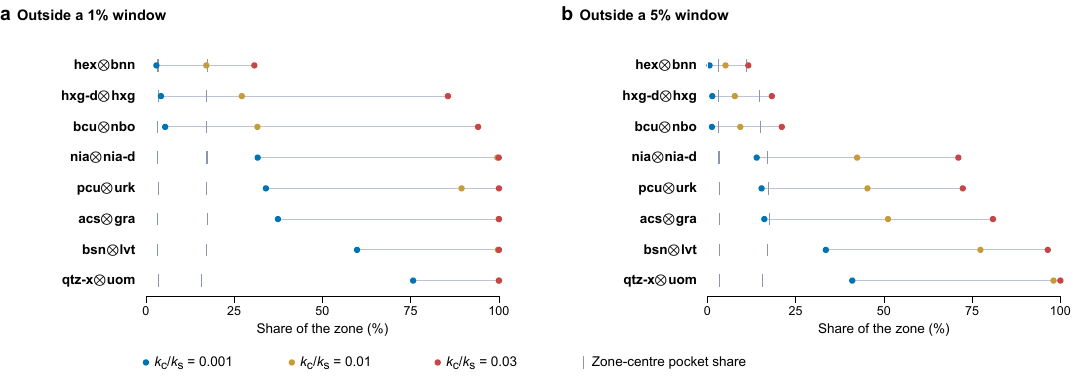}
\caption{\textbf{The zone share in which a lifted band leaves its first-order value grows with $\kc/\ks$ and is set by the soft modes of each pair.} Share of the zone in which any assigned branch of the full composite deviates from its first-order value by more than 1\% (\textbf{a}) or 5\% (\textbf{b}) in $\omega^2$, at $\kc/\ks=10^{-3}$, $10^{-2}$ and $3\times10^{-2}$, for the eight exactly flat pairs whose band splits on a measure-zero set (Section~S3.5). Each point uses 8,192 wavevectors in eight scrambled Sobol blocks. Rows are ordered by the 1\% share at $\kc/\ks=10^{-3}$, which runs from 3.0\% for \net{hex}$\otimes$\net{bnn} to 75.7\% for \net{qtz-x}$\otimes$\net{uom} (Table~\ref{M-tab:atlas}). Grey ticks mark the geometric reference-pocket fractions, approximately 0.1\%, 3\% and 17\% of the zone at the three stiffness ratios. The part that exceeds each deviation threshold depends on the pair and is listed in Table~\ref{tab:full_zerovol}. The pocket is the set $\rho\le2\sqrt{\kc/\ks}$, where $\rho$ is the distance from the wavevector to the nearest reciprocal-lattice point divided by the cube root of the reciprocal-cell volume (Table~\ref{tab:full_zerovol}). Pairs with soft cage or floppy-net modes across the zone, such as \net{bsn}$\otimes$\net{lvt} and \net{qtz-x}$\otimes$\net{uom}, leave the 1\% window almost everywhere already at $\kc/\ks=10^{-2}$, whereas \net{hex}$\otimes$\net{bnn} stays inside it on 69\% of the zone at $3\times10^{-2}$. Table~\ref{tab:zerovol} lists the totals, together with the two pairs that split on no measure-zero set, \net{fcu}$\otimes$\net{flu} and \net{she-d}$\otimes$\net{she}, which were computed later on the same wavevectors and are not drawn here.}
\label{fig:s_zones}
\end{figure*}

\begin{figure*}[t]\centering
\includegraphics[width=\textwidth]{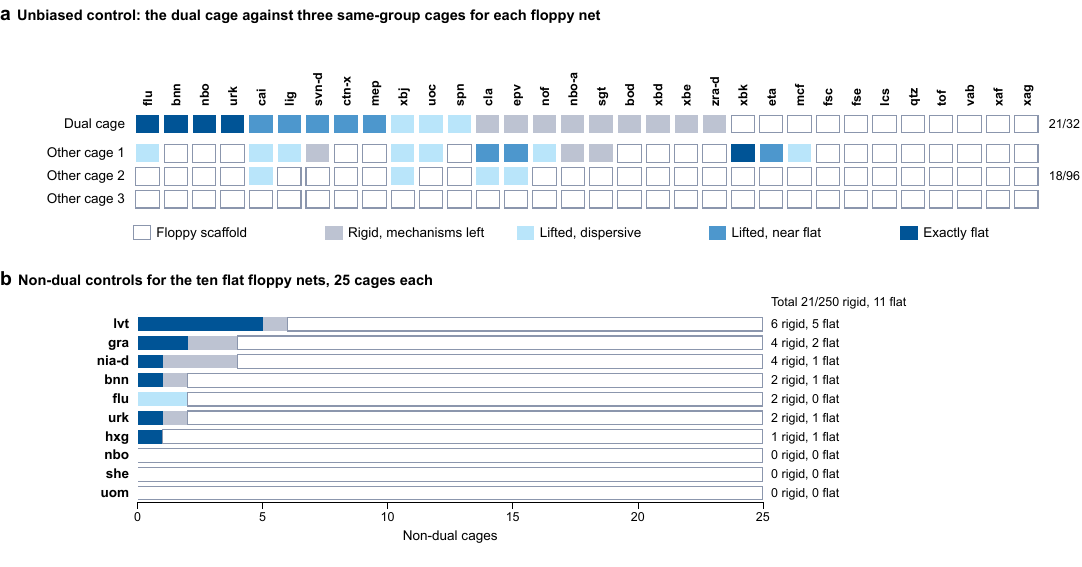}
\caption{\textbf{Dual cages are rigid for most floppy nets, whereas other cages of the same space group are mostly floppy.} \textbf{a}, Unbiased control of Section~S3.4. Each column is one of the 32 floppy nets drawn uniformly from the nets with $\nu>0$ and a same-group dual, caged by its dual (top row) and by three nets of the same space group drawn uniformly (rows below, sorted by outcome). Cell colour gives the outcome at generic wavevectors: floppy scaffold, rigid scaffold that leaves mechanisms unlifted, every mechanism lifted with a dispersive band, near flat (spread below 5\%), or exactly flat. The dual is rigid for 21 of 32 nets and the other cages for 18 of 96, a difference of 47 percentage points with a 95\% cluster bootstrap interval of 31 to 63. Four dual cages and one other cage give an exactly flat band. \textbf{b}, Non-dual controls of Section~S3.3: each floppy net of the ten exactly flat dual pairs caged by 25 nets of its space group, with the roles held fixed. Bars stack the same outcomes; open bars are floppy scaffolds. Only 21 of 250 cages are rigid. The 11 exactly flat ones come from six floppy nets, and five of them hold \net{lvt}. The mean-coordination test of the dual scaffolds is Fig.~\ref{M-fig:atlas}d.}
\label{fig:s_rigidity}
\end{figure*}

\begin{figure*}[t]\centering
\includegraphics[width=\textwidth]{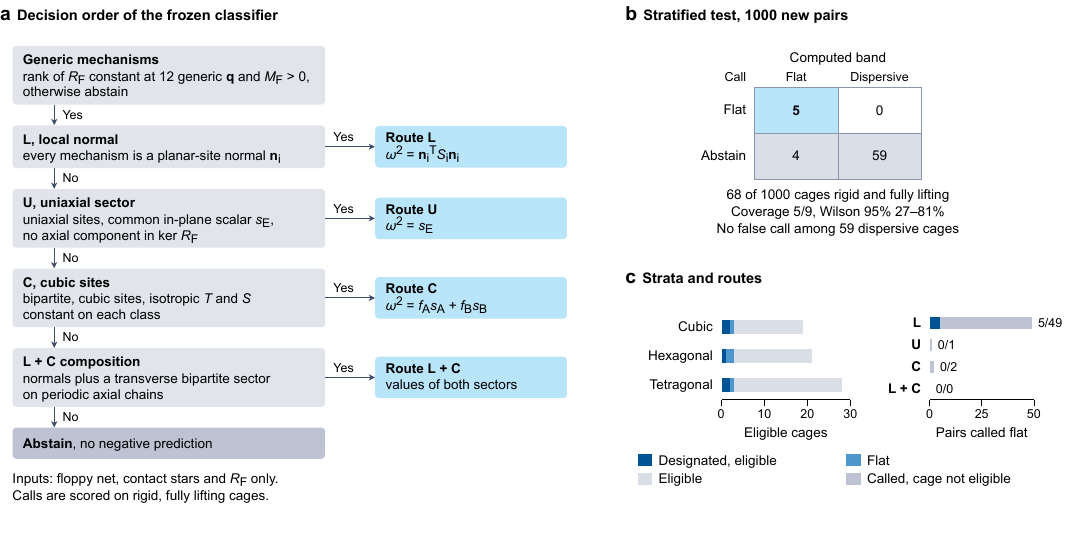}
\caption{\textbf{The frozen classifier designated five of nine flat cages in a stratified draw of 1000 new pairs and made no false call.} \textbf{a}, Decision order of the classifier, which reads only the floppy net, its contact stars and the rank of $R_F$ at 12 generic wavevectors. It tests route L, then route U, then route C, then the composition of local normals with a transverse bipartite sector found in \net{acs}$\otimes$\net{gra}, and abstains otherwise. It makes no negative prediction. Calls are scored only for cages that are rigid at generic wavevectors and lift every mechanism. \textbf{b}, Stratified prospective test: 1000 ordered pairs of distinct three-dimensional non-dual nets sharing a space-group setting, drawn equally from cubic, hexagonal and tetragonal groups, classified and sealed before any spectrum was computed. Of the 168 cages with a rigid scaffold, 68 lift every mechanism. Nine of these are exactly flat; the classifier designated five and abstained on four, and it abstained on all 59 dispersive cages (one-sided 95\% upper bound on the false-call rate among dispersive cages, 5\%). \textbf{c}, Eligible cages per stratum with their flat and designated subsets, and designations by route among all 1000 pairs. Route L made 49 calls, 5 of them on eligible cages, all correct. Routes U and C made one and two calls, all on cages that were not eligible. With the uniform test of Fig.~\ref{M-fig:atlas}b, the prospective tests contain \protEligible{} eligible cages, of which \protFlat{} are flat and \protFlatCalled{} were designated.}
\label{fig:s_classifier}
\end{figure*}

\begin{figure*}[t]\centering
\includegraphics[width=\textwidth]{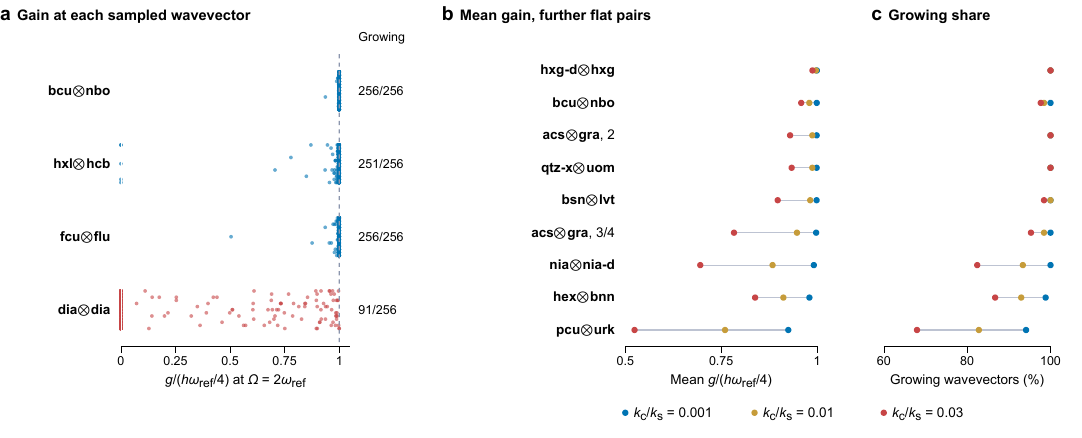}
\caption{\textbf{A drive at twice the flat-band frequency grows almost every sampled wavevector at close to the single-mode rate, and stays effective for every tested pair at small $\kc$.} \textbf{a}, Largest Floquet exponent $g$ at $\Omega=2\omega_{\rm ref}$ for each of the \gainNq{} wavevectors of Fig.~\ref{fig:drive}a, in units of the single Mathieu rate $h\omega_{\rm ref}/4$, with $\kc=10^{-3}$, $h=0.1$ and no damping. Numbers give the growing wavevectors, $g/\omega_{\rm ref}>10^{-5}$. For the three flat pairs the median ratio is 0.9996, 0.9993 and 0.9995, and only a few wavevectors fall well below it. For \net{dia}$\otimes$\net{dia}, referenced to $\omega^2=2\kc$, 165 wavevectors do not grow and the others spread over the whole range. \textbf{b}, Mean of $g/(h\omega_{\rm ref}/4)$ and \textbf{c}, share of growing wavevectors, on 256 scrambled Sobol wavevectors, for each flat value of the eight exactly flat pairs that split on a measure-zero set (two values for \net{acs}$\otimes$\net{gra}) at $\kc/\ks=10^{-3}$, $10^{-2}$ and $3\times10^{-2}$. At $\kc/\ks=10^{-3}$ the mean ratio of the seven pairs other than \net{bcu}$\otimes$\net{nbo} lies between \otherMeanMin{} and \otherMeanMax, lowest for \otherLow. It falls with $\kc/\ks$ in step with the zone shares of Fig.~\ref{fig:s_zones}. Of the ten flat pairs, only \net{she-d}$\otimes$\net{she} was not driven.}
\label{fig:s_gain}
\end{figure*}

\clearpage
\section{Data tables}\label{sec:data}

This section lists, pair by pair, the data behind every count and value quoted in the main text and in the sections above. All entries are generated from the result files of the code release; RCSR coordinates are not reproduced, only net names, symmetry symbols and derived quantities.

\textbf{Atlas.} Table~\ref{tab:full_atlas} gives the structure and the funnel stage of all 76 same-group dual pairs, and the mean scaffold coordination behind the rigidity bound of Section~S3.4. Table~\ref{tab:full_flat} details the seventeen flat and near-flat pairs.

\begingroup\scriptsize
\begin{longtable}{lllrrrrrrrlll}
\caption{Structural data and screening outcome of all 76 same-group dual pairs, listed as scaffold$\otimes$floppy net. Group: common space group. $G_i$: site groups of the floppy vertices in the composite. $n_c$: contacts per conventional cell of the scaffold. $V_F$, $E_F$: floppy vertices and edges in that cell; $\nu=dV_F-E_F$; $N_{\rm ss}$: generic states of self-stress, so that $M_F=\nu+N_{\rm ss}$. $\bar z_S$: mean coordination of the scaffold. The next two columns are the two checks of Section~S3.4. $\chi=V_F-E_F+F-C$, with $C$ scaffold vertices and $F$ scaffold edges in the same cell, is the Euler residual, which vanishes when the scaffold is the dual of a tiling of the floppy net in that cell. $\bar f$: whether the mean face count of the catalogue natural tiling of the floppy net equals $\bar z_S$, as it must when the scaffold is the dual of that particular tiling. Stage: last step of the funnel passed (no mechanisms: floppy net without generic mechanisms, roles not assigned; scaffold floppy: scaffold has generic mechanisms; unlifted: rigid scaffold, some mechanisms unlifted; dispersive, near flat, exact: all mechanisms lifted, with spread at least 0.05, below 0.05 and zero). Route: route of the frozen classifier when it designates the pair flat; $^*$, designation outside the eligible set (cage not rigid or not fully lifting), which is not scored. The mechanism counts and lifted bands of the same pairs are in Table~\ref{tab:screen}.}\label{tab:full_atlas} \\
\hline Pair & Group & $G_i$ & $n_c$ & $V_F$ & $E_F$ & $\nu$ & $N_{\rm ss}$ & $\bar z_S$ & $\chi$ & $\bar f$ & Stage & Route \\ \hline\endfirsthead
\hline Pair & Group & $G_i$ & $n_c$ & $V_F$ & $E_F$ & $\nu$ & $N_{\rm ss}$ & $\bar z_S$ & $\chi$ & $\bar f$ & Stage & Route \\ \hline\endhead
\hline\endfoot
\net{acs}$\otimes$\net{gra} & $P6_{3}/mmc$ & $\bar{6}m2$ & 10 & 4 & 8 & 4 & 0 & 6.00 & $0$ & no & exact & L + C \\
\net{bcu}$\otimes$\net{nbo} & $Im\bar{3}m$ & $4/mmm$ & 12 & 6 & 12 & 6 & 0 & 8.00 & $0$ & yes & exact & L \\
\net{bsn}$\otimes$\net{lvt} & $I4_{1}/amd:2$ & $2/m$ & 16 & 8 & 16 & 8 & 0 & 6.00 & $0$ & yes & exact & L \\
\net{fcu}$\otimes$\net{flu} & $Fm\bar{3}m$ & $\bar{4}3m$, $m\bar{3}m$ & 56 & 12 & 32 & 4 & 0 & 12.00 & $0$ & yes & exact & C \\
\net{hex}$\otimes$\net{bnn} & $P6/mmm$ & $\bar{6}m2$ & 12 & 2 & 5 & 1 & 0 & 8.00 & $0$ & yes & exact & U \\
\net{hxg-d}$\otimes$\net{hxg} & $Pn\bar{3}m:2$ & $\bar{3}m$ & 8 & 4 & 12 & 0 & 4 & 10.00 & $0$ & yes & exact & L \\
\net{nia}$\otimes$\net{nia-d} & $P6_{3}/mmc$ & $\bar{6}m2$ & 10 & 4 & 12 & 0 & 2 & 6.00 & $0$ & yes & exact & L \\
\net{pcu}$\otimes$\net{urk} & $Pm\bar{3}m$ & $3m$, $m$ & 7 & 7 & 18 & 3 & 0 & 6.00 & $-9$ & no & exact & L \\
\net{qtz-x}$\otimes$\net{uom} & $P6_{2}22$ & $2$ & 12 & 6 & 12 & 6 & 0 & 6.00 & $0$ & yes & exact & L \\
\net{she-d}$\otimes$\net{she} & $Im\bar{3}m$ & $\bar{3}m$, $\bar{4}2m$ & 88 & 20 & 48 & 12 & 8 & 9.00 & $0$ & yes & exact & L \\
\net{bet}$\otimes$\net{mep} & $Pm\bar{3}n$ & $\bar{4}2m$, $3$, $m$ & 112 & 46 & 92 & 46 & 0 & 13.50 & $0$ & yes & near flat &  \\
\net{bsn}$\otimes$\net{lig} & $I4_{1}/amd:2$ & $2$ & 32 & 16 & 24 & 24 & 0 & 6.00 & $0$ & yes & near flat &  \\
\net{mgc-x}$\otimes$\net{mtn} & $Fd\bar{3}m:2$ & $\bar{4}3m$, $3m$, $m$ & 320 & 136 & 272 & 136 & 0 & 13.33 & $0$ & yes & near flat &  \\
\net{mgz-x}$\otimes$\net{mgz-x-d} & $P6_{3}/mmc$ & $1$, $3m$, $m$, $mm2$ & 148 & 68 & 136 & 68 & 0 & 13.33 & $0$ & yes & near flat &  \\
\net{nas}$\otimes$\net{nas-d} & $Pa\bar{3}$ & $1$, $3$ & 512 & 232 & 464 & 232 & 0 & 13.60 & $0$ & yes & near flat &  \\
\net{sve}$\otimes$\net{cai} & $P4/mbm$ & $4/m$, $mm2$ & 24 & 6 & 16 & 2 & 0 & 7.00 & $0$ & yes & near flat &  \\
\net{thp}$\otimes$\net{ctn-x} & $I\bar{4}3d$ & $\bar{4}$, $3$ & 96 & 28 & 64 & 20 & 0 & 8.00 & $0$ & yes & near flat &  \\
\net{bcu-x}$\otimes$\net{sod} & $Im\bar{3}m$ & $\bar{4}2m$ & 48 & 12 & 24 & 12 & 0 & 14.00 & $0$ & yes & dispersive &  \\
\net{cal}$\otimes$\net{cal-d} & $I4/mcm$ & $\bar{4}2m$, $1$, $m$ & 112 & 68 & 136 & 68 & 0 & 13.33 & $0$ & yes & dispersive &  \\
\net{ccu}$\otimes$\net{doh} & $P6/mmm$ & $3m$, $m$, $mm2$ & 64 & 34 & 68 & 34 & 0 & 13.33 & $0$ & yes & dispersive &  \\
\net{ddi}$\otimes$\net{jph} & $I4_{1}/amd:2$ & $\bar{4}2m$, $2$ & 40 & 20 & 32 & 28 & 0 & 8.00 & $0$ & yes & dispersive &  \\
\net{deh-d}$\otimes$\net{gis} & $I4_{1}/amd:2$ & $2$ & 32 & 16 & 32 & 16 & 0 & 10.00 & $0$ & yes & dispersive &  \\
\net{hcp}$\otimes$\net{hcp-d} & $P6_{3}/mmc$ & $\bar{3}m$, $3m$ & 16 & 6 & 16 & 2 & 0 & 12.00 & $0$ & yes & dispersive &  \\
\net{jsm}$\otimes$\net{uoc} & $I4_{1}22$ & $2$ & 16 & 8 & 16 & 8 & 0 & 6.00 & $0$ & no & dispersive &  \\
\net{svn}$\otimes$\net{svn-d} & $Pa\bar{3}$ & $\bar{3}$, $1$ & 48 & 28 & 48 & 36 & 0 & 7.00 & $0$ & yes & dispersive &  \\
\net{sxb}$\otimes$\net{nod} & $Cccm$ & $2$ & 32 & 16 & 24 & 24 & 0 & 6.00 & $0$ & yes & dispersive &  \\
\net{tsi}$\otimes$\net{xbj} & $I4_{1}/amd:2$ & $mm2$ & 32 & 8 & 20 & 4 & 0 & 8.00 & $0$ & yes & dispersive &  \\
\net{xbp}$\otimes$\net{xbz} & $F\bar{4}3m$ & $1$ & 48 & 36 & 64 & 44 & 0 & 6.67 & $0$ & no & dispersive &  \\
\net{xbr}$\otimes$\net{spn} & $Fd\bar{3}m:2$ & $1$ & 48 & 48 & 96 & 48 & 0 & 8.00 & $0$ & yes & dispersive &  \\
\net{xbs}$\otimes$\net{bor} & $P\bar{4}3m$ & $2$, $m$, $mm2$ & 15 & 7 & 12 & 9 & 0 & 7.00 & $0$ & yes & dispersive &  \\
\net{alb-x}$\otimes$\net{alb-x-d} & $P6/mmm$ & $3m$, $mm2$ & 28 & 16 & 32 & 16 & 0 & 12.67 & $0$ & yes & unlifted &  \\
\net{ast-d}$\otimes$\net{ast} & $Fm\bar{3}m$ & $\bar{4}3m$, $3m$ & 96 & 40 & 80 & 40 & 0 & 12.00 & $0$ & yes & unlifted &  \\
\net{bct}$\otimes$\net{xbe} & $I4/mmm$ & $m$, $mm2$ & 16 & 12 & 20 & 16 & 0 & 10.00 & $0$ & yes & unlifted &  \\
\net{bcu-x}$\otimes$\net{nbo-a} & $Im\bar{3}m$ & $mm2$ & 48 & 24 & 36 & 36 & 0 & 14.00 & $0$ & yes & unlifted &  \\
\net{bsn}$\otimes$\net{dia-f} & $I4_{1}/amd:2$ & $1$ & 16 & 16 & 24 & 24 & 0 & 6.00 & $0$ & yes & unlifted &  \\
\net{cla-d}$\otimes$\net{cla} & $I\bar{4}3m$ & $\bar{4}3m$, $1$, $3m$, $m$ & 88 & 82 & 164 & 82 & 0 & 13.71 & $0$ & yes & unlifted &  \\
\net{deh-d}$\otimes$\net{deh} & $I4_{1}/amd:2$ & $1$ & 32 & 32 & 48 & 48 & 0 & 10.00 & $0$ & yes & unlifted &  \\
\net{eca}$\otimes$\net{xbd} & $P6_{3}/mmc$ & $\bar{6}m2$, $mm2$ & 18 & 8 & 14 & 10 & 0 & 8.00 & $0$ & yes & unlifted &  \\
\net{ild}$\otimes$\net{epv} & $R\bar{3}m:H$ & $1$, $2$ & 27 & 24 & 54 & 18 & 0 & 12.00 & $-15$ & yes & unlifted &  \\
\net{odf-d}$\otimes$\net{odf} & $Im\bar{3}$ & $1$, $3$, $m$, $mm2$ & 272 & 260 & 520 & 260 & 0 & 13.30 & $0$ & yes & unlifted &  \\
\net{reo-d}$\otimes$\net{cab} & $Pm\bar{3}m$ & $4mm$ & 6 & 6 & 15 & 3 & 0 & 11.00 & $0$ & yes & unlifted & U$^*$ \\
\net{reo-d}$\otimes$\net{nba} & $Pm\bar{3}m$ & $4/mmm$, $mm2$ & 30 & 15 & 24 & 21 & 0 & 11.00 & $0$ & yes & unlifted &  \\
\net{sta}$\otimes$\net{ats} & $P6_{3}/mmc$ & $\bar{3}m$, $\bar{6}m2$ & 26 & 6 & 18 & 0 & 2 & 6.00 & $0$ & no & unlifted & L$^*$ \\
\net{sxb}$\otimes$\net{nof} & $Cccm$ & $1$ & 16 & 16 & 24 & 24 & 0 & 6.00 & $0$ & yes & unlifted &  \\
\net{tsl}$\otimes$\net{sgt} & $I4_{1}/amd:2$ & $2$, $m$ & 96 & 64 & 128 & 64 & 0 & 12.67 & $0$ & yes & unlifted &  \\
\net{wzz}$\otimes$\net{xaa} & $Fm\bar{3}m$ & $3m$, $mm2$ & 176 & 80 & 120 & 120 & 0 & 7.00 & $0$ & yes & unlifted &  \\
\net{xbp}$\otimes$\net{ttt} & $F\bar{4}3m$ & $m$, $mm2$ & 20 & 20 & 48 & 12 & 0 & 6.67 & $0$ & yes & unlifted &  \\
\net{xbs}$\otimes$\net{bod} & $P\bar{4}3m$ & $2$, $m$, $mm2$ & 14 & 10 & 15 & 15 & 0 & 7.00 & $0$ & yes & unlifted &  \\
\net{zra}$\otimes$\net{zra-d} & $P6/mmm$ & $3m$, $m$, $mm2$ & 96 & 40 & 80 & 40 & 0 & 13.43 & $0$ & yes & unlifted &  \\
\net{bcu-x-a}$\otimes$\net{xag} & $Im\bar{3}m$ & $\bar{4}2m$, $m\bar{3}m$, $mm2$ & 112 & 38 & 72 & 42 & 0 & 4.43 & $0$ & no & scaffold floppy &  \\
\net{dmp}$\otimes$\net{utp} & $Pnna$ & $1$ & 8 & 8 & 12 & 12 & 0 & 4.00 & $0$ & yes & scaffold floppy &  \\
\net{fsc}$\otimes$\net{mot} & $P4/mmm$ & $mm2$ & 6 & 3 & 6 & 3 & 0 & 5.00 & $0$ & yes & scaffold floppy & L$^*$ \\
\net{fse}$\otimes$\net{xux} & $I\bar{4}m2$ & $\bar{4}2m$, $m$ & 20 & 10 & 16 & 14 & 0 & 5.00 & $0$ & no & scaffold floppy &  \\
\net{fsm}$\otimes$\net{cnw} & $Pnma$ & $m$ & 24 & 16 & 32 & 16 & 0 & 4.67 & $0$ & yes & scaffold floppy &  \\
\net{gan}$\otimes$\net{lcs} & $Ia\bar{3}d$ & $\bar{4}$ & 96 & 24 & 48 & 24 & 0 & 5.00 & $0$ & yes & scaffold floppy &  \\
\net{gan}$\otimes$\net{vab} & $Ia\bar{3}d$ & $2$ & 96 & 48 & 72 & 72 & 0 & 5.00 & $0$ & yes & scaffold floppy &  \\
\net{gra}$\otimes$\net{etc} & $P6_{3}/mmc$ & $3m$, $m$ & 8 & 8 & 12 & 12 & 0 & 4.00 & $0$ & yes & scaffold floppy &  \\
\net{lon}$\otimes$\net{gra} & $P6_{3}/mmc$ & $\bar{6}m2$ & 16 & 4 & 8 & 4 & 0 & 4.00 & $0$ & yes & scaffold floppy &  \\
\net{mcf}$\otimes$\net{mcd} & $Cmmm$ & $mm2$ & 16 & 8 & 16 & 8 & 0 & 4.00 & $0$ & yes & scaffold floppy &  \\
\net{mcf-d}$\otimes$\net{mcf} & $Cmmm$ & $1$ & 8 & 8 & 16 & 8 & 0 & 4.67 & $0$ & yes & scaffold floppy &  \\
\net{qzd}$\otimes$\net{bto} & $P6_{2}22$ & $2$ & 18 & 6 & 9 & 9 & 0 & 4.00 & $0$ & yes & scaffold floppy &  \\
\net{qzd}$\otimes$\net{eta} & $P6_{2}22$ & $2$ & 24 & 6 & 9 & 9 & 0 & 4.00 & $0$ & yes & scaffold floppy &  \\
\net{qzd}$\otimes$\net{qtz} & $P6_{2}22$ & $222$ & 12 & 3 & 6 & 3 & 0 & 4.00 & $0$ & yes & scaffold floppy &  \\
\net{qzd}$\otimes$\net{qtz-h} & $P6_{2}22$ & $1$ & 12 & 12 & 18 & 18 & 0 & 4.00 & $-3$ & yes & scaffold floppy &  \\
\net{qzd}$\otimes$\net{uog} & $P6_{2}22$ & $2$ & 12 & 6 & 12 & 6 & 0 & 4.00 & $-3$ & yes & scaffold floppy &  \\
\net{uni}$\otimes$\net{uni-d} & $P6_{1}22$ & $2$ & 24 & 12 & 18 & 18 & 0 & 4.00 & $0$ & yes & scaffold floppy &  \\
\net{unj}$\otimes$\net{etd} & $P6_{1}22$ & $1$ & 12 & 12 & 18 & 18 & 0 & 4.00 & $0$ & yes & scaffold floppy &  \\
\net{xaf}$\otimes$\net{xag} & $Im\bar{3}m$ & $\bar{4}2m$, $m\bar{3}m$, $mm2$ & 108 & 38 & 72 & 42 & 0 & 5.40 & $0$ & no & scaffold floppy &  \\
\net{xbk}$\otimes$\net{ntv} & $Fm\bar{3}m$ & $3m$, $4mm$, $m\bar{3}m$ & 176 & 60 & 120 & 60 & 0 & 5.33 & $0$ & yes & scaffold floppy &  \\
\net{xbt}$\otimes$\net{tof} & $Pn\bar{3}m:2$ & $\bar{4}2m$, $3m$ & 32 & 14 & 28 & 14 & 0 & 6.67 & $0$ & yes & scaffold floppy &  \\
\net{crs}$\otimes$\net{crs-d} & $Fd\bar{3}m:2$ & -- & -- & -- & -- & -- & -- & -- & -- & -- & no mechanisms &  \\
\net{hcp}$\otimes$\net{seh} & $P6_{3}/mmc$ & -- & -- & -- & -- & -- & -- & -- & -- & -- & no mechanisms &  \\
\net{kag}$\otimes$\net{kag-d} & $P6/mmm$ & -- & -- & -- & -- & -- & -- & -- & -- & -- & no mechanisms &  \\
\net{keb}$\otimes$\net{xak} & $I4/mmm$ & -- & -- & -- & -- & -- & -- & -- & -- & -- & no mechanisms &  \\
\net{nbo-x}$\otimes$\net{nbo-x-d} & $Im\bar{3}m$ & -- & -- & -- & -- & -- & -- & -- & -- & -- & no mechanisms &  \\
\net{reo}$\otimes$\net{reo-d} & $Pm\bar{3}m$ & -- & -- & -- & -- & -- & -- & -- & -- & -- & no mechanisms &  \\
\end{longtable}
\endgroup

\begingroup\scriptsize
\begin{longtable}{lllllllr}
\caption{First-order lifted bands of the ten exactly flat and seven near-flat dual pairs. Class: exact or near, with the relative spread $s$ of the second implementation. The spread is defined as in Table~\ref{tab:screen}: the first-order eigenvalues are sorted at each wavevector into branches, the range of each branch over the wavevectors is taken, and the mean of these ranges is divided by the mean of all branch values. Contact stars: for each class of floppy vertex, the number of contacts $n_c$ and the eigenvalues of $S_i$ in units of $\kc$, one entry per distinct contact count. First-order $\omega^2/\kc$: branch means over the generic wavevectors of the conventional cell of the scaffold, with multiplicities; for the near-flat pairs, the range over all branches. Width: largest range of a single branch over the sampled wavevectors, in units of $\kc$. Worst: the largest absolute branch range divided by the mean over all branch values. This uses the same denominator as $s$ and is not the width divided by the mean of its own branch. The condition $s<0.05$ classifies an average spread, not every branch. $\Delta$: largest difference between a numerical branch mean and the closed-form value of Table~\ref{M-tab:atlas} or the value designated by the frozen classifier. Spreads and widths come from the second implementation, at an independent set of 40 random wavevectors of the conventional cell of the scaffold; for the near-flat pairs the spreads of Table~\ref{tab:screen} differ from them by at most 0.002. Fractions are printed only where the number equals the fraction to $10^{-9}$; the values $1.645$ and $1.471$ are embedding values.}\label{tab:full_flat} \\
\hline Pair & Class & $G_i$ & \parbox[t]{3.0cm}{\strut\raggedright Contact stars: $n_c$: eig $S_i/\kc$\strut} & \parbox[t]{2.6cm}{\strut\raggedright First-order $\omega^2/\kc$\strut} & Width & Worst & $\Delta$ \\ \hline\endfirsthead
\hline Pair & Class & $G_i$ & \parbox[t]{3.0cm}{\strut\raggedright Contact stars: $n_c$: eig $S_i/\kc$\strut} & \parbox[t]{2.6cm}{\strut\raggedright First-order $\omega^2/\kc$\strut} & Width & Worst & $\Delta$ \\ \hline\endhead
\hline\endfoot
\net{acs}$\otimes$\net{gra} & \parbox[t]{1.25cm}{\strut\raggedright exact, $1.4\times10^{-15}$\strut} & $\bar{6}m2$ & \parbox[t]{3.0cm}{\strut\raggedright 2: ($0$, $0$, $2$); 3: ($0$, $\tfrac{3}{2}$, $\tfrac{3}{2}$)\strut} & \parbox[t]{2.6cm}{\strut\raggedright $\tfrac{3}{4}\,(\times2)$, $2\,(\times2)$\strut} & $2.7\times10^{-15}$ & $1.9\times10^{-15}$ & $4.4\times10^{-16}$ \\
\net{bcu}$\otimes$\net{nbo} & \parbox[t]{1.25cm}{\strut\raggedright exact, $7.8\times10^{-16}$\strut} & $4/mmm$ & \parbox[t]{3.0cm}{\strut\raggedright 2: ($0$, $0$, $2$)\strut} & \parbox[t]{2.6cm}{\strut\raggedright $2\,(\times6)$\strut} & $2.2\times10^{-15}$ & $1.1\times10^{-15}$ & $4.4\times10^{-16}$ \\
\net{hex}$\otimes$\net{bnn} & \parbox[t]{1.25cm}{\strut\raggedright exact, $1.4\times10^{-15}$\strut} & $\bar{6}m2$ & \parbox[t]{3.0cm}{\strut\raggedright 6: ($\tfrac{12}{7}$, $\tfrac{12}{7}$, $\tfrac{18}{7}$)\strut} & \parbox[t]{2.6cm}{\strut\raggedright $\tfrac{12}{7}$\strut} & $2.4\times10^{-15}$ & $1.4\times10^{-15}$ & $1.4\times10^{-11}$ \\
\net{bsn}$\otimes$\net{lvt} & \parbox[t]{1.25cm}{\strut\raggedright exact, $4.4\times10^{-15}$\strut} & $2/m$ & \parbox[t]{3.0cm}{\strut\raggedright 2: ($0$, $0$, $2$)\strut} & \parbox[t]{2.6cm}{\strut\raggedright $1.645\,(\times8)$\strut} & $2.3\times10^{-14}$ & $1.4\times10^{-14}$ & $3.5\times10^{-11}$ \\
\net{fcu}$\otimes$\net{flu} & \parbox[t]{1.25cm}{\strut\raggedright exact, $1.2\times10^{-15}$\strut} & $\bar{4}3m$, $m\bar{3}m$ & \parbox[t]{3.0cm}{\strut\raggedright 4: ($\tfrac{4}{3}$, $\tfrac{4}{3}$, $\tfrac{4}{3}$); 6: ($2$, $2$, $2$)\strut} & \parbox[t]{2.6cm}{\strut\raggedright $\tfrac{14}{9}\,(\times4)$\strut} & $2.2\times10^{-15}$ & $1.4\times10^{-15}$ & $4.4\times10^{-11}$ \\
\net{hxg-d}$\otimes$\net{hxg} & \parbox[t]{1.25cm}{\strut\raggedright exact, $1.6\times10^{-15}$\strut} & $\bar{3}m$ & \parbox[t]{3.0cm}{\strut\raggedright 2: ($0$, $0$, $2$)\strut} & \parbox[t]{2.6cm}{\strut\raggedright $2\,(\times4)$\strut} & $3.8\times10^{-15}$ & $1.9\times10^{-15}$ & $8.9\times10^{-16}$ \\
\net{nia}$\otimes$\net{nia-d} & \parbox[t]{1.25cm}{\strut\raggedright exact, $1.1\times10^{-15}$\strut} & $\bar{6}m2$ & \parbox[t]{3.0cm}{\strut\raggedright 2: ($0$, $0$, $2$); 3: ($0$, $\tfrac{3}{2}$, $\tfrac{3}{2}$)\strut} & \parbox[t]{2.6cm}{\strut\raggedright $2\,(\times2)$\strut} & $2.4\times10^{-15}$ & $1.2\times10^{-15}$ & 0 \\
\net{pcu}$\otimes$\net{urk} & \parbox[t]{1.25cm}{\strut\raggedright exact, $1.3\times10^{-12}$\strut} & $3m$, $m$ & \parbox[t]{3.0cm}{\strut\raggedright 1: ($0$, $0$, $1$)\strut} & \parbox[t]{2.6cm}{\strut\raggedright $\tfrac{1}{3}\,(\times3)$\strut} & $9.3\times10^{-13}$ & $2.8\times10^{-12}$ & $3.3\times10^{-11}$ \\
\net{qtz-x}$\otimes$\net{uom} & \parbox[t]{1.25cm}{\strut\raggedright exact, $1.2\times10^{-12}$\strut} & $2$ & \parbox[t]{3.0cm}{\strut\raggedright 2: ($0$, $0$, $2$)\strut} & \parbox[t]{2.6cm}{\strut\raggedright $1.471\,(\times6)$\strut} & $5.3\times10^{-12}$ & $3.6\times10^{-12}$ & $1.0\times10^{-12}$ \\
\net{she-d}$\otimes$\net{she} & \parbox[t]{1.25cm}{\strut\raggedright exact, $1.3\times10^{-15}$\strut} & $\bar{3}m$, $\bar{4}2m$ & \parbox[t]{3.0cm}{\strut\raggedright 2: ($0$, $0$, $2$); 8: ($\tfrac{8}{3}$, $\tfrac{8}{3}$, $\tfrac{8}{3}$)\strut} & \parbox[t]{2.6cm}{\strut\raggedright $2\,(\times12)$, $\tfrac{8}{3}\,(\times8)$\strut} & $4.0\times10^{-15}$ & $1.8\times10^{-15}$ & $3.3\times10^{-11}$ \\
\net{bet}$\otimes$\net{mep} & near, 0.033 & $\bar{4}2m$, $3$, $m$ & \parbox[t]{3.0cm}{\strut\raggedright 1: ($0$, $0$, $1$); 3: ($0.361$, $1.115$, $1.525$); 4: ($1$, $1$, $2$)\strut} & \parbox[t]{2.6cm}{\strut\raggedright $0.435$--$1.393$ (46 branches)\strut} & 0.096 & 0.117 & -- \\
\net{bsn}$\otimes$\net{lig} & near, 0.026 & $2$ & \parbox[t]{3.0cm}{\strut\raggedright 2: ($0$, $0.270$, $1.730$)\strut} & \parbox[t]{2.6cm}{\strut\raggedright $0.000$--$1.730$ (24 branches)\strut} & 0.138 & 0.117 & -- \\
\net{sve}$\otimes$\net{cai} & near, 0.010 & $4/m$, $mm2$ & \parbox[t]{3.0cm}{\strut\raggedright 2: ($0$, $0.983$, $1.017$); 8: ($2.667$, $2.667$, $2.667$)\strut} & \parbox[t]{2.6cm}{\strut\raggedright $1.210$--$1.225$ (2 branches)\strut} & 0.013 & 0.010 & -- \\
\net{thp}$\otimes$\net{ctn-x} & near, 0.019 & $\bar{4}$, $3$ & \parbox[t]{3.0cm}{\strut\raggedright 3: ($0.099$, $1.451$, $1.451$); 4: ($\tfrac{2}{3}$, $\tfrac{5}{3}$, $\tfrac{5}{3}$)\strut} & \parbox[t]{2.6cm}{\strut\raggedright $1.319$--$1.557$ (20 branches)\strut} & 0.106 & 0.071 & -- \\
\net{mgc-x}$\otimes$\net{mtn} & near, 0.018 & $\bar{4}3m$, $3m$, $m$ & \parbox[t]{3.0cm}{\strut\raggedright 2: ($0$, $0.806$, $1.194$); 3: ($0.479$, $1.260$, $1.260$); 4: ($\tfrac{4}{3}$, $\tfrac{4}{3}$, $\tfrac{4}{3}$)\strut} & \parbox[t]{2.6cm}{\strut\raggedright $0.302$--$1.194$ (136 branches)\strut} & 0.041 & 0.053 & -- \\
\net{mgz-x}$\otimes$\net{mgz-x-d} & near, 0.025 & $1$, $3m$, $m$, $mm2$ & \parbox[t]{3.0cm}{\strut\raggedright 1: ($0$, $0$, $1$); 2: ($0$, $0.732$, $1.268$); 3: ($0$, $0.767$, $1.233$)\strut} & \parbox[t]{2.6cm}{\strut\raggedright $0.203$--$1.199$ (68 branches)\strut} & 0.052 & 0.073 & -- \\
\net{nas}$\otimes$\net{nas-d} & near, 0.010 & $1$, $3$ & \parbox[t]{3.0cm}{\strut\raggedright 1: ($0$, $0$, $1$); 2: ($0$, $0.517$, $1.483$); 3: ($0$, $0.647$, $1.353$); 4: ($0.270$, $1.287$, $1.443$)\strut} & \parbox[t]{2.6cm}{\strut\raggedright $0.188$--$1.643$ (232 branches)\strut} & 0.035 & 0.046 & -- \\
\end{longtable}
\endgroup

\textbf{Exactness and counting.} Table~\ref{tab:full_exact} lists the fixed-cage widths behind Lemma~\ref{lem:exact} and the second-order coefficients of two sampled route-L pairs with non-eigenvector normals. Table~\ref{tab:counting} compares every flat pair with the counting rule.

\begin{table}[h]\centering\footnotesize
\caption{Lifted bands with the scaffold held fixed, at every order in $\kc$. Top: sampled squared-frequency width $\Delta\omega^2$, in units of $k_F$, of the exact bands of the three pairs covered by Lemma~\ref{lem:exact}. Columns give $\kc/k_F=0.1, 1, 3$ and the number $n_q$ of generic wavevectors. Bottom: two sampled route-L pairs with non-eigenvector normals. Mixing is $\|(\one-\mathbf n_i\mathbf n_i^{\mathsf T})S_i\mathbf n_i\|/\kc$, the off-normal part of $S_i\mathbf n_i$. Entries give the largest second-order branch-width coefficient, in units of $\kc^2/k_F$, for each stated gap cutoff. Gaps are in units of the floppy bond stiffness $k_F$. Parentheses give the number of retained wavevectors. The third non-eigenvector pair, \net{pcu}$\otimes$\net{urk}, was computed separately (Section~S2.6). The other four flat dual pairs are covered by Lemma~\ref{lem:exact}(i) or Lemma~\ref{lem:balance_exact}, and the post hoc calculation of Section~S3.7 samples \net{acs}$\otimes$\net{gra}.}\label{tab:full_exact}
\begin{tabular}{llrccc}
\hline
Pair & Route & $n_q$ & $\kc/k_F=0.1$ & $\kc/k_F=1$ & $\kc/k_F=3$ \\
\hline
\net{bcu}$\otimes$\net{nbo} & L & 1032 & $2.7\times10^{-15}$ & $7.5\times10^{-15}$ & $2.0\times10^{-14}$ \\
\net{hex}$\otimes$\net{bnn} & U & 1032 & $3.3\times10^{-15}$ & $6.2\times10^{-15}$ & $1.2\times10^{-14}$ \\
\net{fcu}$\otimes$\net{flu} lower root & C & 1032 & $9.2\times10^{-15}$ & $1.2\times10^{-14}$ & $2.1\times10^{-14}$ \\
\net{fcu}$\otimes$\net{flu} upper root & C & 1032 & $1.7\times10^{-14}$ & $2.1\times10^{-14}$ & $3.9\times10^{-14}$ \\
\hline
\end{tabular}\\[4pt]
\begin{tabular}{llcccccc}
\hline
Pair & Route & mixing & gap $\ge0$ & gap $\ge0.01$ & gap $\ge0.05$ & gap $\ge0.1$ & gap $\ge0.3$ \\
\hline
\net{bsn}$\otimes$\net{lvt} & L & 0.764 & $1.2\times10^{8}$ (1024) & 32.4 (586) & 5.31 (268) & 2.8 (147) & 0.419 (18) \\
\net{qtz-x}$\otimes$\net{uom} & L & 0.882 & $2.71\times10^{7}$ (1024) & 73.1 (651) & 11.9 (319) & 4.38 (161) & -- \\
\hline
\end{tabular}
\end{table}

\begingroup\scriptsize
\begin{longtable}{lllrrlllrrll}
\caption{Counting rule against the flat bands found. The rule of Ref.~\cite{calugaru_general_2022} (App.~I.4 therein) guarantees $b_a=\max(0,n_a-E)$ flat bands at each eigenvalue $a$ of the on-site tensor $S$ of multiplicity $n_a$, with $E$ the number of floppy edges; $dV$ is the number of displacement components. The rank-resolved count of Section~S2.7 replaces $E$ by the rank of $R_F$ on the eigenspace: $d_a=\dim(\mathcal E_a\cap\mathcal K)=n_a-\operatorname{rank}(R_FB_a)$, listed in the order of the eigenvalues and evaluated at 12 generic wavevectors, at which it does not vary. In every row $\sum d_a$ equals the number of flat bands whose value is an eigenvalue of $S$, so the bands left over are exactly those marked $^*$. Counts are per common primitive cell; $m$ is the number of primitive cells in the conventional cell of the scaffold. Flat: flat first-order values with multiplicities, at 64 generic wavevectors; $^*$, the value is not an eigenvalue of $S$. Close but distinct eigenvalues, such as $1.3333$ and $1.3334$ in \net{mob}$\otimes$\net{scp}, are printed with as many digits as separate them. Set: dual, the ten flat dual pairs; test 1, the five flat pairs of the first prospective test; planar, \net{hxl}$\otimes$\net{hcb}. Route: route of the frozen classifier (abstain: no designation). Class is defined by $b_a$. In (i), the bound accounts for every flat band. In (ii), it is positive but accounts for only some; no pair has this class. In (iii), the bound is zero and every flat value is an eigenvalue of $S$. In (iv), the bound is zero and at least one flat value lies outside that spectrum.}\label{tab:counting} \\
\hline Pair & Set & Route & $dV$ & $E$ & eig $S/\kc$: $n_a\to b_a$ & $d_a$ & Flat $\omega^2/\kc$ & $\sum b_a$ & $\sum d_a$ & Class & $m$ \\ \hline\endfirsthead
\hline Pair & Set & Route & $dV$ & $E$ & eig $S/\kc$: $n_a\to b_a$ & $d_a$ & Flat $\omega^2/\kc$ & $\sum b_a$ & $\sum d_a$ & Class & $m$ \\ \hline\endhead
\hline\endfoot
\net{acs}$\otimes$\net{gra} & dual & L + C & 12 & 8 & \parbox[t]{4.6cm}{\strut\raggedright $0$: 6$\to$0; $\tfrac{3}{2}$: 4$\to$0; $2$: 2$\to$0\strut} & \parbox[t]{1.1cm}{\strut\raggedright 0; 0; 2\strut} & \parbox[t]{2.6cm}{\strut\raggedright $\tfrac{3}{4}\,(\times2)$$^*$; $2\,(\times2)$\strut} & 0 & 2 & (iv) & 1 \\
\net{bcu}$\otimes$\net{nbo} & dual & L & 9 & 6 & \parbox[t]{4.6cm}{\strut\raggedright $0$: 6$\to$0; $2$: 3$\to$0\strut} & \parbox[t]{1.1cm}{\strut\raggedright 0; 3\strut} & \parbox[t]{2.6cm}{\strut\raggedright $2\,(\times3)$\strut} & 0 & 3 & (iii) & 2 \\
\net{hex}$\otimes$\net{bnn} & dual & U & 6 & 5 & \parbox[t]{4.6cm}{\strut\raggedright $\tfrac{12}{7}$: 4$\to$0; $\tfrac{18}{7}$: 2$\to$0\strut} & \parbox[t]{1.1cm}{\strut\raggedright 1; 0\strut} & \parbox[t]{2.6cm}{\strut\raggedright $\tfrac{12}{7}\,(\times1)$\strut} & 0 & 1 & (iii) & 1 \\
\net{bsn}$\otimes$\net{lvt} & dual & L & 12 & 8 & \parbox[t]{4.6cm}{\strut\raggedright $0$: 8$\to$0; $2$: 4$\to$0\strut} & \parbox[t]{1.1cm}{\strut\raggedright 0; 0\strut} & \parbox[t]{2.6cm}{\strut\raggedright $1.645\,(\times4)$$^*$\strut} & 0 & 0 & (iv) & 2 \\
\net{fcu}$\otimes$\net{flu} & dual & C & 9 & 8 & \parbox[t]{4.6cm}{\strut\raggedright $\tfrac{4}{3}$: 6$\to$0; $2$: 3$\to$0\strut} & \parbox[t]{1.1cm}{\strut\raggedright 0; 0\strut} & \parbox[t]{2.6cm}{\strut\raggedright $\tfrac{14}{9}\,(\times1)$$^*$\strut} & 0 & 0 & (iv) & 4 \\
\net{hxg-d}$\otimes$\net{hxg} & dual & L & 12 & 12 & \parbox[t]{4.6cm}{\strut\raggedright $0$: 8$\to$0; $2$: 4$\to$0\strut} & \parbox[t]{1.1cm}{\strut\raggedright 0; 4\strut} & \parbox[t]{2.6cm}{\strut\raggedright $2\,(\times4)$\strut} & 0 & 4 & (iii) & 1 \\
\net{nia}$\otimes$\net{nia-d} & dual & L & 12 & 12 & \parbox[t]{4.6cm}{\strut\raggedright $0$: 6$\to$0; $\tfrac{3}{2}$: 4$\to$0; $2$: 2$\to$0\strut} & \parbox[t]{1.1cm}{\strut\raggedright 0; 0; 2\strut} & \parbox[t]{2.6cm}{\strut\raggedright $2\,(\times2)$\strut} & 0 & 2 & (iii) & 1 \\
\end{longtable}
\endgroup

\textbf{Classifier.} Table~\ref{tab:full_retro} lists the 58 eligible pairs of the retrospective test, and Table~\ref{tab:full_prospective} the 88 eligible pairs of the two prospective tests together with the three route-U and route-C designations made outside the eligible set.

\begingroup\footnotesize
\begin{longtable}{llllll}
\caption{Retrospective classification of every eligible pair (rigid cage, all mechanisms lifted), counting each ordered pair once. Sets: D, dual pairs (Table~\ref{tab:full_atlas}); N, non-dual controls of Section~S3.3; Ud and Un, dual and non-dual cages of the unbiased control of Section~S3.4. Route: frozen-classifier route (--, abstention). Outcome: flat values with multiplicities, or the dispersive range of $\omega^2/\kc$, at 32 generic wavevectors of the common primitive cell. A pair is flat when the largest sorted-branch range is at most $10^{-8}\|S\|_2$. For a dispersive pair, $s$ is the spread of Tables~\ref{tab:screen} and \ref{tab:full_flat}, the mean branch range over the mean branch value, and worst is the largest branch range over the same mean. Because this cell is primitive and the sample differs, $s$ can differ from the conventional-cell value of Table~\ref{tab:screen} for the same pair. Both the designated and the observed multiplicities are per common primitive cell. Correct: the designation agrees with the outcome, an abstention counting as correct for a dispersive pair. Of the 58 pairs, 22 are flat and 22 of them are designated; 0 of the 36 dispersive pairs are designated.}\label{tab:full_retro} \\
\hline Pair & Sets & Route & Designated $\omega^2/\kc$ & First-order outcome & Correct \\ \hline\endfirsthead
\hline Pair & Sets & Route & Designated $\omega^2/\kc$ & First-order outcome & Correct \\ \hline\endhead
\hline\endfoot
\net{acs}$\otimes$\net{gra} & D & L + C & \parbox[t]{2.4cm}{\strut\raggedright $\tfrac{3}{4}\,(\times2)$, $2\,(\times2)$\strut} & \parbox[t]{5.0cm}{\strut\raggedright flat: $\tfrac{3}{4}\,(\times2)$, $2\,(\times2)$\strut} & yes \\
\net{bcu}$\otimes$\net{nbo} & D, Ud & L & \parbox[t]{2.4cm}{\strut\raggedright $2\,(\times3)$\strut} & \parbox[t]{5.0cm}{\strut\raggedright flat: $2\,(\times3)$\strut} & yes \\
\net{bsn}$\otimes$\net{lvt} & D & L & \parbox[t]{2.4cm}{\strut\raggedright $1.645\,(\times4)$\strut} & \parbox[t]{5.0cm}{\strut\raggedright flat: $1.645\,(\times4)$\strut} & yes \\
\net{fcu}$\otimes$\net{flu} & D, Ud & C & \parbox[t]{2.4cm}{\strut\raggedright $\tfrac{14}{9}$\strut} & \parbox[t]{5.0cm}{\strut\raggedright flat: $\tfrac{14}{9}$\strut} & yes \\
\net{gis-e}$\otimes$\net{lvt} & N & L & \parbox[t]{2.4cm}{\strut\raggedright $1.504\,(\times4)$\strut} & \parbox[t]{5.0cm}{\strut\raggedright flat: $1.504\,(\times4)$\strut} & yes \\
\net{hex}$\otimes$\net{bnn} & D, Ud & U & \parbox[t]{2.4cm}{\strut\raggedright $\tfrac{12}{7}$\strut} & \parbox[t]{5.0cm}{\strut\raggedright flat: $\tfrac{12}{7}$\strut} & yes \\
\net{hxg-d}$\otimes$\net{hxg} & D & L & \parbox[t]{2.4cm}{\strut\raggedright $2\,(\times4)$\strut} & \parbox[t]{5.0cm}{\strut\raggedright flat: $2\,(\times4)$\strut} & yes \\
\net{kag}$\otimes$\net{bnn} & N & U & \parbox[t]{2.4cm}{\strut\raggedright $\tfrac{12}{7}$\strut} & \parbox[t]{5.0cm}{\strut\raggedright flat: $\tfrac{12}{7}$\strut} & yes \\
\end{longtable}
\endgroup

\begingroup\scriptsize
\begin{longtable}{lrllllll}
\caption{Every eligible pair of the two prospective tests of the frozen classifier, whose source file has SHA-256 \texttt{cdb601e1c1c00d92\allowbreak b18a33aa3576f498\allowbreak 1b5535a88978aeb0\allowbreak 77a8e5e44c8711d7}. Test 1: 300 ordered pairs of distinct three-dimensional nets sharing a space-group setting, drawn uniformly (seed 20261121). Test 2: 1,000 further pairs, drawn uniformly within the cubic (334 pairs), hexagonal (333 pairs) and tetragonal (333 pairs) space groups (seed 20261131), excluding every pair screened before. In both tests the predictions were sealed before any spectrum was computed. Eligible: rigid cage that lifts every mechanism. Outcome: flat or dispersive at 32 generic wavevectors of the common primitive cell, and whether the classifier designated it; flatness, $s$ and worst are defined as in Table~\ref{tab:full_retro}. Flat values that differ beyond the grouping tolerance of $10^{-6}$ are printed with as many digits as separate them, as for the two bands of \net{tca}$\otimes$\net{hyb} at $1.93066\kc$ and $1.93073\kc$. The last three rows are the route-U and route-C designations of test 2, which fall outside the eligible set because the scaffold has $M_S$ generic mechanisms; their spectrum is that of the fixed scaffold and they are not scored. ID: index in the sealed sample.}\label{tab:full_prospective} \\
\hline Test & ID & Stratum & Pair & Route & Designated $\omega^2/\kc$ & Outcome & First-order spectrum \\ \hline\endfirsthead
\hline Test & ID & Stratum & Pair & Route & Designated $\omega^2/\kc$ & Outcome & First-order spectrum \\ \hline\endhead
\hline\endfoot
1 & 187 & uniform & \net{mob}$\otimes$\net{scp} & L & \parbox[t]{2.2cm}{\strut\raggedright $1.092\,(\times4)$, $1.333\,(\times2)$\strut} & flat, designated & \parbox[t]{4.6cm}{\strut\raggedright flat: $1.092\,(\times4)$, $1.333\,(\times2)$\strut} \\
1 & 263 & uniform & \net{xfs}$\otimes$\net{nfc} & L & \parbox[t]{2.2cm}{\strut\raggedright $0.134\,(\times8)$\strut} & flat, designated & \parbox[t]{4.6cm}{\strut\raggedright flat: $0.134\,(\times8)$\strut} \\
1 & 270 & uniform & \net{npo-e}$\otimes$\net{ybh} & L & \parbox[t]{2.2cm}{\strut\raggedright $5.708\,(\times2)$\strut} & flat, designated & \parbox[t]{4.6cm}{\strut\raggedright flat: $5.708\,(\times2)$\strut} \\
1 & 106 & uniform & \net{pdb}$\otimes$\net{xav} & -- & \parbox[t]{2.2cm}{\strut\raggedright --\strut} & flat, abstained & \parbox[t]{4.6cm}{\strut\raggedright flat: $4.59\times10^{-6}\,(\times3)$, $1.598$\strut} \\
1 & 206 & uniform & \net{ith-e}$\otimes$\net{urg} & -- & \parbox[t]{2.2cm}{\strut\raggedright --\strut} & flat, abstained & \parbox[t]{4.6cm}{\strut\raggedright flat: $0.085\,(\times6)$, $1.906\,(\times4)$\strut} \\
1 & 32 & uniform & \net{sxq}$\otimes$\net{lcs-a} & -- & \parbox[t]{2.2cm}{\strut\raggedright --\strut} & dispersive, abstained & \parbox[t]{4.6cm}{\strut\raggedright disp.: $0.000$--$0.620$, $s=0.077$, worst 0.613\strut} \\
1 & 37 & uniform & \net{tca}$\otimes$\net{fnr} & -- & \parbox[t]{2.2cm}{\strut\raggedright --\strut} & dispersive, abstained & \parbox[t]{4.6cm}{\strut\raggedright disp.: $0.140$--$0.824$, $s=0.214$, worst 0.584\strut} \\
1 & 43 & uniform & \net{mhg}$\otimes$\net{rtw} & -- & \parbox[t]{2.2cm}{\strut\raggedright --\strut} & dispersive, abstained & \parbox[t]{4.6cm}{\strut\raggedright disp.: $1.211$--$2.667$, $s=0.048$, worst 0.106\strut} \\
\end{longtable}
\endgroup

\textbf{Unbiased control.} Table~\ref{tab:full_h5} lists all 128 cages of Section~S3.4.

\begingroup\footnotesize
\begin{longtable}{lrllll}
\caption{Every row of the unbiased control of Section~S3.4. Each of the 32 floppy nets, drawn uniformly from the 78 nets with $\nu>0$ and a same-group dual (seed 20261111), is caged by its dual and by three nets of the same space group drawn uniformly. Each cage is listed with its generic mechanism count $M_S$ in parentheses and the flags R, rigid ($M_S=0$); L, every mechanism lifted; F, exactly flat at first order; --, none. Totals: dual cages 21 R, 12 L, 4 F of 32; non-dual cages 18 R, 15 L, 1 F of 96.}\label{tab:full_h5} \\
\hline Floppy net & $\nu_F$ & Dual cage & \multicolumn{3}{c}{Same-group non-dual cages} \\ \hline\endfirsthead
\hline Floppy net & $\nu_F$ & Dual cage & \multicolumn{3}{c}{Same-group non-dual cages} \\ \hline\endhead
\hline\endfoot
\net{bnn} & 1 & \net{hex} (0) RLF & \net{alc-a} (60) -- & \net{cel-a} (60) -- & \net{ltl} (36) -- \\
\net{bod} & 15 & \net{xbs} (0) R & \net{fud} (15) -- & \net{bor} (9) -- & \net{ftl} (24) -- \\
\net{cai} & 2 & \net{sve} (0) RL & \net{tbr} (0) RL & \net{mhg} (0) RL & \net{wxs-a} (192) -- \\
\net{cla} & 82 & \net{cla-d} (0) R & \net{smp} (0) RL & \net{ncb} (0) RL & \net{hcz} (10) -- \\
\net{ctn-x} & 20 & \net{thp} (0) RL & \net{fnn} (12) -- & \net{thp-a} (24) -- & \net{ctn-d} (60) -- \\
\net{epv} & 18 & \net{ild} (0) R & \net{cfe} (18) -- & \net{nlr} (0) RL & \net{ato-e} (0) RL \\
\net{eta} & 9 & \net{qzd} (3) -- & \net{qtz-g} (18) -- & \net{oso} (9) -- & \net{sml} (0) RL \\
\net{flu} & 4 & \net{fcu} (0) RLF & \net{xbk} (24) -- & \net{gdl-a} (504) -- & \net{dnc} (0) RL \\
\end{longtable}
\endgroup

\textbf{Finite stiffness and drive.} Table~\ref{tab:full_zerovol} gives the sampling intervals of the zone fractions of Table~\ref{tab:zerovol}, and Table~\ref{tab:full_gain} the gain statistics of every driven pair.

\begingroup\footnotesize
\begin{longtable}{lrrcrcr}
\caption{Uncertainty and composition of the zone fractions of Table~\ref{tab:zerovol}. 95\% CI: interval over the eight scrambled Sobol blocks of 1,024 wavevectors. $\Gamma$ pocket: part of the counted fraction that lies in the zone-centre pocket where the lifted band meets the acoustic branch, defined as $\rho\le2\sqrt{\kc/\ks}$. The pocket itself covers about 0.1\%, 3\% and 17\% of the zone at the three stiffness ratios. The column lists only the part of it that exceeds the threshold, which depends on the pair. Here $\rho$ is the distance from the wavevector to the nearest reciprocal-lattice point, in Cartesian coordinates of the primitive reciprocal basis (without the factor $2\pi$), divided by the cube root of the volume of the reciprocal cell, so that the radius scales as the pocket, as $\sqrt{\kc}$. Outside $\Gamma$: interval of the remainder. The last twelve rows, \net{fcu}$\otimes$\net{flu} and \net{she-d}$\otimes$\net{she}, were computed later with the same script and wavevectors. Weight: mean overlap of the assigned eigenstates with the mechanism projector.}\label{tab:full_zerovol} \\
\hline Pair & $\kc/\ks$ & $\delta$ (\%) & 95\% CI (\%) & in $\Gamma$ pocket (\%) & outside $\Gamma$, 95\% CI (\%) & weight \\ \hline\endfirsthead
\hline Pair & $\kc/\ks$ & $\delta$ (\%) & 95\% CI (\%) & in $\Gamma$ pocket (\%) & outside $\Gamma$, 95\% CI (\%) & weight \\ \hline\endhead
\hline\endfoot
\net{acs}$\otimes$\net{gra} & 0.001 & 1 & 36.9--37.9 & 0.09 & 36.8--37.8 & 0.991 \\
\net{acs}$\otimes$\net{gra} & 0.001 & 5 & 15.7--16.6 & 0.09 & 15.6--16.5 & 0.991 \\
\net{acs}$\otimes$\net{gra} & 0.01 & 1 & 99.9--100.0 & 3.41 & 96.4--96.7 & 0.973 \\
\net{acs}$\otimes$\net{gra} & 0.01 & 5 & 50.6--51.8 & 3.41 & 47.1--48.4 & 0.973 \\
\net{acs}$\otimes$\net{gra} & 0.03 & 1 & 100.0--100.0 & 17.54 & 82.1--82.8 & 0.951 \\
\net{acs}$\otimes$\net{gra} & 0.03 & 5 & 80.3--81.5 & 17.54 & 62.8--64.0 & 0.951 \\
\net{bcu}$\otimes$\net{nbo} & 0.001 & 1 & 5.1--5.8 & 0.11 & 5.0--5.7 & 1.000 \\
\net{bcu}$\otimes$\net{nbo} & 0.001 & 5 & 1.1--1.5 & 0.11 & 1.0--1.4 & 1.000 \\
\end{longtable}
\endgroup

\begingroup\footnotesize
\begin{longtable}{lrlrrrrl}
\caption{Linear gain of the parametric drive at $\Omega=2\omega_{\rm ref}$, $h=0.1$, $\ks=1$, no damping, for every pair and contact stiffness computed. $\omega_{\rm ref}$: flat first-order frequency of the band addressed (for \net{dia}$\otimes$\net{dia}, $\omega^2=2\kc$). Grow: wavevectors with $g_{\max}/\omega_{\rm ref}>10^{-5}$. Window: wavevectors within 4\% of $h/4$ (computed for the first four rows). Mean: the average of $g_{\max}/\omega_{\rm ref}$ divided by the single-mode Mathieu value $h/4$. The first four rows use 256 uniform wavevectors of the primitive zone, the others 256 scrambled Sobol wavevectors. \net{bcu}$\otimes$\net{nbo} at $\kc=0.001$ therefore appears twice, once on each wavevector set; the second row belongs to the stiffness series of the eight splitting pairs, and the two rows agree to the printed digits in the mean and differ in the quantiles only by sampling.}\label{tab:full_gain} \\
\hline Pair & $\kc$ & $\omega_{\rm ref}^2/\kc$ & $n_q$ & grow & window & mean$/(h/4)$ & $g_{\max}/\omega_{\rm ref}$: 5, 50, 95\% \\ \hline\endfirsthead
\hline Pair & $\kc$ & $\omega_{\rm ref}^2/\kc$ & $n_q$ & grow & window & mean$/(h/4)$ & $g_{\max}/\omega_{\rm ref}$: 5, 50, 95\% \\ \hline\endhead
\hline\endfoot
\net{bcu}$\otimes$\net{nbo} & 0.001 & $2$ & 256 & 256 & 255 & 0.999 & 0.02496, 0.02499, 0.02499 \\
\net{fcu}$\otimes$\net{flu} & 0.001 & $\tfrac{14}{9}$ & 256 & 256 & 252 & 0.996 & 0.02479, 0.02499, 0.02499 \\
\net{hxl}$\otimes$\net{hcb} & 0.001 & $\tfrac{3}{2}$ & 256 & 251 & 244 & 0.974 & 0.02418, 0.02498, 0.02499 \\
\net{dia}$\otimes$\net{dia} & 0.001 & $2$ & 256 & 91 & 21 & 0.256 & 0.00000, 0.00000, 0.02456 \\
\net{acs}$\otimes$\net{gra} & 0.001 & $\tfrac{3}{4}$ & 256 & 256 & -- & 0.998 & 0.02490, 0.02496, 0.02498 \\
\net{acs}$\otimes$\net{gra} & 0.001 & $2$ & 256 & 256 & -- & 0.999 & 0.02496, 0.02497, 0.02499 \\
\net{acs}$\otimes$\net{gra} & 0.01 & $\tfrac{3}{4}$ & 256 & 252 & -- & 0.948 & 0.02282, 0.02443, 0.02475 \\
\net{acs}$\otimes$\net{gra} & 0.01 & $2$ & 256 & 256 & -- & 0.988 & 0.02435, 0.02467, 0.02499 \\
\end{longtable}
\endgroup
\end{document}